\documentclass{aastex62}

\accepted{by \textit{Astrophysical Journal} in April, 2019}

\shorttitle{Droplets I}
\shortauthors{Chen et al.}

\begin{document}

\title{Droplets I: Pressure-Dominated Coherent Structures in L1688 and B18}

\correspondingauthor{Hope How-Huan Chen}
\email{hopechen@utexas.edu}

\author[0000-0001-6222-1712]{Hope How-Huan Chen}
\affil{Department of Astronomy, The University of Texas, Austin, TX 78712, USA}

\author{Jaime E. Pineda}
\affil{Max-Planck-Institut f\"ur extraterrestrische Physik, Giesenbachstrasse 1, D-85748 Garching, Germany}

\author{Alyssa A. Goodman}
\affil{Harvard-Smithsonian Center for Astrophysics, 60 Garden St., Cambridge, MA 02138, USA}

\author{Andreas Burkert}
\affil{University Observatory Munich (USM), Scheinerstrasse 1, 81679 Munich, Germany}

\author{Stella S. R. Offner}
\affil{Department of Astronomy, The University of Texas, Austin, TX 78712, USA}

\author{Rachel K. Friesen}
\affil{National Radio Astronomy Observatory, 520 Edgemont Rd., Charlottesville, VA, 22903, USA}

\author{Philip C. Myers}
\affil{Harvard-Smithsonian Center for Astrophysics, 60 Garden St., Cambridge, MA 02138, USA}

\author{Felipe Alves}
\affil{Max-Planck-Institut f\"ur extraterrestrische Physik, Giesenbachstrasse 1, D-85748 Garching, Germany}

\author{H\'ector G. Arce}
\affil{Department of Astronomy, Yale University, P.O. Box 208101, New Haven, CT 06520-8101, USA}

\author{Paola Caselli}
\affil{Max-Planck-Institut f\"ur extraterrestrische Physik, Giesenbachstrasse 1, D-85748 Garching, Germany}

\author{Ana Chac\'on-Tanarro}
\affil{Max-Planck-Institut f\"ur extraterrestrische Physik, Giesenbachstrasse 1, D-85748 Garching, Germany}

\author{Michael Chun-Yuan Chen}
\affil{Department of Physics and Astronomy, University of Victoria, 3800 Finnerty Road, Victoria, BC V8P 5C2, Canada}

\author{James Di Francesco}
\affil{Department of Physics and Astronomy, University of Victoria, 3800 Finnerty Road, Victoria, BC V8P 5C2, Canada}
\affil{Herzberg Astronomy and Astrophysics, National Research Council of Canada, 5071 West Saanich Road, Victoria, BC V9E 2E7, Canada}

\author{Adam Ginsburg}
\affil{National Radio Astronomy Observatory, Socorro, NM 87801, USA}

\author{Jared Keown}
\affil{Department of Physics and Astronomy, University of Victoria, 3800 Finnerty Road, Victoria, BC V8P 5C2, Canada}

\author{Helen Kirk}
\affil{Department of Physics and Astronomy, University of Victoria, 3800 Finnerty Road, Victoria, BC V8P 5C2, Canada}
\affil{Herzberg Astronomy and Astrophysics, National Research Council of Canada, 5071 West Saanich Road, Victoria, BC V9E 2E7, Canada}

\author{Peter G. Martin}
\affil{Canadian Institute for Theoretical Astrophysics, University of Toronto, 60 St.\ George St., Toronto, ON M5S 3H8, Canada}

\author{Christopher Matzner}
\affil{Department of Astronomy \& Astrophysics, University of Toronto, 50 St.\ George Street, Toronto, ON M5S 3H4, Canada}

\author{Anna Punanova}
\affil{Ural Federal University, 620002, 19 Mira Street, Yekaterinburg, Russia}

\author{Elena Redaelli}
\affil{Max-Planck-Institut f\"ur extraterrestrische Physik, Giesenbachstrasse 1, D-85748 Garching, Germany}

\author{Erik Rosolowsky}
\affil{Department of Physics, 4-181 CCIS, University of Alberta, Edmonton, AB T6G 2E1, Canada}

\author{Samantha Scibelli}
\affil{Steward Observatory, 933 North Cherry Avenue, Tucson, AZ 85721, USA}

\author{Youngmin Seo}
\affil{Jet Propulsion Laboratory, NASA, 4800 Oak Grove Drive, Pasadena, CA 91109, USA}

\author{Yancy Shirley}
\affil{Steward Observatory, 933 North Cherry Avenue, Tucson, AZ 85721, USA}

\author{Ayushi Singh}
\affil{Department of Astronomy \& Astrophysics, University of Toronto, 50 St.\ George Street, Toronto, ON M5S 3H4, Canada}

\collaboration{(The GAS Collaboration)}

\begin{abstract}


We present the observation and analysis of newly discovered coherent structures in the L1688 region of Ophiuchus and the B18 region of Taurus.  Using data from the Green Bank Ammonia Survey \citep[GAS;][]{GAS_DR1}, we identify regions of high density and near-constant, almost-thermal, velocity dispersion.  Eighteen coherent structures are revealed, twelve in L1688 and six in B18, each of which shows  a sharp ``transition to coherence'' in velocity dispersion around its periphery.  The identification of these structures provides a chance to study the coherent structures in molecular clouds statistically.  The identified coherent structures have a typical radius of 0.04 pc and a typical mass of 0.4 M$_\sun$, generally smaller than previously known coherent cores identified by  \citet{Goodman_1998}, \citet{Caselli_2002}, and \citet{Pineda_2010}.  We call these structures ``droplets.''  We find that unlike previously known coherent cores, these structures are not virially bound by self-gravity and are instead predominantly confined by ambient pressure.  The droplets have density profiles shallower than a critical Bonnor-Ebert sphere, and they have a velocity ($V_\mathrm{LSR}$) distribution consistent with the dense gas motions traced by NH$_3$ emission.  These results point to a potential formation mechanism through pressure compression and turbulent processes in the dense gas.  We present a comparison with a magnetohydrodynamic simulation of a star-forming region, and we speculate on the relationship of droplets with larger, gravitationally bound coherent cores, as well as on the role that droplets and other coherent structures play in the star formation process.

\end{abstract}

\keywords{ISM: clouds --- ISM: kinematics and dynamics --- ISM: structure --- stars: formation --- radio lines: ISM --- magnetohydrodynamics (MHD) --- ISM: individual (L1688, B18)}

\section{Introduction}
\label{sec:intro}
In the early 1980s, NH$_3$ was identified as an excellent tracer of the cold, dense gas associated with highly extinguished compact regions.  These regions were named ``dense cores'' by \citet{Myers_1983a}, and their properties were studied and documented in a series of papers throughout the 1980s and 1990s whose titles began with ``Dense Cores in Dark Clouds'' \citep{Myers_1983a,Myers_1983b,Myers_1983c,Benson_1983,Fuller_1992,Goodman_1993,Benson_1998,Caselli_2002}.  Since the start of that series, astronomers have used the ``dense core'' paradigm as a way to think about the small \citep[0.1 pc, with the smallest being $\sim$ 0.03 pc;][]{Myers_1983b, Jijina_1999}, prolate but roundish \citep[aspect ratio near 2;][]{Myers_1991}, quiescent \citep[velocity dispersion nearly thermal;][]{Fuller_1992}, blobs of gas that can form stars like the Sun.  Whether these cores also exist in clusters where more massive stars form \citep{Evans_1999,Garay_1999,Tan_2006,Li_2015}, how long-lived and/or transient these cores might be \citep{Bertoldi_1992,BallesterosParedes_1999,Elmegreen_2000,Enoch_2008}, and how they relate to the ubiquitous filamentary structure inside star-forming regions \citep{McKee_2007,Andre_2014,Padoan_2014,Hacar_2013,Tafalla_2015} are still open questions.  Nonetheless, a gravitationally collapsing ``dense core'' remains the central theme in discussions of star-forming material.


\citet{Barranco_1998} made observations of NH$_3$ hyperfine line emission of four ``dense cores'' and found that the linewidths in the interior of a dense core are roughly constant at a value slightly higher than a purely thermal linewidth, and that the linewidths start to increase near the edge of the dense core.  Using observations of OH and C$^{18}$O line emission, \citet{Goodman_1998} proposed a characteristic radius where the scaling law between the linewidth and the size changes, marking the ``transition to coherence.''  \citet{Goodman_1998} found that the characteristic radius is $\sim$ 0.1 pc and that within $\sim$ 0.1 pc from the center of a dense core the linewidth is virtually constant.  This gave birth to the idea of the existence of ``coherent cores'' at the densest part of previously identified ``dense cores.''  The coherence is defined by a transition from supersonic to subsonic turbulent velocity dispersion which is found to accompany a sharp change in the scaling law between the velocity dispersion and the size scale.  \citet{Goodman_1998} hypothesized that the coherent core provides the needed ``calmness,'' or low turbulence, environment for further star formation dominated by gravitational collapse.


Using GBT observations of NH$_3$ hyperfine line emission, \citet{Pineda_2010} made the first direct observation of a coherent core, resolving the transition to coherence across the boundary from a ``Larson's Law''-like (turbulent) regime to a coherent (thermal) one.  The observed coherent core sits in the B5 region in Perseus and has an elongated shape with a characteristic radius of $\sim$ 0.2 pc.  The interior linewidths are almost constant and subsonic but are not purely thermal.  Later VLA observations by \citet{Pineda_2011} of the interior of B5 show that there are finer structures inside the coherent core, and \citet{Pineda_2015} found that these sub-structures are forming stars in a free-fall time of $\sim$ 40,000 years.  The gravitationally collapsing sub-structures inside the coherent core are consistent with the picture of star formation within the ``calmness'' of a coherent core.

The coherent core in B5 has remained the only known example where the transition to coherence is spatially resolved with a single tracer.  In search of other coherent structures in nearby molecular clouds, we follow the same procedure adopted by \citet{Pineda_2010} and identify a total of 18 coherent structures, 12 in the L1688 region in Ophiuchus and 6 in the B18 region in Perseus, using data from the Green Bank Ammonia Survey \citep[GAS;][]{GAS_DR1}.  Although many of these structures may be associated with previously known cores or density features, this is the first time ``transitions to coherence'' are captured using a single tracer.  The 18 coherent structures identified within a total projected area on the plane of the sky of $\sim$ 0.6 pc$^2$ suggest the ubiquity of coherent structures in nearby molecular clouds.  This catalogue allows statistical analyses of coherent structures for the first time.

In the analyses presented in this paper, we find that these newly identified coherent structures have small sizes, $\sim$ 0.04 pc, and masses, $\sim$ 0.4 M$_\sun$\footnote{Like many of the dense cores observed by \citet{Myers_1983c}, a coherent region has a thermally dominated velocity dispersion.  The identification of these coherent structures are ``new'' in the sense that ``transitions to coherence'' are captured in a single tracer for the first time for many of these structures and that the identified coherent structures form a previously omitted population of \emph{gravitationally unbound and pressure confined coherent structures}, as shown in the analyses below.  We acknowledge that many of the coherent structures examined in this paper might be associated with previously known cores or density features.  See Appendix \ref{sec:appendix_overlap} for discussion.}.  Unlike previously known coherent cores, the coherent structures identified in this paper are mostly gravitationally unbound and are instead predominantly bound by pressure provided by the ambient gas motions, in spite of the subsonic velocity dispersions found in these structures\footnote{In this paper, the adjectives ``supersonic,'' ``transonic,'' and ``subsonic'' indicate levels of turbulence.  A supersonic/transonic/subsonic velocity dispersion has a turbulent (non-thermal) component larger than/comparable to/smaller than the sonic velocity.  See Equation \ref{eq:sigmaTot} below for a definition of the thermal and non-thermal components of velocity dispersion}.  We term this newly discovered population of \emph{gravitationally unbound and pressure confined coherent structures} ``droplets'' and examine their relation to the known gravitationally bound and likely star-forming coherent cores and other dense cores.



In this paper, we present a full description of the physical properties of the droplets and discuss their potential formation mechanism.  In \S\ref{sec:data}, we describe the data used in this paper, including data from the GAS DR1 \citep[\S\ref{sec:data_GAS};][]{GAS_DR1}, maps of column density and dust temperature based on SED fitting of observations made by the Herschel Gould Belt Survey \citep[\S\ref{sec:data_Herschel};][]{Andre_2010}, and the catalogues of previously known NH$_3$ cores \citep[\S\ref{sec:data_catalogs};][]{Goodman_1993,Pineda_2010}.  In \S\ref{sec:analysis}, we present our analysis of the droplets, including their identification (\S\ref{sec:analysis_id}), basic properties (\S\ref{sec:analysis_basic}), and a virial analysis including an ambient gas pressure term (\S\ref{sec:analysis_virial}).  In the discussion, we further examine the nature of their pressure confinement in \S\ref{sec:discussion_confinement}, by comparing the radial density and pressure profiles to the Bonnor-Ebert model (\S\ref{sec:discussion_confinement_BE}) and the logotropic spheres (\S\ref{sec:discussion_confinement_logo}).  We examine the relation between the droplets and the host molecular cloud by looking into the velocity distributions (\S\ref{sec:discussion_confinement_PPV}).  We then demonstrate that formation of droplets is possible in a magnetohydrodynamic (MHD) simulation and speculate on the formation mechanism of the droplets in \S\ref{sec:discussion_simulation}, and we discuss their relation to coherent cores and their evolution in \S\ref{sec:discussion_definition}.  Lastly in \S\ref{sec:conclusion}, we summarize this work and outline future projects that might shed more light on how droplets form, their relationship with structures at different size scales, and the role they might play in star formation.

\section{Data}
\label{sec:data}

\subsection{Green Bank Ammonia Survey (GAS)}
\label{sec:data_GAS}
The Green Bank Ammonia Survey \citep[GAS;][]{GAS_DR1} is a Large Program at the \textit{Green Bank Telescope} (GBT) to map most Gould Belt star-forming regions with A$_\mathrm{V}$ $\geq$ 7 mag visible from the northern hemisphere in emission from NH$_3$ and other key molecules\footnote{The data from the first data release are public and can be found at \url{https://dataverse.harvard.edu/dataverse/GAS_Project}.}.  The data used in this work are from the first data release (DR1) of GAS that includes four nearby star-forming regions: L1688 in Ophiuchus, B18 in Taurus, NGC1333 in Perseus, and Orion A.

To achieve better physical resolution, only the two closest regions in the GAS DR1 are used in our present study.  L1688 in Ophiuchus sits at a distance of $137.3\pm6$ pc \citep{OrtizLeon_2017}, and B18 in Taurus sits at a distance of $126.6\pm1.7$ pc \citep[notice this is updated from the distance adopted by \citealt{GAS_DR1}, which was taken from \citealt{Schlafly_2014};][]{Galli_2018}.  At these distances, the GBT FWHM beam size of 32\arcsec at 23 GHz corresponds to $\sim$ 4350 AU (0.02 pc).  The GBT beam size at 23 GHz also matches well with the Herschel SPIRE 500 $\micron$ FWHM beam size of 36\arcsec \citep[see \S\ref{sec:data_Herschel} and discussions in][]{GAS_DR1}.  The GBT observations have a spectral resolution of 5.7 kHz, or $\sim$ 0.07 km s$^{-1}$ at 23 GHz.

\subsubsection{Fitting the NH$_3$ Line Profile}
\label{sec:data_GAS_fitting}
\begin{figure}[ht!]
\plotone{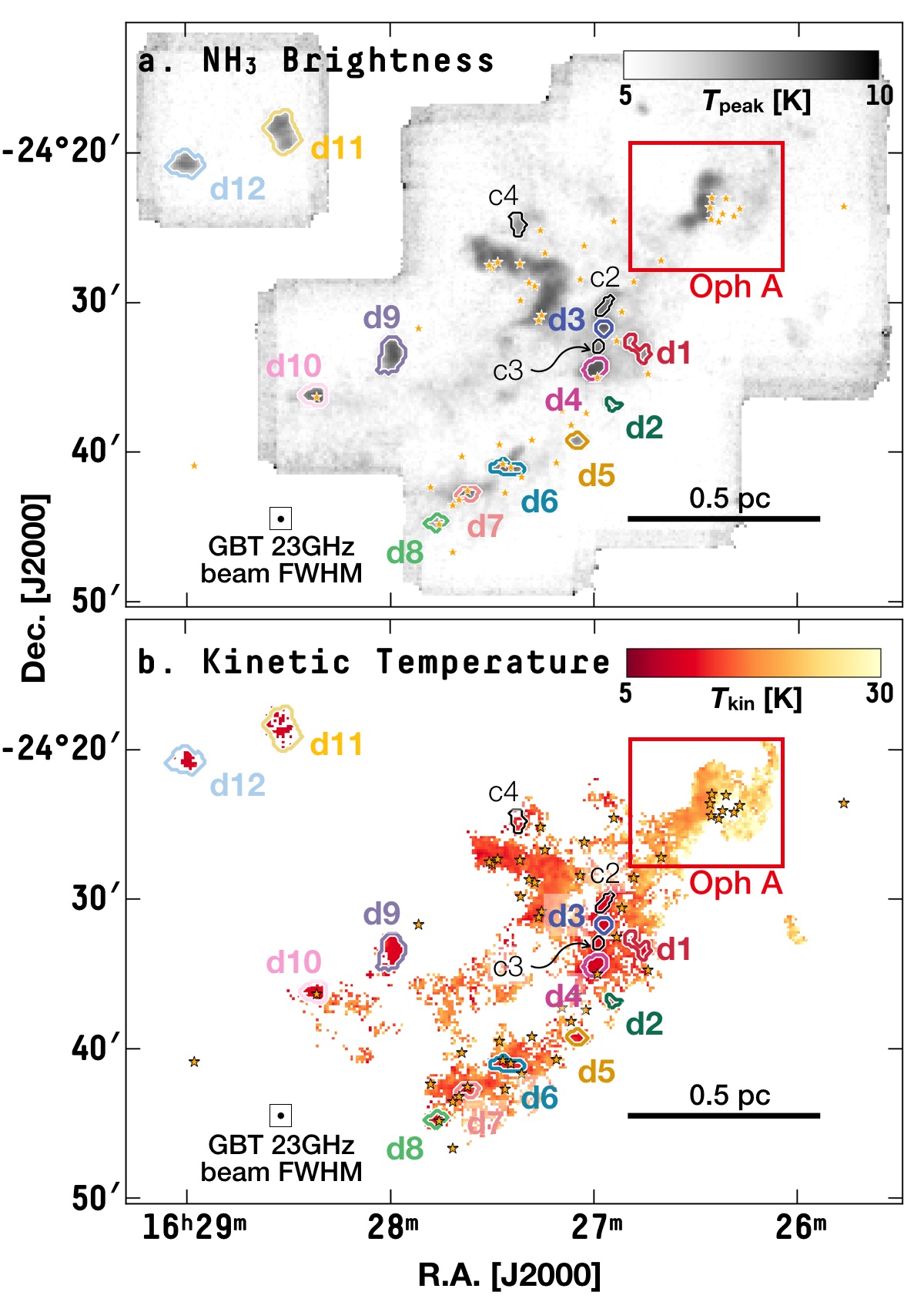}
\caption{\label{fig:L1688_TpeakTkin} L1688 in Ophiuchus: Maps of \textbf{(a)} peak NH$_3$ (1, 1) brightness in the unit of main-beam temperature, $T_\mathrm{peak}$, and \textbf{(b)} kinetic temperature, $T_\mathrm{kin}$.  The colored contours mark the boundaries of droplets, and the black contours mark the boundaries of droplet candidates.  Because L1688-c1E and L1688-c1W overlap with L1688-d1, they are not shown in this figure or Fig.\ \ref{fig:L1688_VlsrSigma} (see \S\ref{sec:analysis_id}).  The stars mark the positions of Class 0/I and flat-spectrum protostars from \citet{Dunham_2015}.  The scale bar at the bottom right corner corresponds to 0.5 pc at the distance of Ophiuchus.  The black circle at the bottom left corner of each panel shows the beam FWHM of the GBT observations at 23 GHz.  See Appendix \ref{sec:appendix_gallery} for a gallery of the close-up views of the droplets.}
\end{figure}

\begin{figure}[ht!]
\plotone{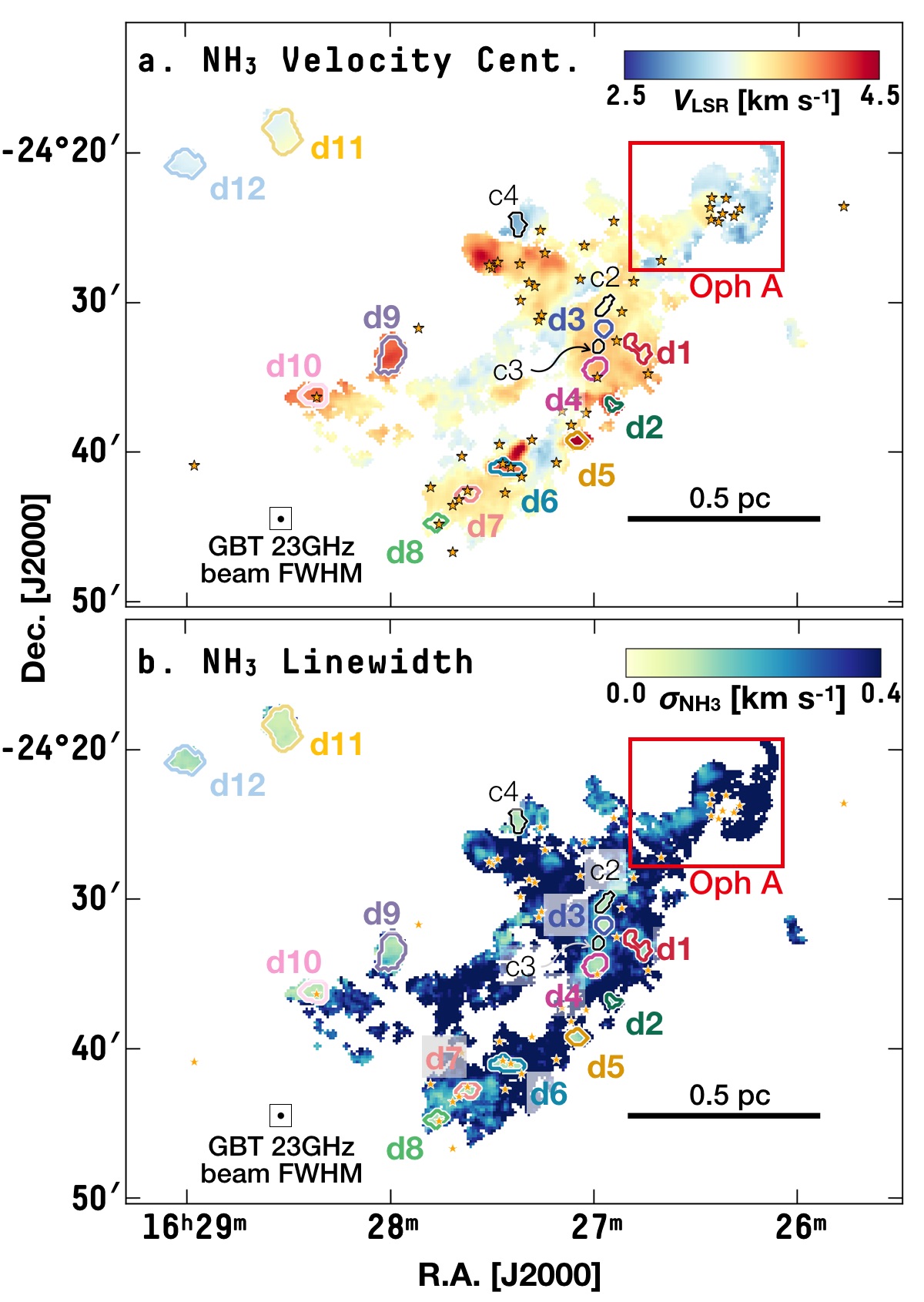}
\caption{\label{fig:L1688_VlsrSigma} Like Fig.\ \ref{fig:L1688_TpeakTkin} but for maps of \textbf{(a)} velocity centroid, $V_\mathrm{LSR}$, and \textbf{(b)} velocity dispersion, $\sigma_{\mathrm{NH}_3}$.}
\end{figure}

\begin{figure}[ht!]
\plotone{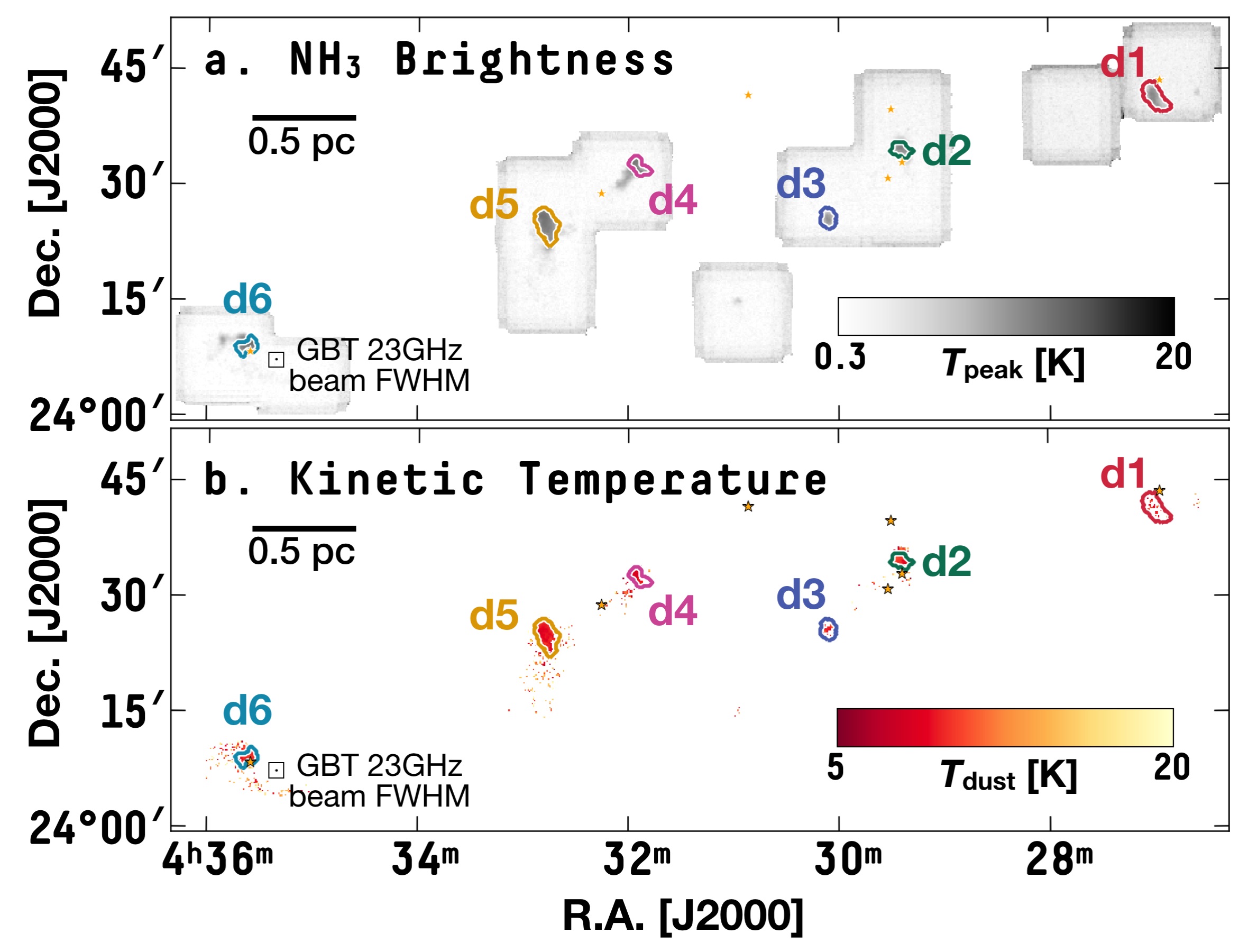}
\caption{\label{fig:B18_TpeakTkin} Like Fig.\ \ref{fig:L1688_TpeakTkin} but for B18 in Taurus, showing maps of \textbf{(a)} peak NH$_3$ (1, 1) brightness in the unit of main-beam temperature, $T_\mathrm{peak}$, and \textbf{(b)} kinetic temperature, $T_\mathrm{kin}$.  Here, the stars mark the positions of Class 0/I and flat-spectrum protostars with a reliability grade of A- or higher from \citet{Rebull_2010}.  The scale bar at the bottom right corner corresponds to 0.5 pc at the distance of Taurus.}
\end{figure}

\begin{figure}[ht!]
\plotone{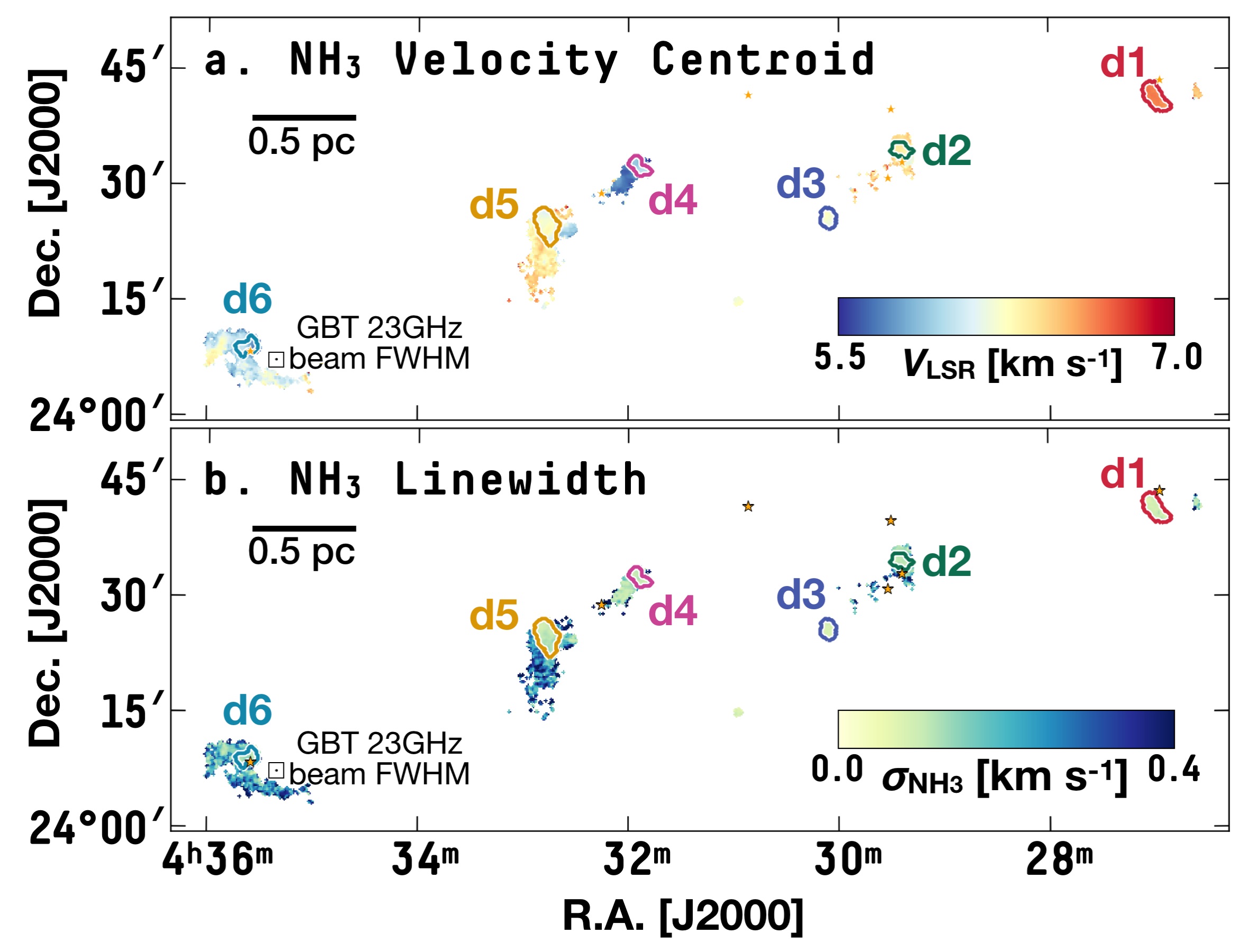}
\caption{\label{fig:B18_VlsrSigma} Like Fig.\ \ref{fig:B18_TpeakTkin} but for maps of \textbf{(a)} velocity centroid, $V_\mathrm{LSR}$, and \textbf{(b)} velocity dispersion, $\sigma_{\mathrm{NH}_3}$.}
\end{figure}

In the GAS DR1, a (single) Gaussian line shape is assumed in fitting spectra of NH$_3$ (1, 1) and (2, 2) hyperfine line emission \citep[see \S3.1 in][]{GAS_DR1}.  The fitting is carried out using the ``cold-ammonia'' model and a forward-modeling approach in the \texttt{PySpecKit} package \citep{pyspeckit}, which was developed in \citet{GAS_DR1} and built upon the results from \citet{Rosolowsky_2008a} and \citet{Friesen_2009} in the theoretical framework laid out by \citet{Mangum_2015}.  No fitting of multiple velocity components or non-Gaussian profiles was attempted in GAS DR1, but the single-component fitting produced good quality results in $\gtrsim$ 95\% of detections in all regions included in the GAS DR1.  From the fit, we obtain the velocity centroid of emission along each line of sight (Gaussian mean of the best fit) and the velocity dispersion (Gaussian $\sigma$), where we have sufficient signal-to-noise in NH$_3$ (1, 1) emission.  For lines of sight where we detect both NH$_3$ (1, 1) and (2, 2), the model described in \citet{GAS_DR1} provides estimates of parameters including the kinetic temperature and the NH$_3$ column density.  Figs.\ \ref{fig:L1688_TpeakTkin} to \ref{fig:B18_VlsrSigma} show the parameters derived from the fitting of the NH$_3$ hyperfine line profiles.

\subsection{Herschel Column Density Maps}
\label{sec:data_Herschel}
\begin{figure}[ht!]
\plotone{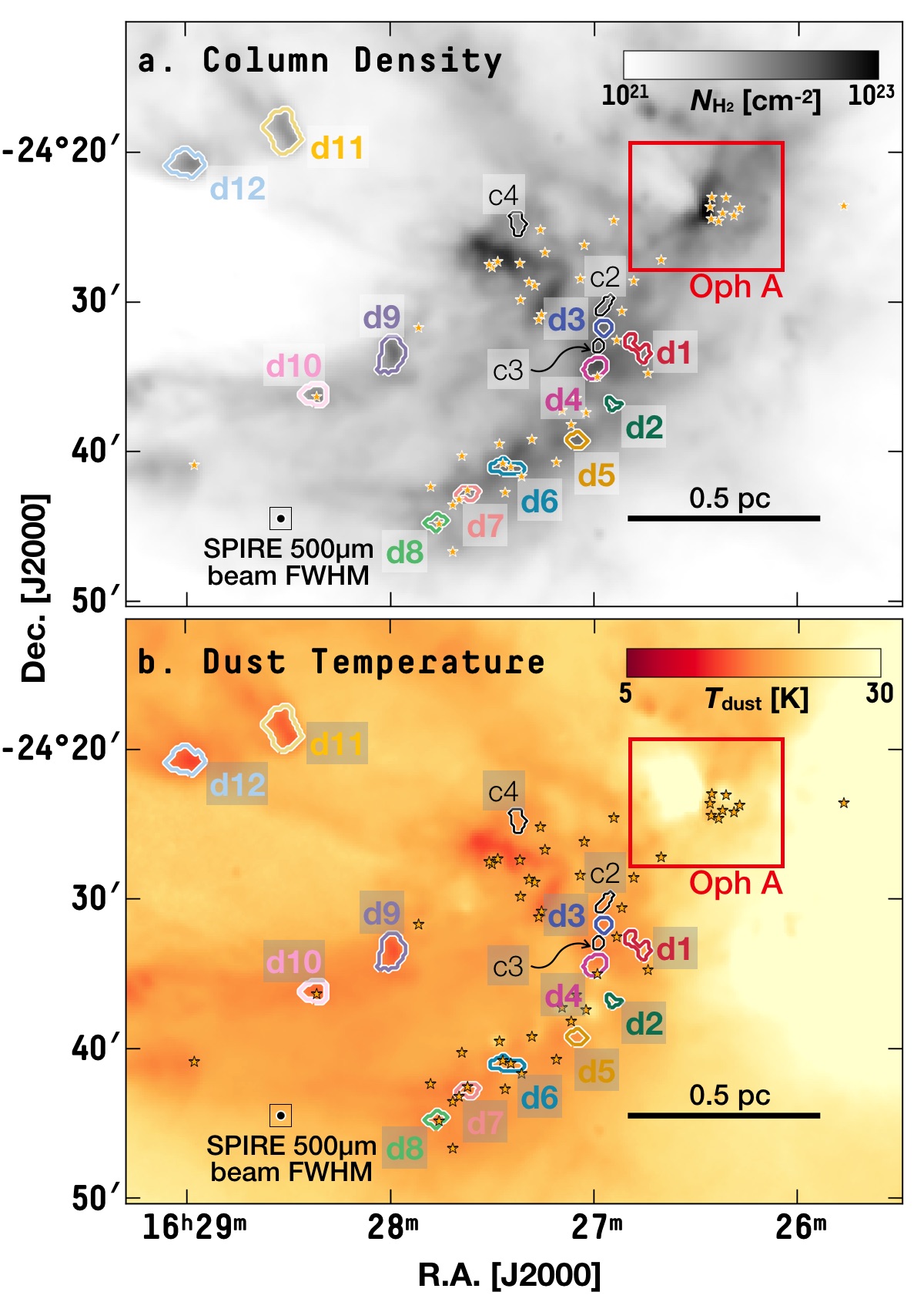}
\caption{\label{fig:L1688_Herschel} Like Fig.\ \ref{fig:L1688_TpeakTkin} but for maps of \textbf{(a)} total column density, $N_{\mathrm{H}_2}$, and \textbf{(b)} dust temperature, $T_\mathrm{dust}$, derived from Herschel observations.}
\end{figure}

\begin{figure}[ht!]
\plotone{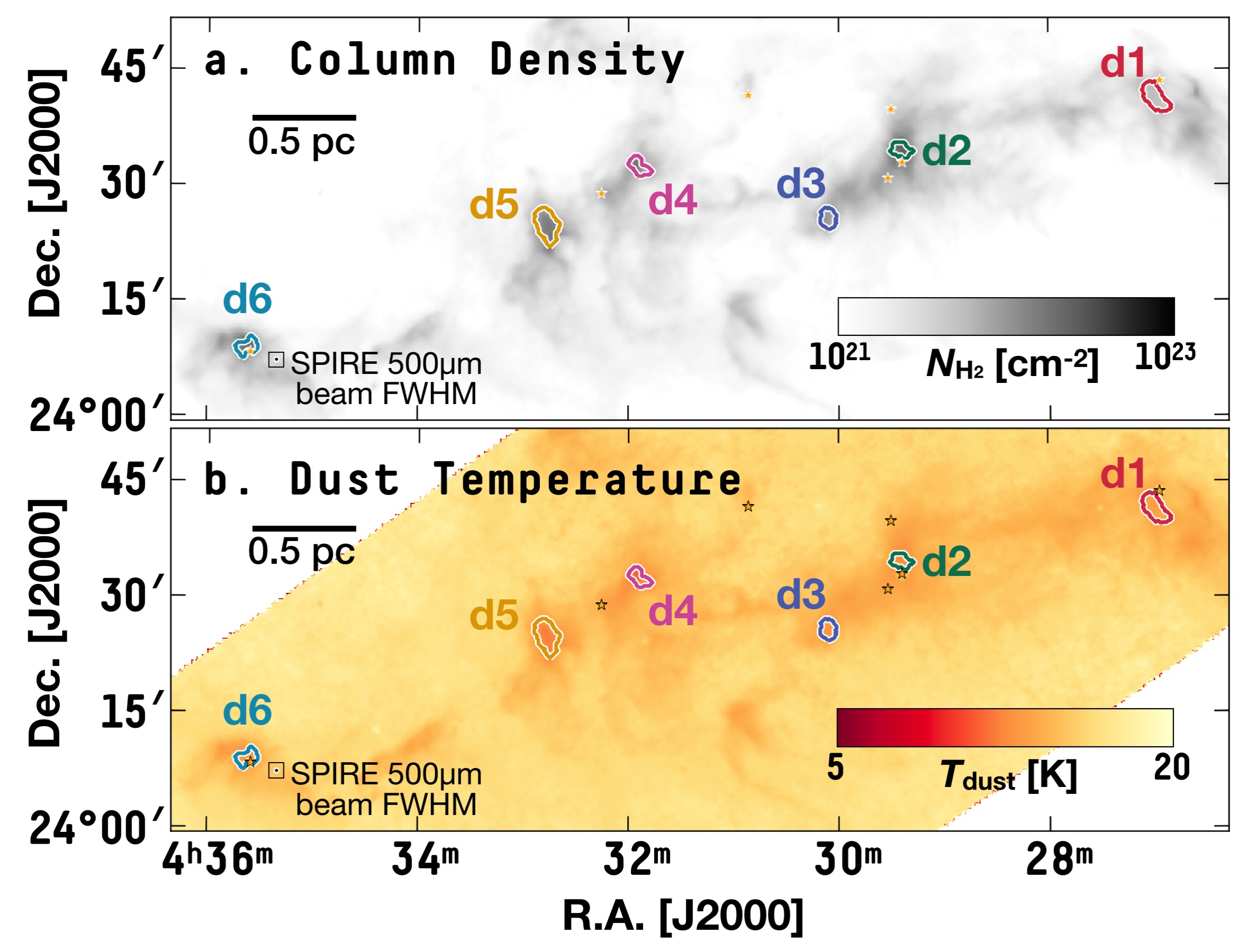}
\caption{\label{fig:B18_Herschel} Like Fig.\ \ref{fig:B18_TpeakTkin} but for maps of \textbf{(a)} total column density, $N_{\mathrm{H}_2}$, and \textbf{(b)} dust temperature, $T_\mathrm{dust}$, derived from Herschel observations.}
\end{figure}

The \textit{Herschel} column density maps are derived from archival Herschel PACS 160 and SPIRE 250/350/500 $\micron$ observations of dust emission, observed as part of the Herschel Gould Belt Survey \citep[HGBS][]{Andre_2010}.  We establish the zero point of emission at each wavelength using \textit{Planck} observations of the same regions \citep{PlanckXI}.  The emission maps are then convolved to match the SPIRE 500 $\micron$ beam FWHM of 36\arcsec and passed to a least squares fitting routine, where we assume that the emission at these wavelengths follow a modified blackbody emission function, $I_\nu = (1-\exp^{-\tau_\nu}) B_\nu(T)$, where $B_\nu(T)$ is the blackbody radiation, and $\tau$ is the frequency-dependent opacity.  The opacity can be written as a function of the mass column density, $\tau_\nu = \kappa_\nu \Sigma$, where $\kappa_\nu$ is the opacity coefficient.  At these wavelengths, $\kappa_\nu$ can be described by a power-law function of frequency, $\kappa_\nu = \kappa_{\nu_0} \left(\frac{\nu}{\nu_0}\right)^\beta$, where $\beta$ is the emissivity index, and $\kappa_{\nu_0}$ is the opacity coefficient at frequency $\nu_0$.  Here we adopt $\kappa_{\nu_0}$ of 0.1 cm$^2$ g$^{-1}$ at $\nu_0$ = 1000 GHz \citep{Hildebrand_1983} and a fixed $\beta$ of 1.62 \citep{PlanckXI}.  The resulting $I_\nu$ is a function of the temperature and the dust column density, the latter of which can be further converted to the total number column density by assuming a dust-to-gas ratio (100, for the maps we derive) and defining a mean molecular weight\footnote{In this paper, we use the mean molecular weight per H$_2$ molecule (2.8 u; $\mu_{\mathrm{H}_2}$ in \citealt{Kauffmann_2008}) in the calculation of the mass and other density related quantities, and we use the mean molecular weight \emph{per free particle} (2.37 u; $\mu_\mathrm{p}$ in \citealt{Kauffmann_2008}) in the calculation of the velocity dispersion and pressure.  Both numbers are derived assuming a hydrogen mass ratio of M$_\mathrm{H}$/M$_\mathrm{total}$ $\approx$ 0.71, a helium mass ratio of M$_\mathrm{He}$/M$_\mathrm{total}$ $\approx$ 0.27, and a metal mass ratio of M$_\mathrm{Z}$/M$_\mathrm{total}$ $\approx$ 0.02 \citep{Cox_2000}.  See Appendix A.1 in \citet{Kauffmann_2008}.} \citep[2.8 u; $\mu_{\mathrm{H}_2}$ in][]{Kauffmann_2008}.  The resulting column density map has an angular resolution of 36\arcsec (the SPIRE 500 $\micron$ beam FWHM), which matches well with the GBT beam FWHM at 23 GHz (32\arcsec).  In the following analyses, we do not apply convolution to further match the resolutions of the Herschel and GBT observations, before regridding the maps onto the same projection and gridding (Nyquist-sampled).   Resulting maps column density and dust temperature are shown in Fig.\ \ref{fig:L1688_Herschel} and Fig.\ \ref{fig:B18_Herschel} for L1688 in Ophiuchus and B18 in Taurus, respectively.

\subsection{Source Catalogs}
\label{sec:data_catalogs}
To understand \textit{droplets} in context, we need compilations of the physical properties of previously identified dense cores.  \citet{Goodman_1993} (see \S\ref{sec:data_catalogs_Goodman93}) present a summary of cores from the observational surveys described in \citet{Benson_1989} and \citet{Ladd_1994}.  The cores in \citet{Goodman_1993} have low, nearly thermal velocity dispersions, and some of them are known to be ``coherent'' based on an apparent abrupt spatial transition from supersonic (in OH and C$^{18}$O) to subsonic (in NH$_3$) velocity dispersion \citep{Goodman_1998,Caselli_2002}.  We also include the coherent core in the B5 region in Perseus, as observed in NH$_3$ \citep{Pineda_2010}, the only coherent structure known before this work where the spatial change in linewidth is captured in a single tracer.

\subsubsection{Dense Cores Measured in NH$_3$}
\label{sec:data_catalogs_Goodman93}
\citet{Goodman_1993} presented a survey of 43 sources with observations of NH$_3$ line emission (see Table 1 and Table 2 in \citealt{Goodman_1993}; see also the \href{http://simbad.harvard.edu/simbad/sim-ref?querymethod=bib&simbo=on&submit=submit+bibcode&bibcode=1993ApJ...406..528G}{SIMBAD object list}), based on observations made by \citet{Benson_1989} and \citet{Ladd_1994}.  The observations were carried out at the 37 m telescope of the Haystack Observatory and the 43 m telescope of the National Radio Astronomy Observatory (NRAO), resulting in a spatial resolution coarser than the modern GBT observations by a factor of $\sim$ 2.5.  The velocity resolution of observations done by \citet{Benson_1989} and \citet{Ladd_1994} ranges from 0.07 to 0.20 km s$^{-1}$.  For comparison with the kinematic properties of the droplets measured using the GAS observations of NH$_3$ emission \citep{GAS_DR1}, we adopt values that were also measured using observations of NH$_3$ hyperfine line emission, presented by \citet{Goodman_1993}.  We correct the physical properties summarized in \citet{Goodman_1993} with the modern measurement of the distance to each region.  The updated distances are summarized in Appendix \ref{sec:appendix_distances}.

The updated distances affect the physical properties listed in Table 1 in \citet{Goodman_1998}.  The size scales with the distance, $D$, by a linear relation, $R \propto D$.  Since the mass was calculated from the number density derived from NH$_3$ hyperfine line fitting, it scales with the volume of the structure, and thus $M \propto D^3$.  The updated distances also affect the velocity gradient and related quantities listed in Table 1 and Table 2 in \citet{Goodman_1998}, which we do not use for the analyses presented in this work.

Besides the updated distances, we combine the measurements of the kinetic temperature and the NH$_3$ linewidth, originally presented by \citet{Benson_1989} and \citet{Ladd_1994}, to derive the thermal and the non-thermal components of the velocity dispersion.  See Equation \ref{eq:sigmaTot} below for the definitions of the velocity dispersion components.

Among the 43 sources examined by \citet{Goodman_1993}, eight sources were later confirmed by \citet{Goodman_1998} and/or \citet{Caselli_2002} to be ``coherent cores,'' using a combination of gas tracers of various critical densities (OH, C$^{18}$O, NH$_3$, and N$_2$H$^{+}$).  The interiors of these eight sources show signs of a uniform and nearly thermal distribution of velocity dispersion.  However, unlike B5 and the newly identified coherent structures in this paper, the ``transition to coherence'' was not spatially resolved with a single tracer for these eight coherent cores.  For the ease of discussion, we refer to the entire sample of 43 sources as the ``dense cores,'' as they were originally referred to by \citet{Goodman_1993}.  However, note that some of the 43 sources have masses and sizes up to $\sim$ 100 M$_\sun$ and $\sim$ 1 pc, respectively.  These larger-scale structures do not strictly fit in the definition of a dense core (with a small size and a nearly thermal velocity dispersion; see \S\ref{sec:discussion_definition} for more discussions) and might be better categorized as ``dense clumps'' \citep[as in][]{McKee_2007}.

\subsubsection{Coherent Core in B5}
\label{sec:data_catalogs_Pineda10}
Using GBT observations of NH$_3$ hyperfine line emission with a setup similar to GAS, \citet{Pineda_2010} observed a coherent core in the B5 region in Perseus and spatially resolved the ``transition to coherence''---NH$_3$ linewidths changing from supersonic values outside the core to subsonic values inside---for the first time.  The coherent core sits in the eastern part of the molecular cloud in Perseus, at a distance of $315\pm 32$ pc \citep[the quantities measured by \citealt{Pineda_2010} assuming a distance of 250 pc are updated according to the new distance measurement;][]{Schlafly_2014}.  At 315 pc, the GBT resolution at 23 GHz corresponds to a spatial resolution of $\sim$ 0.05 pc.  The coherent core has an elongated shape, with a size of $\sim$ 0.2 pc.

\citet{Pineda_2010} identified the coherent core in B5 as a peak in NH$_3$ brightness surrounded by an abrupt change in NH$_3$ velocity dispersion ($\sim$ 4 km s$^{-1}$ pc$^{-1}$).  In the following analysis, we search the new GAS data for coherent structures reminiscent of the B5 core, looking for abrupt drops in NH$_3$ linewidth to nearly thermal values around local concentrations of dense gas traced by NH$_3$ (see \S\ref{sec:analysis_id} for details).  Below in the comparison between B5 and the newly identified coherent structures, we consistently follow the same methods adopted by \citet{Pineda_2010} to derive the basic physical properties using GBT observations of NH$_3$ hyperfine line emission and Herschel column density maps derived from SED fitting (\S\ref{sec:data_Herschel}; see also \S\ref{sec:analysis_basic} for details on the measurements of the physical properties).

\section{Analysis}
\label{sec:analysis}

\subsection{Identification of the Droplets}
\label{sec:analysis_id}
In this work, we look for coherent structures defined by abrupt drops in NH$_3$ linewidth and an interior with uniform, nearly thermal velocity dispersion\footnote{The data and the codes used for the analyses presented in this work are made public on \textit{GitHub} at the \href{https://github.com/hopehhchen/Droplets/tree/master/Droplets/}{repository, \texttt{hopehhchen/Droplets}}.}, reminiscent of previously known coherent cores examined by \citet{Goodman_1998}, \citet{Caselli_2002}, and \citet{Pineda_2010}.  We identify the coherent structures using data from the Green Bank Ammonia Survey \citep[see \S\ref{sec:data_GAS}][]{GAS_DR1} and the Herschel maps of column density and dust temperature derived in \S\ref{sec:data_Herschel}, to enable a statistical analysis of coherent structures in two of the closest molecular clouds, Ophiuchus and Taurus.

By eye, one can already recognize many small plateaus of subsonic velocity dispersion associated with NH$_3$-bright structures throughout L1688 and B18 in the maps of observed velocity dispersion ($\sigma_{\mathrm{NH}_3}$) and NH$_3$ brightness (Figs.\ \ref{fig:L1688_TpeakTkin} to \ref{fig:B18_VlsrSigma}).  To identify these coherent structures quantitatively, we follow the procedure adopted by \citet{Pineda_2010} to identify the coherent core region in B5.  The set of criteria we use in this work to define the boundaries of coherent structures starts with the transition in velocity dispersion, $\sigma_{\mathrm{NH}_3}$, from a supersonic to a subsonic value, and continues with the spatial distribution of NH$_3$ brightness, $T_\mathrm{peak}$, and the velocity centroid, $V_\mathrm{LSR}$.  A set of quantitative prescriptions for defining the boundary of a coherent structure is given below as a step-by-step procedure:



\begin{enumerate}
\item \label{id:step1} We start with the intersection of areas enclosed by two contours: one of the NH$_3$ velocity dispersion and one of the NH$_3$ brightness.  First, we find the contour where the NH$_3$ velocity dispersion ($\sigma_{\mathrm{NH}_3}$) has a non-thermal component equal to the thermal component at the median kinetic temperature measured in the targeted region.  (See \S\ref{sec:analysis_basic} and Equation \ref{eq:sigmaTot} for details on the definition of velocity dispersion components.)  Second, we select the contour that corresponds to the \textit{10-$\sigma$ level}, where the NH$_3$ brightness ($T_\mathrm{peak}$) is equal to 10 times the local rms noise, to match the extents of the contiguous regions where successful fits to the NH$_3$ (1, 1) profiles were found in \citet{GAS_DR1}.  The intersection of the areas enclosed by these two contours is then used to define an initial mask.  By this definition, the initial mask encloses a region where we have subsonic velocity dispersion \emph{and} a signal-to-noise ratio larger than 10.

\item \label{id:step2} We expect the pixels within the mask defined in Step \ref{id:step1} to have a continuous distribution of velocity centroids ($V_\mathrm{LSR}$).  In this step, we remove pixels with $V_\mathrm{LSR}$ that leads to local velocity gradients (between the targeted pixel and its neighboring pixels within the mask) larger than the overall velocity gradient found for all pixels within the mask by a factor of $\sim$ 2.  This procedure generally removes pixels with local velocity gradients greater than 20 to 30 km s$^{-1}$ pc$^{-1}$, which is larger than the velocity gradients known to exist because of realistic physical processes in these regions.  The mask editing is done with the aid of Glue\footnote{A GUI Python library built to explore relationships within and among related datasets, including image arrays \citep{glue, glue_2017}.  See \url{http://glueviz.org/} for documentation.}.

\item \label{id:step3} We then check whether the mask from Step \ref{id:step2} contains a single local peak in NH$_3$ brightness.  If there are more than one NH$_3$ brightness peaks, we find the contour level that corresponds to the saddle point between the peaks.  This contour level is then used to separate the mask from Step \ref{id:step2} into regions, each of which has a single NH$_3$ brightness peak.  However, if a region has an NH$_3$ brightness peak no more than 3 times the local rms noise level above the saddle point, the region is excluded, and only its \textit{sibling} region with the brighter peak is kept.  We examine and categorize the regions excluded in this step as \textit{candidates} (see below).


\item \label{id:step4} The Herschel maps of column density and dust temperature are then used to make sure that the defined structure (a \textit{region} from Step \ref{id:step3}) is centered around a local rise in column density and a dip in dust temperature, consistent with the expectation of dense cores \citep{Crapsi_2007}.

\item \label{id:step5} Lastly, we make sure that the resulting structure is resolved by the GBT beam at 23 GHz (32\arcsec).  We impose two criteria: 1) the projected area needs to be larger than a beam, and 2) the effective radius (the geometric mean of the major and minor axes; see \S\ref{sec:analysis_basic}) needs to be larger than the beam FWHM.
\end{enumerate}

Using these criteria, we identify 12 coherent structures in L1688 and 6 coherent structures in B18.  In Figs.\ \ref{fig:L1688_TpeakTkin} to \ref{fig:B18_Herschel}, the boundaries of the identified coherent structures in L1688 and B18 are shown as colored contours.  Although the criteria are consistent with those used by \citet{Pineda_2010} to define the coherent core in B5 and do not impose any limits on size, the newly identified coherent structures in L1688 and B18 are generally smaller than previously known coherent cores (see \S\ref{sec:analysis_basic}).  As mentioned in \S\ref{sec:intro}, we refer to the newly identified coherent structures as ``droplets'' for ease of discussion.

\begin{figure}[ht!]
\plotone{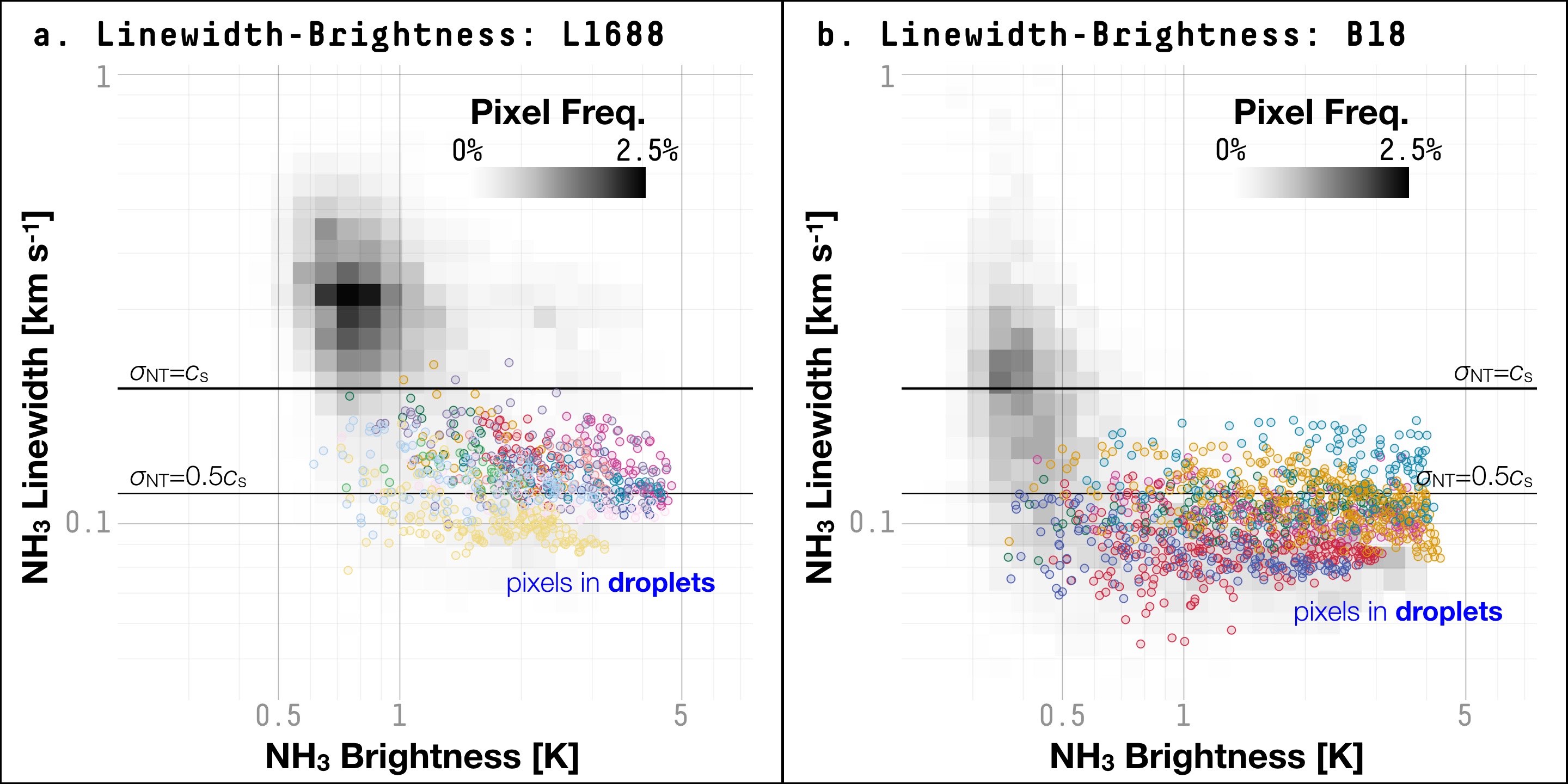}
\caption{\label{fig:TpeakSigma} Distributions of NH$_3$ linewidths and peak NH$_3$ brightness in main-beam units, for every pixel with significant detection of NH$_3$ (1, 1) emission \textbf{(a)} in L1688 and \textbf{(b)} in B18.  The 2D histogram in each panel shows the distribution of pixels in the entire map, with the pixel frequency defined as the percentage of pixels on the map falling in each 2D bin in the 2D histogram.  The colored dots are individual pixels inside droplets, with colors matching the contours in Figs.\ \ref{fig:L1688_TpeakTkin}, \ref{fig:L1688_VlsrSigma}, and \ref{fig:L1688_Herschel} for L1688, and Figs.\ \ref{fig:B18_TpeakTkin}, \ref{fig:B18_VlsrSigma}, and \ref{fig:B18_Herschel} for B18.  The horizontal lines are the expected NH$_3$ linewidths when the non-thermal component of velocity dispersion is respectively equal to the sonic speed (thicker line) and half the sonic speed (thinner line), for a medium with an average particle mass of 2.37 u and a temperature of 10 K.}
\end{figure}

As the criteria indicate, each droplet has a high NH$_3$ peak brightness and a subsonic velocity dispersion, in contrast to the ambient region, where if NH$_3$ emission is detected, we find a mostly supersonic velocity dispersion and a moderate distribution of NH$_3$ brightness.  Fig.\ \ref{fig:TpeakSigma} shows the distributions of NH$_3$ linewidths and peak NH$_3$ brightness in main-beam units, for all pixels where there is significant detection of NH$_3$ emission and for pixels within the droplet boundaries \citep[see][for criteria used to determine the significance of detection]{GAS_DR1}.  We observe an overall anti-correlation between the observed NH$_3$ linewidth and NH$_3$ brightness, and the relation between the two quantities flattens toward the high NH$_3$ brightness end when the NH$_3$ linewidth approaches a thermally dominated value.  The droplets are found in this regime of high NH$_3$ brightness and thermally dominated NH$_3$ linewidths.

\begin{figure}[ht!]
\plotone{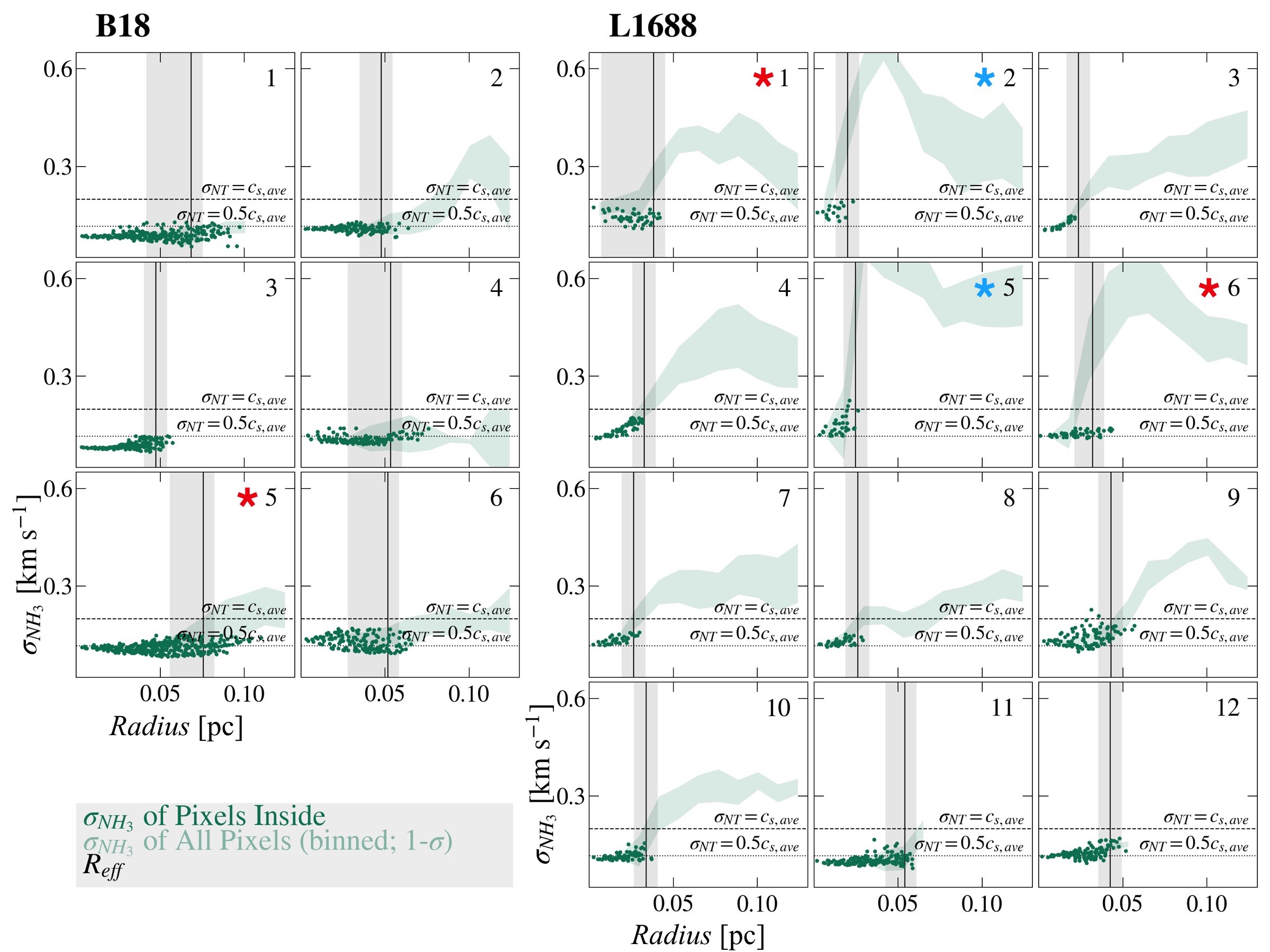}
\caption{\label{fig:sigmas} The NH$_3$ velocity dispersion as a function of distance from the center of each droplet.  The dark green dots represent individual pixels inside the boundary of each droplet.  The transparent green band shows the 1-$\sigma$ distribution of pixels in each distance bin, with a bin size equal to the beam FWHM of GAS observations.  The dashed and dotted lines show the expected NH$_3$ linewidths when the velocity dispersion non-thermal component is equal to the sonic speed and half the sonic speed, respectively.  The vertical black line marks the effective radius, $R_\mathrm{eff}$, and the gray vertical band marks the uncertainty in $R_\mathrm{eff}$.  A red asterisk indicates that the droplet has an elongated shape with an aspect ratio larger than 2 that could bias the measurements using equidistant annuli (L1688-d1, L1688-d6, and B18-d5), and a blue asterisk indicates that the droplet sits near the edge of the region where NH$_3$ emission is detected, resulting in the measurements at larger radii being dominated by fewer pixels (L1688-d2 and L1688-d5).}
\end{figure}

Fig.\ \ref{fig:sigmas} shows the radial profile of NH$_3$ velocity dispersion; the virtually constant NH$_3$ velocity dispersion in the interiors is consistent with what \citet{Goodman_1998} found for coherent cores \citep[see also][]{Pineda_2010}.  See Appendix \ref{sec:appendix_gallery} for a gallery of the close-up views of the droplets.

%


Two of the 18 droplets, L1688-d11 and B18-d4, are found at the positions of the dense cores analyzed by \citet{Goodman_1993}, L1696A and TMC-2A, respectively.  The two droplets correspond to the central parts of the corresponding dense cores and have radii a factor of $\sim$ 0.7 times the radii measured for these dense cores \citep{Benson_1989, Goodman_1993, Ladd_1994}.  See Appendix \ref{sec:appendix_overlap} for a comparison of measured properties.



In Figs.\ \ref{fig:L1688_TpeakTkin} to \ref{fig:B18_Herschel}, we also plot the positions of Class 0/I and flat spectrum protostars in the catalogues presented by \citet{Dunham_2015} and \citet{Rebull_2010}, for L1688 and B18, respectively.  Within the boundaries of six (out of 18) droplets---L1688-d4, L1688-d6, L1688-d7, L1688-d8, L1688-d10, and B18-d6, we find at least one protostar along the line of sight.  Consistent with the results presented by \citet{Seo_2015} and \citet{Friesen_2009}, none of the six droplets where we find protostar(s) within the boundaries shows a strong signature of increased $T_\mathrm{kin}$ or $\sigma_{\mathrm{NH}_3}$ around the protostar(s).  While the existence of YSOs within the boundary of a droplet in the plane of the sky does not necessarily indicate actual associations of these six droplets with protostars, it is possible that some of the droplets are associated with at least one YSO.  See below in \S\ref{sec:discussion_definition} for more discussion on the association between cores and YSOs and how it might be used as a way to define subsets of cores.


\subsubsection{Droplet Candidates}
\label{sec:analysis_id_candidates}

Besides the total of 18 droplets identified in L1688 and B18, we also include 5 \textit{droplet candidates} in L1688 (black contours in Fig.\ \ref{fig:L1688_TpeakTkin}, \ref{fig:L1688_VlsrSigma}, and \ref{fig:L1688_Herschel}).  Each droplet candidate is identified by a spatial change from supersonic velocity dispersion outside the boundary to subsonic velocity dispersion inside.  However, they do not meet at least one criterion listed above.  The detailed reasons why each of these coherent structures is identified as a \textit{droplet candidate}, instead of a droplet, are listed below:

\begin{enumerate}
\item L1688-c1E and L1688-c1W: These two droplet candidates are the eastern and western parts of the droplet L1688-d1, each of which has a local peak in NH$_3$ brightness.  However, neither peak is more than 3 times the local rms noise level above the saddle point between them, i.e., neither satisfies the criterion described in Step \ref{id:step3}.  Thus, we identify the entire region as a single droplet, L1688-d1, and include the eastern and the western parts of L1688-d1 as two droplet candidates.

\item L1688-c2: This droplet candidate shows a local dip in NH$_3$ velocity dispersion and a local peak in NH$_3$ brightness.  However, the local peak in NH$_3$ brightness cannot be separated from the emission in the droplet L1688-d3 by more than 3 times the local rms noise in NH$_3$ (1, 1) observations.  Nor do we find an independent local peak corresponding to L1688-c2 on the Herschel column density map.  (That is, L1688-c2 does not meet the criteria described in Steps \ref{id:step3} and \ref{id:step4} above.)

\item L1688-c3: Similar to L1688-c2, L1688-c3 shows a local dip in NH$_3$ velocity dispersion and a local peak in NH$_3$ brightness.  However, the local peak in NH$_3$ brightness cannot be separated from the emission in the droplet L1688-d4 by more than 3 times the local rms noise in NH$_3$ (1, 1) observations.  Nor do we find an independent local peak corresponding to L1688-c3 on the Herschel column density map. While the projected area of L1688-c3 is larger than a beam, its effective radius is only $\sim$2.6 times the beam FWHM.  (That is, L1688-c3 does not meet the criteria described in Steps \ref{id:step3}, \ref{id:step4}, and \ref{id:step5} above.)

\item L1688-c4: While L1688-c4 does show a significant dip in NH$_3$ velocity dispersion and an independent peak in NH$_3$ brightness, it sits close to the edge of the region where we have enough signal-to-noise of NH$_3$ (1, 1) emission to obtain a confident fit to the hyperfine line profile \citep{GAS_DR1}.  We do not find a strong and independent local peak corresponding to L1688-c4 on the Herschel column density map, either.  Thus, we classify L1688-c4 as a droplet candidate.  (That is, L1688-c4 does not meet the criterion described in Step \ref{id:step4} above.)
\end{enumerate}

In the following analyses, when we discuss the properties of the droplets or, together with previously known coherent cores, the coherent structures, we exclude the droplet candidates.  The droplet candidates are included on the plots to show the distributions of physical properties of potential coherent structures at even smaller scales, which are only marginally resolved by the GAS observations.  The Oph A region (marked by the red rectangles in Fig.\ \ref{fig:L1688_TpeakTkin}, \ref{fig:L1688_VlsrSigma}, and \ref{fig:L1688_Herschel}) could potentially host more droplets/droplet candidates.  However, Oph A is known to also host a cluster of young stellar objects (YSOs), and as Fig.\ \ref{fig:L1688_VlsrSigma}b and \ref{fig:L1688_Herschel}b show, the extent of cold and subsonic dense gas identifiable on the maps of dust temperature and NH$_3$ velocity dispersion is limited.  No coherent structure that satisfies the above criteria can be identified.

The same methods devised here to identify the boundaries and derived the physical properties of the coherent structures in L1688 and in B18 are applied on the data obtained by \citet{Pineda_2010} to derive the physical properties of the coherent core in Perseus B5 in the following analyses.

%


\begin{longrotatetable}
\begin{deluxetable*}{lcccccccc}
\tablecaption{Physical Properties of Droplets and Droplet Candidates\label{table:basic}}
\tablehead{\colhead{ID\tablenotemark{a}} & \multicolumn{2}{c}{Position} & \colhead{Mass\tablenotemark{b}} & \colhead{Effective Radius\tablenotemark{c}} & \colhead{NH$_3$ Linewidth\tablenotemark{d}} & \colhead{NH$_3$ Kinetic Temp.} & \colhead{Total Vel. Dispersion\tablenotemark{e}} & \colhead{YSO(s)\tablenotemark{f}} \\ \colhead{} & \multicolumn{2}{c}{[J2000]} & \colhead{($M$)} & \colhead{($R_\mathrm{eff}$)} & \colhead{($\sigma_{\mathrm{NH}_3}$)} & \colhead{($T_\mathrm{kin}$)} & \colhead{($\sigma_\mathrm{tot}$)} & \colhead{} \\ \cline{2-3} \colhead{} & \colhead{R.A.} & \colhead{Dec.} & \colhead{M$_\sun$} & \colhead{pc} & \colhead{km s$^{-1}$} & \colhead{K} & \colhead{km s$^{-1}$} & \colhead{} }
\startdata
L1688-d1 & 16$^\mathrm{h}$26$^\mathrm{m}$47$^\mathrm{s}$.07 & -24\arcdeg33\arcmin8\farcs3 & $0.17\pm 0.03$ & $0.038^{+0.007}_{-0.031}$ & $0.14\pm 0.01$ & $12.0\pm 0.6$ & $0.24\pm 0.01$ & N \\
L1688-d2 & 16$^\mathrm{h}$26$^\mathrm{m}$54$^\mathrm{s}$.54 & -24\arcdeg36\arcmin52\arcsec.4 & $0.03\pm 0.01$ & $0.020^{+0.007}_{-0.008}$ & $0.16\pm 0.01$ & $12.8\pm 0.9$ & $0.25\pm 0.01$ & N \\
L1688-d3 & 16$^\mathrm{h}$26$^\mathrm{m}$57$^\mathrm{s}$.07 & -24\arcdeg31\arcmin44\arcsec.8 & $0.08\pm 0.03$ & $0.024^{+0.007}_{-0.007}$ & $0.12\pm 0.01$ & $10.2\pm 0.3$ & $0.21\pm 0.01$ & N \\
L1688-d4 & 16$^\mathrm{h}$26$^\mathrm{m}$59$^\mathrm{s}$.59 & -24\arcdeg34\arcmin28\arcsec.8 & $0.73\pm 0.05$ & $0.033^{+0.007}_{-0.007}$ & $0.14\pm 0.01$ & $10.6\pm 0.2$ & $0.23\pm 0.01$ & Y \\
L1688-d5 & 16$^\mathrm{h}$27$^\mathrm{m}$4$^\mathrm{s}$.96 & -24\arcdeg39\arcmin17\arcsec.6 & $0.13\pm 0.03$ & $0.025^{+0.007}_{-0.007}$ & $0.14\pm 0.01$ & $12.4\pm 0.6$ & $0.24\pm 0.01$ & N \\
L1688-d6 & 16$^\mathrm{h}$27$^\mathrm{m}$25$^\mathrm{s}$.50 & -24\arcdeg41\arcmin6\arcsec.2 & $0.22\pm 0.04$ & $0.032^{+0.018}_{-0.011}$ & $0.12\pm 0.01$ & $12.6\pm 0.4$ & $0.23\pm 0.01$ & Y \\
L1688-d7 & 16$^\mathrm{h}$27$^\mathrm{m}$37$^\mathrm{s}$.27 & -24\arcdeg42\arcmin50\arcsec.2 & $0.10\pm 0.02$ & $0.026^{+0.009}_{-0.007}$ & $0.13\pm 0.01$ & $13.2\pm 0.4$ & $0.24\pm 0.01$ & Y \\
L1688-d8 & 16$^\mathrm{h}$27$^\mathrm{m}$46$^\mathrm{s}$.44 & -24\arcdeg44\arcmin45\arcsec.4 & $0.10\pm 0.01$ & $0.026^{+0.009}_{-0.007}$ & $0.13\pm 0.01$ & $12.7\pm 0.7$ & $0.23\pm 0.01$ & Y \\
L1688-d9 & 16$^\mathrm{h}$27$^\mathrm{m}$59$^\mathrm{s}$.43 & -24\arcdeg33\arcmin33\arcsec.0 & $0.55\pm 0.03$ & $0.043^{+0.021}_{-0.008}$ & $0.14\pm 0.01$ & $11.2\pm 0.5$ & $0.23\pm 0.01$ & N \\
L1688-d10 & 16$^\mathrm{h}$28$^\mathrm{m}$22$^\mathrm{s}$.12 & -24\arcdeg36\arcmin16\arcsec.8 & $0.22\pm 0.02$ & $0.034^{+0.009}_{-0.007}$ & $0.11\pm 0.01$ & $11.5\pm 0.5$ & $0.22\pm 0.01$ & Y \\
L1688-d11 & 16$^\mathrm{h}$28$^\mathrm{m}$31$^\mathrm{s}$.53 & -24\arcdeg18\arcmin36\arcsec.1 & $0.46\pm 0.02$ & $0.054^{+0.010}_{-0.012}$ & $0.10\pm 0.01$ & $10.1\pm 0.8$ & $0.20\pm 0.01$ & N \\
L1688-d12 & 16$^\mathrm{h}$28$^\mathrm{m}$59$^\mathrm{s}$.99 & -24\arcdeg20\arcmin45\arcsec.2 & $0.38\pm 0.02$ & $0.042^{+0.014}_{-0.007}$ & $0.12\pm 0.01$ & $10.1\pm 0.5$ & $0.21\pm 0.01$ & N \\
\hline
L1688-c1E\tablenotemark{g} & 16$^\mathrm{h}$26$^\mathrm{m}$49$^\mathrm{s}$.36 & -24\arcdeg32\arcmin39\arcsec.0 & $0.02\pm 0.02$ & $0.020^{+0.007}_{-0.007}$ & $0.14\pm 0.01$ & $12.4\pm 0.7$ & $0.24\pm 0.01$ & N \\
L1688-c1W\tablenotemark{h} & 16$^\mathrm{h}$26$^\mathrm{m}$45$^\mathrm{s}$.22 & -24\arcdeg33\arcmin30\arcsec.5 & $0.07\pm 0.02$ & $0.022^{+0.007}_{-0.007}$ & $0.15\pm 0.01$ & $11.9\pm 0.5$ & $0.24\pm 0.01$ & N \\
L1688-c2 & 16$^\mathrm{h}$26$^\mathrm{m}$56$^\mathrm{s}$.89 & -24\arcdeg30\arcmin18\arcsec.7 & $0.10\pm 0.02$ & $0.024^{+0.011}_{-0.010}$ & $0.12\pm 0.01$ & $11.3\pm 0.4$ & $0.22\pm 0.01$ & N \\
L1688-c3 & 16$^\mathrm{h}$26$^\mathrm{m}$58$^\mathrm{s}$.74 & -24\arcdeg33\arcmin1\arcsec.4 & $0.06\pm 0.02$ & $0.019^{+0.007}_{-0.007}$ & $0.15\pm 0.01$ & $11.7\pm 0.4$ & $0.24\pm 0.01$ & N \\
L1688-c4 & 16$^\mathrm{h}$27$^\mathrm{m}$22$^\mathrm{s}$.28 & -24\arcdeg24\arcmin52\arcsec.2 & $0.05\pm 0.02$ & $0.028^{+0.007}_{-0.007}$ & $0.12\pm 0.01$ & $12.8\pm 0.9$ & $0.23\pm 0.01$ & N \\
\hline
\hline
B18-d1 & 4$^\mathrm{h}$26$^\mathrm{m}$58$^\mathrm{s}$.95 & 24\arcdeg41\arcmin16\arcsec.6 & $0.34\pm 0.02$ & $0.064^{+0.027}_{-0.025}$ & $0.09\pm 0.01$ & $10.5\pm 1.2$ & $0.20\pm 0.01$ & N \\
B18-d2 & 4$^\mathrm{h}$29$^\mathrm{m}$24$^\mathrm{s}$.13 & 24\arcdeg34\arcmin42\arcsec.2 & $1.24\pm 0.05$ & $0.045^{+0.021}_{-0.012}$ & $0.11\pm 0.01$ & $10.0\pm 0.4$ & $0.21\pm 0.01$ & N \\
B18-d3 & 4$^\mathrm{h}$30$^\mathrm{m}$5$^\mathrm{s}$.71 & 24\arcdeg25\arcmin40\arcsec.6 & $0.49\pm 0.02$ & $0.044^{+0.014}_{-0.007}$ & $0.08\pm 0.01$ & $9.8\pm 0.9$ & $0.19\pm 0.01$ & N \\
B18-d4 & 4$^\mathrm{h}$31$^\mathrm{m}$54$^\mathrm{s}$.48 & 24\arcdeg32\arcmin28\arcsec.2 & $0.56\pm 0.03$ & $0.050^{+0.022}_{-0.024}$ & $0.11\pm 0.01$ & $9.1\pm 0.3$ & $0.20\pm 0.01$ & N \\
B18-d5 & 4$^\mathrm{h}$32$^\mathrm{m}$46$^\mathrm{s}$.54 & 24\arcdeg24\arcmin51\arcsec.9 & $1.87\pm 0.05$ & $0.071^{+0.034}_{-0.019}$ & $0.11\pm 0.01$ & $9.5\pm 0.4$ & $0.20\pm 0.01$ & N \\
B18-d6 & 4$^\mathrm{h}$35$^\mathrm{m}$36$^\mathrm{s}$.32 & 24\arcdeg9\arcmin0\arcsec.7 & $0.72\pm 0.04$ & $0.048^{+0.017}_{-0.022}$ & $0.13\pm 0.01$ & $9.9\pm 0.3$ & $0.22\pm 0.01$ & Y
\enddata
\tablenotetext{a}{L1688-c1E to L1688-c4 are droplet candidates.}
\tablenotetext{b}{Based on the column density map derived from SED fitting of Herschel observations.  See \S\ref{sec:data_Herschel}.}
\tablenotetext{c}{The geometric mean of the $T_\mathrm{peak}$ weighted spatial dispersions along the major and the minor axes.  See \S\ref{sec:analysis_basic}.  See also Appendix \ref{sec:appendix_radius} for details on determining the uncertainties.}
\tablenotetext{d}{The best-fit Gaussian $\sigma$.}
\tablenotetext{e}{Derived from NH$_3$ linewidths and kinetic temperatures.  See Equation \ref{eq:sigmaTot}.}
\tablenotetext{f}{A value of ``Y'' means that there is at least one YSO within the droplet boundary defined on the plane of the sky (see \S\ref{sec:analysis_id}), and a value of ``N'' means that there is no YSO within the droplet boundary.  The YSO positions are taken from the catalogue presented by \citet{Rebull_2010} for B18 and the catalogue presented by \citet{Dunham_2015} for L1688.  Since we are interested in the association between cores/droplets and the YSOs potentially forming inside, only Class 0/I and flat spectrum protostars are considered here.}
\tablenotetext{g}{The eastern part of L1688-d1.}
\tablenotetext{h}{The western part of L1688-d1.}
\end{deluxetable*}
\end{longrotatetable}

\subsubsection{Contrast with Velocity Coherent Filaments}
\label{sec:analysis_id_fibers}

We note that \citet{Hacar_2013} and \citet{Tafalla_2015} used the term ``coherent'' to describe continuous structures in the position-position-velocity space, with continuous distributions of line-of-sight velocity ($V_\mathrm{LSR}$).  The method they adopted is a friend-of-friend clustering algorithm and does not impose any criteria on the velocity dispersion.  Since in Step \ref{id:step2}, we require a coherent structure to have a continuous distribution of $V_\mathrm{LSR}$, the newly identified coherent structures could theoretically be parts of ``velocity coherent filaments,'' but the same can be said of any structures that are identified to have continuous structures on the plane of the sky and continuous distributions of line-of-sight velocity.  We do \emph{not} recommend equating the coherent structures, including the newly identified droplets in this work and the coherent cores previously analyzed by \citet{Goodman_1998}, \citet{Caselli_2002}, and \citet{Pineda_2010}, to ``velocity coherent filaments'' identified by \citet{Hacar_2013}.  Specifically, the droplets and other coherent structures are defined by abrupt drops in velocity dispersion from supersonic to subsonic values around their boundaries, which none of the ``velocity coherent filaments'' examined by \citet{Hacar_2013} show.  Moreover, in contrast to the elongated shapes of the ``velocity coherent filaments'' examined by \citet{Hacar_2013}, the droplets are mostly round, with aspect ratios generally between 1 and 2 (with the exceptions of L1688-d1 with an aspect ratio of $\sim$ 2.50, L1688-d6 with an aspect ratio of $\sim$ 2.52, and B18-d5 with an aspect ratio of $\sim$ 2.03; these exceptions are marked with red asterisks on Fig.\ \ref{fig:sigmas}).

\subsection{Mass, Size, and Velocity Dispersion}
\label{sec:analysis_basic}


With the droplet boundary defined in \S\ref{sec:analysis_id}, we calculate the mass of each droplet using the column density map derived from SED fitting of Herschel observations (see \S\ref{sec:data_Herschel}).  To remove the contribution of line-of-sight material, the minimum column density within the droplet boundary is used as a baseline and subtracted off.  The mass is then estimated by summing column density (after baseline subtraction) within the droplet boundary.  This baseline subtraction method is similar to the ``clipping paradigm'' studied by \citet{Rosolowsky_2008b}, and has been applied by \citet{Pineda_2015} to estimate the mass of structures within the coherent core in B5.  For the droplets, we find a typical mass\footnote{Unless otherwise noted, the typical value of each physical property presented in this work is the median value of the entire sample of 18 droplets---excluding the droplet candidates---with the upper and lower bounds being the values measured at the 84th and 16th percentiles, which would correspond to $\pm$1 standard deviation around the median value if the distribution is Gaussian.} of $0.4^{+0.4}_{-0.3}$ M$_\sun$.  Table \ref{table:basic} lists the mass of each droplet.  In Appendix \ref{sec:appendix_baseline}, we discuss the reasons for adopting the clipping method and the uncertainty therein, and in Appendix \ref{sec:appendix_bias}, we examine the uncertainty in mass measurements due to the potential bias in SED fitting.


We define the radius of each droplet based on the NH$_3$ brightness weighted second moments along the major and minor axes.  We designate the \textit{major} axis direction as the one with the greatest dispersion in $T_\mathrm{peak}$ according to a principal component analysis (PCA), and the minor axis is oriented perpendicular to the major axis\footnote{The same process is used to define the major and minor axes in the \textit{Python} package for computing the dendrogram, \texttt{astrodendro}.  See \url{http://dendrograms.org/} for documentation.}.  The effective radius is then the geometric mean of sizes along the major and minor axes, $R_\mathrm{eff} = \sqrt{r_\mathrm{maj} r_\mathrm{min}}$, where $r_\mathrm{maj}$ and $r_\mathrm{min}$ are derived by multiplying the NH$_3$ brightness weighted second moments by a factor of $2\sqrt{2 \ln{2}}$, the scaling factor between the second moment and the full width at half maximum (FWHM) for a Gaussian shape.  The multiplication of the scaling factor of $2 \sqrt{2\ln{2}}$ is done in the same way as the method applied by \citet{Benson_1989} and \citet{Goodman_1993} to estimate the radii of dense cores and is applied to approximate the ``true radius'' of the droplet.

The resulting effective radii of droplets are listed in Table \ref{table:basic} and have a typical value of $0.04\pm 0.01$ pc.  The effects of the resolution and the irregular shape of the boundary are included in the uncertainties listed in Table \ref{table:basic}.  Fig.\ \ref{fig:sigmas} shows that the effective radius, $R_\mathrm{eff}$, plotted on top of the radial profile of velocity dispersion, $\sigma_{\mathrm{NH}_3}$, of each droplet, well characterizes the change from supersonic to subsonic velocity dispersion.  See Appendix \ref{sec:appendix_gallery} for a comparison between a circle with a radius equal to $R_\mathrm{eff}$ and the actual boundary of a droplet on the plane of the sky, and see Appendix \ref{sec:appendix_radius} for details on estimating the uncertainty and for a discussion on other common ways to derive the ``effective radius.''





From the GAS observations, we derive the NH$_3$ velocity dispersion, $\sigma_{\mathrm{NH}_3}$, and the gas kinetic temperature, $T_\mathrm{kin}$ (Figs.\ \ref{fig:L1688_TpeakTkin} to \ref{fig:B18_VlsrSigma}; see \S\ref{sec:data_GAS_fitting} for details).  Assuming that the bulk molecular component is in thermal equilibrium with the NH$_3$ component and assuming also that the non-thermal component of the velocity dispersion is independent of the chemical species observed, we can estimate a \textit{total velocity dispersion}, $\sigma_\mathrm{tot}$, from the thermal component, $\sigma_\mathrm{T}$, and the non-thermal (turbulent) component, $\sigma_\mathrm{NT}$:

\begin{eqnarray}
\label{eq:sigmaTot}
\sigma_\mathrm{tot}^2 &= &\sigma_\mathrm{NT}^2 + \sigma_\mathrm{T}^2 \\
 &= &\left(\sigma_{\mathrm{NH}_3}^2-\frac{k_\mathrm{B} T_\mathrm{kin}}{m_{\mathrm{NH}_3}}\right) + \frac{k_\mathrm{B} T_\mathrm{kin}}{m_\mathrm{ave}}\ \mathrm{,}  \nonumber
\end{eqnarray}

\noindent where $k_\mathrm{B}$ is the Boltzmann constant, and $m_{\mathrm{NH}_3}$ and $m_\mathrm{ave}$ are the molecular weight of NH$_3$ and the mean molecular weight in molecular clouds, respectively.  Note that by definition, the thermal component, $\sigma_\mathrm{T}$, is equal to the sonic speed, $c_\mathrm{s}$, in a medium with a particle mass of $m_\mathrm{ave}$ at a temperature of $T_\mathrm{kin}$.  Following \citet{Kauffmann_2008}, we use the mean molecular weight per free particle of 2.37 u ($\mu_\mathrm{p}$ in \citealt{Kauffmann_2008}).

For each droplet, we obtain characteristic values of the NH$_3$ velocity dispersion, $\sigma_{\mathrm{NH}_3}$, and the kinetic temperature, $T_\mathrm{kin}$, by taking the median value for the pixels within the droplet boundary on the parameter maps.  Following Equation \ref{eq:sigmaTot}, we then estimate $\sigma_\mathrm{NT}$, $\sigma_\mathrm{T}$, and $\sigma_\mathrm{tot}$, for each droplet.  Note that $\sigma_\mathrm{tot}$ is sometimes referred to as the ``1D velocity dispersion,'' concerning the motions along the line of sight, as opposed to the ``3D velocity dispersion,'' which cannot be observed but can be estimated by multiplying the 1D velocity dispersion by a factor of $\sqrt{3}$ assuming isotropy.  We find a typical $\sigma_\mathrm{tot}$ of $0.22\pm 0.02$ km s$^{-1}$ for the droplets (see Table \ref{table:basic}).  For reference, the purely thermal velocity dispersion at 10 K is 0.19 km s$^{-1}$.

\begin{figure}[ht!]
\plotone{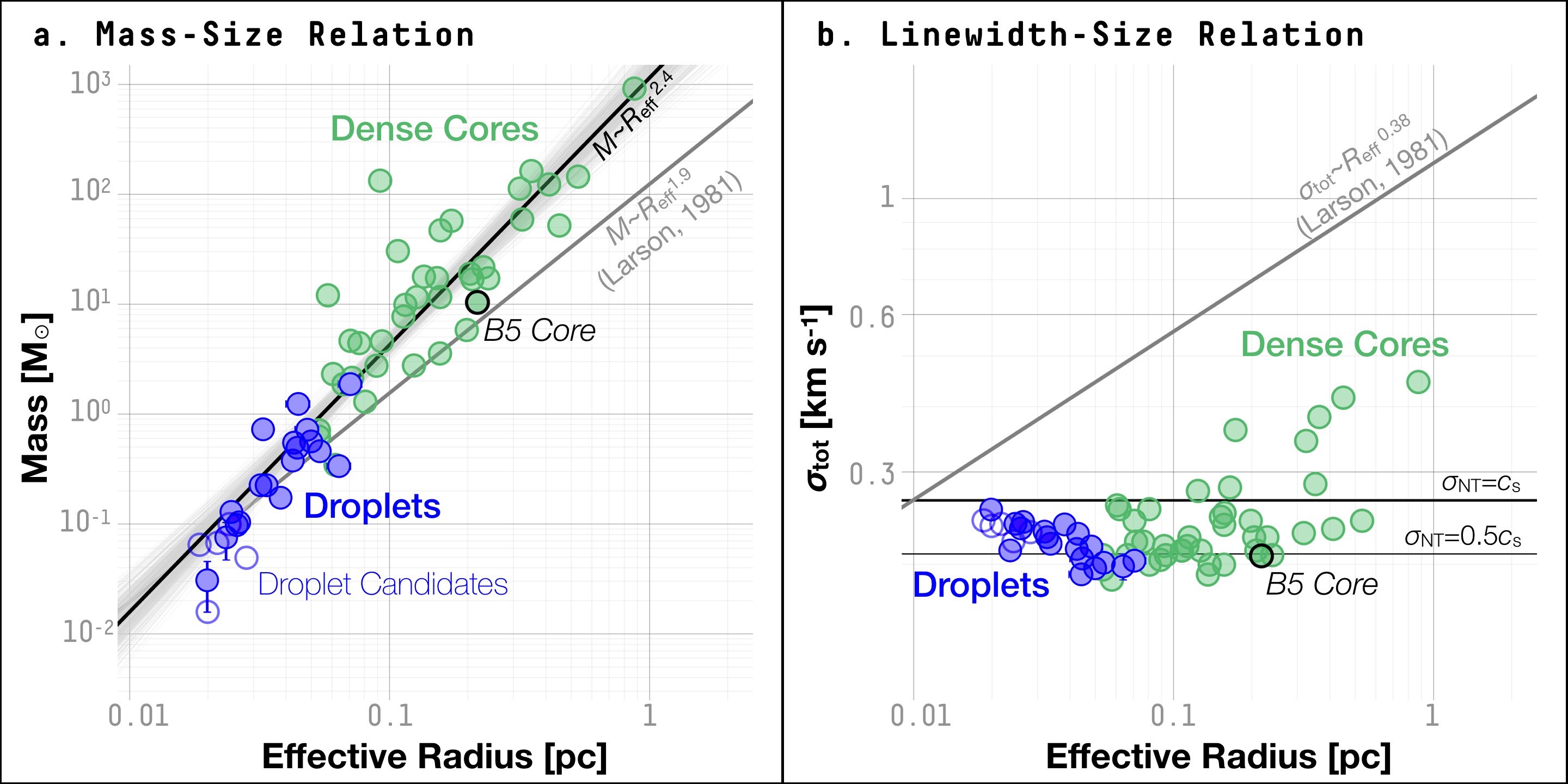}
\caption{\label{fig:basic} \textbf{(a)} The mass, $M$, plotted against the effective radius, $R_\mathrm{eff}$, for dense cores (green circles), the coherent core in B5 (a green circle marked with a black edge), and the newly identified coherent structures: droplets (filled blue circles) and droplet candidates (empty blue circles).  The black line shows a power-law relation between the mass and the effective radius, found for both the dense cores (including B5) and the droplets (excluding droplet candidates) by a gradient-based MCMC sampler.  Randomly selected 10\% of the accepted parameters in the MCMC chain are plotted as transparent lines for reference.  The solid gray line shows the empirical relation based on observations of larger-scale structures examined by \citet{Larson_1981}.  \textbf{(b)} The total velocity dispersion, $\sigma_\mathrm{tot}$, plotted against the effective radius, $R_\mathrm{eff}$, for the same structures as in (a).  The horizontal lines show $\sigma_\mathrm{tot}$ expected for structures where the non-thermal component is equal to the sonic speed ($c_\mathrm{s}$; thicker line) and half the sonic speed (thinner line) of a medium with an mean molecular weight of 2.37 u at a temperature of 10 K.  The gray line shows an empirical relation adopted from \citet{Larson_1981}.  Here we convert the linewidth in the relation presented by \citet{Larson_1981} to $\sigma_\mathrm{tot}$ by assuming that the linewidth was measured from the CO (1-0) line emission with a gas temperature of 10 K.}
\end{figure}

Fig.\ \ref{fig:basic} shows the distributions of mass, $M$, and total velocity dispersion, $\sigma_\mathrm{tot}$, plotted against the effective radius, $R_\mathrm{eff}$, of droplets/droplet candidates in comparison with previously known coherent cores as well as other dense cores (see \S\ref{sec:data_catalogs} for details on how the physical properties were estimated for the dense cores).  Fig.\ \ref{fig:basic}a shows that droplets seem to fall along the same mass-radius relation as the dense/coherent cores.  Using a gradient-based MCMC sampler to find a power-law relation between the mass and effective radius, $M \propto R_\mathrm{eff}^p$, for all the previously known dense/coherent cores (including B5) and the droplets (excluding droplet candidates), we find a power-law index, $p = 2.4\pm 0.1$\footnote{The gradient-based MCMC sampling is implemented using the \textit{Python} package, \texttt{PyMC3}.  See \url{http://docs.pymc.io/index.html} for documentation.}.  This exponent lies between those expected for structures with constant surface density, $M \propto R^{2}$, and structures with constant volume density, $M \propto R^{3}$.  As a reference, \citet{Larson_1981} found a scaling law, $M \propto R^{1.9}$, for larger-scale molecular structures (with sizes of 0.1 to 100 pc and masses of 1 M$_\sun$ to $3\times10^5$ M$_\sun$), using a compilation of observations of molecular line emission from species including $^{12}$CO, $^{13}$CO, H$_2$CO, and for a few objects, NH$_3$ and other N-bearing species.

Fig.\ \ref{fig:basic}b shows the relationship between $\sigma_\mathrm{tot}$ and $R_\mathrm{eff}$.  At scales below 0.1 pc, all structures shown have a subsonic velocity dispersion.  The continuity of the distribution of $M$, $R_\mathrm{eff}$, and $\sigma_\mathrm{tot}$ between the newly identified coherent structures---\textit{droplets}---and the previously known coherent cores as well as other dense cores suggests that the identification of droplets is robust, and that droplets fall toward the small-size end of a potentially continuous population of coherent structures across different size scales.  We discuss this continuity in details in \S\ref{sec:discussion_definition}.

%




\subsection{Virial Analysis: Kinetic Support, Self-Gravity, and Ambient Gas Pressure}
\label{sec:analysis_virial}
To investigate the stability of the coherent structures, we follow \citet{Pattle_2015} to consider the balance between internal kinetic energy, self-gravity, and the ambient gas pressure, with respect to the equilibrium expression:


\begin{equation}
\label{eq:virial}
2\Omega_\mathrm{K} = -(\Omega_\mathrm{G}+\Omega_\mathrm{P})\ \mathrm{,}
\end{equation}

\noindent where $\Omega_\mathrm{K}$ is the internal kinetic energy; $\Omega_\mathrm{G}$ is the gravitational potential energy; and $\Omega_\mathrm{P}$ is the energy term representing the confinement provided by the ambient gas pressure acting on the structure.  The ``external pressure'' comes from thermal and non-thermal (turbulent) motions of the ambient gas (see the analysis below in \S\ref{sec:analysis_virial_P}).  Since we do not have the observations needed to estimate magnetic energy, the magnetic energy term, $\Omega_\mathrm{M}$, is omitted \citep[compared to Equation 27 in][]{Pattle_2015}.  Here we focus on pressure exerted on a structure by thermal and non-thermal (turbulent) motions of the ambient gas for $\Omega_\mathrm{P}$, and we ignore any contribution of ionizing photons to pressure \citep[see discussions in][]{WardThompson_2006, Pattle_2015}.


\subsubsection{Internal Kinetic Energy, $\Omega_\mathrm{K}$}
\label{sec:analysis_virial_K}
The internal kinetic energy, $\Omega_\mathrm{K}$, is given by:

\begin{equation}
\label{eq:virial_K}
\Omega_\mathrm{K} = \frac{3}{2}M \sigma_\mathrm{tot}^2\ \mathrm{,}
\end{equation}

\noindent where $M$ is the mass and $\sigma_\mathrm{tot}$ is the total velocity dispersion, estimated from the observed NH$_3$ velocity dispersion, $\sigma_{\mathrm{NH}_3}$, and gas kinetic temperature, $T_\mathrm{kin}$, following Equation \ref{eq:sigmaTot} (see \S\ref{sec:analysis_basic} for details).  The factor of 3 stands for the correction applied to the ``1D velocity dispersion,'' $\sigma_\mathrm{tot}$, to obtain an estimate of the 3D velocity dispersion, assuming isotropy (see \S\ref{sec:analysis_basic}).  For droplets, we measure a typical kinetic energy of $4.5^{+5.8}_{-2.8}\times 10^{41}$ erg. Table \ref{table:virial} gives results for each droplet.

\begin{deluxetable*}{lcccc}
\tablecaption{Virial Properties of Droplets and Droplet Candidates\label{table:virial}}
\tablehead{\colhead{ID\tablenotemark{a}} & \colhead{Internal Kinetic Energy\tablenotemark{b}} & \colhead{Gravitational Potential Energy\tablenotemark{c}} & \colhead{Ambient Gas Pressure\tablenotemark{d}} & \colhead{Energy Term for Ambient Pressure\tablenotemark{e}} \\ \colhead{} & \colhead{($\Omega_\mathrm{K}$)} & \colhead{($\left|\Omega_\mathrm{G}\right|$)} & \colhead{($P_\mathrm{amb}/k_\mathrm{B}$)} & \colhead{($\left|\Omega_\mathrm{P}\right|$)} \\ \colhead{} & \colhead{erg} & \colhead{erg} & \colhead{K cm$^{-3}$} & \colhead{erg}}
\startdata
L1688-d1 & $2.9\pm 0.5\times 10^{41}$ & $4.0^{+2.7}_{-2.0}\times 10^{40}$ & $7.0\pm 0.7\times 10^5$ & $2.0\pm 0.3\times 10^{42}$ \\
L1688-d2 & $5.9\pm 2.9\times 10^{40}$ & $2.4^{+1.6}_{-2.4}\times 10^{39}$ & $8.4\pm 1.3\times 10^5$ & $3.3\pm 0.6\times 10^{41}$ \\
L1688-d3 & $1.0\pm 0.4\times 10^{41}$ & $1.2^{+0.8}_{-0.9}\times 10^{40}$ & $6.8\pm 0.7\times 10^5$ & $4.5\pm 0.6\times 10^{41}$ \\
L1688-d4 & $1.1\pm 0.1\times 10^{42}$ & $8.3^{+5.5}_{-4.1}\times 10^{41}$ & $6.7\pm 0.8\times 10^5$ & $1.2\pm 0.2\times 10^{42}$ \\
L1688-d5 & $2.2\pm 0.5\times 10^{41}$ & $3.4^{+2.3}_{-1.7}\times 10^{40}$ & $1.5\pm 0.2\times 10^6$ & $1.1\pm 0.2\times 10^{42}$ \\
L1688-d6 & $3.6\pm 0.6\times 10^{41}$ & $8.1^{+5.4}_{-4.1}\times 10^{40}$ & $9.7\pm 1.4\times 10^5$ & $1.6\pm 0.3\times 10^{42}$ \\
L1688-d7 & $1.8\pm 0.3\times 10^{41}$ & $2.1^{+1.4}_{-1.0}\times 10^{40}$ & $4.1\pm 0.6\times 10^5$ & $3.8\pm 0.6\times 10^{41}$ \\
L1688-d8 & $1.6\pm 0.2\times 10^{41}$ & $1.9^{+1.3}_{-0.9}\times 10^{40}$ & $2.7\pm 0.4\times 10^5$ & $2.4\pm 0.4\times 10^{41}$ \\
L1688-d9 & $8.5\pm 0.6\times 10^{41}$ & $3.6^{+2.4}_{-1.8}\times 10^{41}$ & $2.6\pm 0.5\times 10^5$ & $1.1\pm 0.2\times 10^{42}$ \\
L1688-d10 & $3.2\pm 0.3\times 10^{41}$ & $7.7^{+5.1}_{-3.8}\times 10^{40}$ & $2.8\pm 0.5\times 10^5$ & $5.4\pm 1.0\times 10^{41}$ \\
L1688-d11 & $5.5\pm 0.5\times 10^{41}$ & $2.0^{+1.3}_{-1.0}\times 10^{41}$ & $5.0\pm 1.0\times 10^4$ & $4.0\pm 0.9\times 10^{41}$ \\
L1688-d12 & $5.1\pm 0.4\times 10^{41}$ & $1.7^{+1.1}_{-0.9}\times 10^{41}$ & $9.7\pm 1.3\times 10^4$ & $3.8\pm 0.6\times 10^{41}$ \\
\hline
L1688-c1E\tablenotemark{f} & $2.6\pm 3.3\times 10^{40}$ & $6.5^{+4.3}_{-6.5}\times 10^{38}$ & $7.7\pm 0.8\times 10^5$ & $3.1\pm 0.4\times 10^{41}$ \\
L1688-c1W\tablenotemark{g} & $1.1\pm 0.3\times 10^{41}$ & $1.1^{+0.7}_{-0.6}\times 10^{40}$ & $7.8\pm 0.8\times 10^5$ & $4.1\pm 0.6\times 10^{41}$ \\
L1688-c2 & $1.5\pm 0.3\times 10^{41}$ & $2.1^{+1.4}_{-1.1}\times 10^{40}$ & $6.4\pm 0.7\times 10^5$ & $4.6\pm 0.7\times 10^{41}$ \\
L1688-c3 & $1.1\pm 0.4\times 10^{41}$ & $1.2^{+0.8}_{-0.9}\times 10^{40}$ & $7.9\pm 0.8\times 10^5$ & $2.6\pm 0.4\times 10^{41}$ \\
L1688-c4 & $7.8\pm 3.0\times 10^{40}$ & $4.4^{+2.9}_{-3.3}\times 10^{39}$ & $1.2\pm 0.1\times 10^6$ & $1.3\pm 0.2\times 10^{42}$ \\
\hline
\hline
B18-d1 & $3.9\pm 0.5\times 10^{41}$ & $9.1^{+6.0}_{-4.5}\times 10^{40}$ & $6.8\pm 1.2\times 10^4$ & $9.0\pm 3.3\times 10^{41}$ \\
B18-d2 & $1.5\pm 0.1\times 10^{42}$ & $1.8^{+1.2}_{-0.9}\times 10^{42}$ & $1.8\pm 0.3\times 10^5$ & $8.2\pm 3.0\times 10^{41}$ \\
B18-d3 & $5.4\pm 0.5\times 10^{41}$ & $2.8^{+1.9}_{-1.4}\times 10^{41}$ & $1.2\pm 0.1\times 10^5$ & $5.2\pm 1.7\times 10^{41}$ \\
B18-d4 & $6.5\pm 0.4\times 10^{41}$ & $3.3^{+2.2}_{-1.6}\times 10^{41}$ & $6.6\pm 0.9\times 10^4$ & $4.2\pm 1.4\times 10^{41}$ \\
B18-d5 & $2.3\pm 0.1\times 10^{42}$ & $2.5^{+1.7}_{-1.3}\times 10^{42}$ & $1.4\pm 0.3\times 10^5$ & $2.5\pm 1.0\times 10^{42}$ \\
B18-d6 & $1.0\pm 0.1\times 10^{42}$ & $5.5^{+3.7}_{-2.8}\times 10^{41}$ & $1.8\pm 0.4\times 10^5$ & $1.0\pm 0.4\times 10^{42}$
\enddata
\tablenotetext{a}{L1688-c1E to L1688-c4 are droplet candidates.}
\tablenotetext{b}{See Equation \ref{eq:virial_K}.}
\tablenotetext{c}{A potential energy, with the zero point defined at infinity.  The effects of various assumptions regarding the geometry are considered in error estimation.  Absolute values are listed in this table.  See Equation \ref{eq:virial_G} and the text.}
\tablenotetext{d}{Measured in the region immediately outside each droplet.  See Equation \ref{eq:Pext}.}
\tablenotetext{e}{A potential energy, with the zero point defined at equilibrium.  Absolute values are listed in this table.  See Equation \ref{eq:virial_P}.}
\tablenotetext{f}{The eastern part of L1688-d1.}
\tablenotetext{g}{The western part of L1688-d1.}
\end{deluxetable*}

\subsubsection{Gravitational Potential Energy, $\Omega_\mathrm{G}$}
\label{sec:analysis_virial_G}
Assuming spherical geometry, gravitational potential energy, $\Omega_\mathrm{G}$, can be estimated from total mass and an effective radius; we adopt a gravitational potential energy expression:

\begin{equation}
\label{eq:virial_G}
\Omega_\mathrm{G} = \frac{-3}{5}\frac{GM^2}{R_\mathrm{eff}}\ \mathrm{,}
\end{equation}

\noindent where we assume that the sphere of material has a uniform density distribution.  In comparison, a sphere of material with a power-law density distribution, $\rho \propto r^{-2}$, has an absolute value of gravitational potential energy, $\left|\Omega_\mathrm{G}\right|$, a factor of $\sim$ 1.7 larger than that expressed in Equation \ref{eq:virial_G}, and a sphere with a Gaussian density distribution has $\left|\Omega_\mathrm{G}\right|$ a factor of $\sim$ 2 smaller than that expressed in Equation \ref{eq:virial_G} \citep{Pattle_2015, Kirk_2017b}.  In the following analysis, we include the deviation in $\Omega_\mathrm{G}$ due to different assumptions of density distributions in the estimated errors.  In \S\ref{sec:discussion_confinement_BE}, we show that the density distributions in droplets are nearly uniform at small radii with relatively shallow drops toward the outer edges, validating the assumption of a uniform density distribution used to derive Equation \ref{eq:virial_G}.

\begin{figure}[ht!]
\plotone{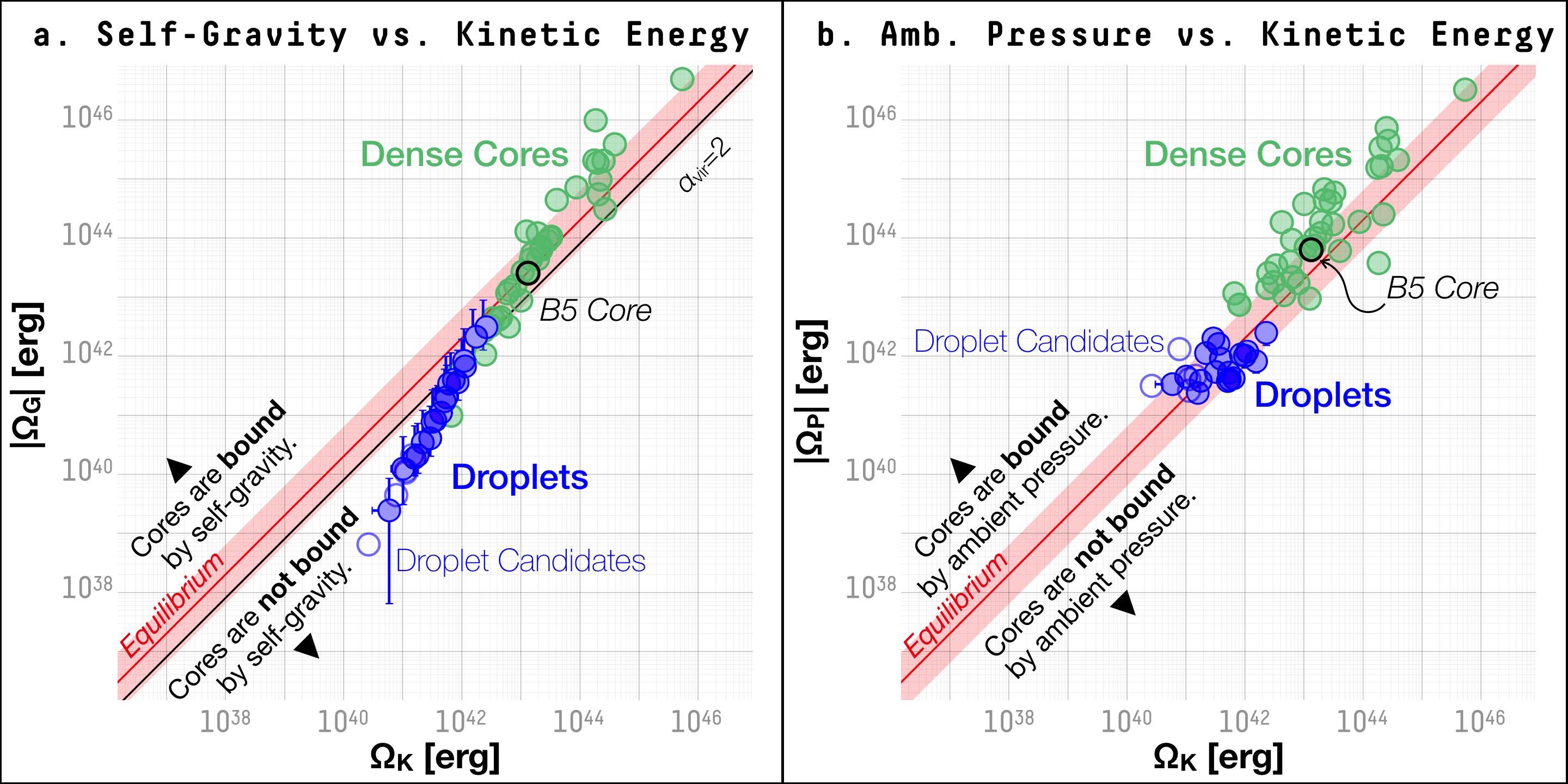}
\caption{\label{fig:virialGP} \textbf{(a)} Gravitational potential energy, $\Omega_\mathrm{G}$, plotted against internal kinetic energy, $\Omega_\mathrm{K}$, for dense cores (green circles), the coherent core in B5 (a green circle marked with a black edge), and the newly identified coherent structures: droplets (filled blue circles) and droplet candidates (empty blue circles).  The red band from the lower left to the top right marks the equilibrium between $\Omega_\mathrm{G}$ and $\Omega_\mathrm{K}$ (solid red line) within an order of magnitude (pink band).  The black line marks where the conventional virial parameter, $\alpha_\mathrm{vir}$, has a value of 2.  \textbf{(b)} The energy term representing the confinement provided by the ambient gas pressure, $\Omega_\mathrm{P}$, plotted agains the internal kinetic energy, $\Omega_\mathrm{K}$, for the same structures shown in (a).  Similarly, the red band from the lower left to the top right marks an equilibrium between $\Omega_\mathrm{P}$ and $\Omega_\mathrm{K}$ (solid red line) within an order of magnitude (pink red band).}
\end{figure}

For droplets, we measure a typical gravitational potential energy of $1.3^{+5.0}_{-1.1}\times 10^{41}$ erg (absolute value; see Table \ref{table:virial}).  Fig.\ \ref{fig:virialGP}a shows that most of the dense cores, including previously known coherent cores such as the one in B5, are close to an equilibrium between the gravitational potential energy and the internal kinetic energy.  This indicates that the self-gravity of these coherent cores is substantial and may provide the binding force needed to keep the cores from dispersing.  On the other hand, gravity in the newly identified droplets appears to be less dominant compared to the internal kinetic energy.  For most of the droplets, the internal kinetic energy is close to an order of magnitude larger than the gravitational potential energy.


That larger structures have more dominant gravitational potential energies than smaller structures is expected for structures with a nearly flat $\sigma_\mathrm{tot}$-size relation and a steep mass-size relation (Fig.\ \ref{fig:basic}).  For the coherent structures under discussion, we observe a power-law mass-size relation, $M \propto R_\mathrm{eff}^{2.4}$, and with a constant $\sigma_\mathrm{tot}$, we would expect a power-law relation between the gravitational potential energy and the size, $\left|\Omega_\mathrm{G}\right| \propto R_\mathrm{eff}^{3.8}$, and a power-law relation between the internal kinetic energy and the size, $\Omega_\mathrm{K} \propto R_\mathrm{eff}^{2.4}$.  Consequently, a smaller coherent structure would have a smaller ratio between the gravitational potential energy and the internal kinetic energy, $\left|\Omega_\mathrm{G}\right|/\Omega_\mathrm{K}$.  For reference, structures with a constant $\left|\Omega_\mathrm{G}\right|/\Omega_\mathrm{K}$ are expected to have a mass-size relation of $M \propto R_\mathrm{eff}$.

The above comparison between the gravitational potential energy and the internal kinetic energy, without considering the ambient turbulent pressure, is analogous to an analysis of stability using a virial parameter, $\alpha_\mathrm{vir} = \frac{a \sigma_\mathrm{tot}^2 R_\mathrm{eff}}{GM}$, where the leading factor, $a$, varies according to the assumption of the density distribution \citep[e.g., $a = 5$ for a spherical structure with a uniform density, and $a = 3$ for a spherical structure with a power-law density profile with an index of 2, $\rho \propto r^{-2}$; see][]{Bertoldi_1992}.  Conventionally, structures with $\alpha_\mathrm{vir} \leq 2$ would be considered ``gravitationally bound.''  By this measure, only the most massive droplets (with masses on the order of 1 M$_\sun$) along with most of the dense cores are ``gravitationally bound'' (Fig.\ \ref{fig:virialGP}a).

\subsubsection{Energy Term Representing Ambient Pressure Confinement, $\Omega_\mathrm{P}$}
\label{sec:analysis_virial_P}
The pressure term, $\Omega_\mathrm{P}$, in the virial equation (Equation \ref{eq:virial}) is characteristic of the pressure exerted on a structure by thermal and non-thermal (turbulent) motions of the ambient gas.  To avoid the impression that there is a clear-cut boundary between the interior and the exterior of the targeted structure, we call the pressure provided by the ambient gas motions the ``ambient gas pressure,'' $P_\mathrm{amb}$, which is sometimes called the ``external pressure'' and denoted by $P_\mathrm{ext}$ in previous works \citep{WardThompson_2007, Pattle_2015, Kirk_2017b}.

For a spherical structure with a radius of $R_\mathrm{eff}$, the pressure term is given by:

\begin{equation}
\label{eq:virial_P}
\Omega_\mathrm{P} = -3 P_\mathrm{amb} V = -4\pi P_\mathrm{amb} R_\mathrm{eff}^3\ \mathrm{,}
\end{equation}

\noindent where $P_\mathrm{amb}$ is the ambient gas pressure, and $V$ is the volume of the structure under discussion \citep{WardThompson_2006, Pattle_2015}.  The pressure exerted on the structure can be estimated from:

\begin{equation}
\label{eq:Pext}
P_\mathrm{amb} = \rho_\mathrm{amb} \sigma_\mathrm{tot,amb}^2\ \mathrm{,}
\end{equation}

\noindent where $\rho_\mathrm{amb}$ is the volume density of the ambient gas, and $\sigma_\mathrm{tot,amb}$ is the total velocity dispersion, including both thermal and non-thermal motions of the ambient gas (same as $\sigma_\mathrm{tot}$ defined in Equation \ref{eq:sigmaTot} for the gas in the core).  The leading factor of $3$ in Equation \ref{eq:virial_P} is applied to estimate the effects of gas motions in the 3D space, since for $\sigma_\mathrm{tot,amb}$, we use the ``1D (line-of-sight) velocity dispersion'' measured from observations.  See the discussion in \S\ref{sec:analysis_basic}.

We base our calculation of the pressure, $P_\mathrm{amb}$, on the maps of $\sigma_{\mathrm{NH}_3}$ and $T_\mathrm{kin}$ from fitting the NH$_3$ hyperfine line profiles (for estimating $\sigma_\mathrm{tot,amb}$; Figs.\ \ref{fig:L1688_TpeakTkin} to \ref{fig:B18_VlsrSigma}) and the Herschel column density maps (for estimating $\rho_\mathrm{amb}$; Fig.\ \ref{fig:L1688_Herschel} and Fig.\ \ref{fig:B18_Herschel}).  The former is possible, because there is significant detection of NH$_3$ (1, 1) emission in regions surrounding the droplets and the coherent core in B5, which appear embedded in the dense gas components of the clouds (see Fig.\ \ref{fig:L1688_VlsrSigma} and Fig.\ \ref{fig:B18_VlsrSigma}).  We use the region (on the plane of the sky) immediately outside the targeted structure but within $(R_\mathrm{eff}+0.1)$ pc from the center of the structure to obtain an estimate of the ambient gas pressure.  Since the typical sonic scale in nearby molecular clouds is roughly 0.1 pc \citep{Federrath_2013}, the hope is that the selected region represents the projection of the volume within a sonic scale from the surface of the structure and that the estimated pressure is from the motions of the gas relevant in confining the structure.  The volume density of the ambient gas is estimated in the same fashion as demonstrated above in \S\ref{sec:analysis_basic} and Fig.\ \ref{fig:cartoon}, by taking the difference between the mass measured within the core boundary and the mass measured within $(R_\mathrm{eff}+0.1)$ pc from the core center, $\Delta M = M(r < (R_\mathrm{eff}+0.1\ \mathrm{pc})) - M_\mathrm{core}$, and dividing it by the difference in volume assuming a spherical geometry, $\Delta V = \frac{4}{3}\pi ((R_\mathrm{eff}+0.1\ \mathrm{pc})^3-R_\mathrm{eff}^3)$.  The total velocity dispersion of the ambient gas, $\sigma_\mathrm{tot,amb}$, is estimated by taking the median value of $\sigma_\mathrm{tot}$ measured at pixels within the same projected region (outside the core, but within $(R_\mathrm{eff}+0.1)$ pc from the core center).  For cores where we do not have significant detection toward every pixel within this projected region, we estimate an uncertainty up to $\lesssim$ 50\%.  We emphasize that the measurement of the ambient gas pressure and the energy term representing the ambient gas pressure, $\Omega_\mathrm{P}$, using this method is independent of the measurement of the kinetics within the core (e.g.\ $\sigma_\mathrm{tot}$ and the internal kinetic energy, $\Omega_\mathrm{K}$), since non-overlapping projected regions are used for the measurements.  We also note that, in contrast to previous works, it is possible to measure the local variation in ambient gas pressure through this method with the GAS observations \citep[][see also discussions in \citealt{Kirk_2017b}]{GAS_DR1}.


Plugging the measured $\rho_\mathrm{amb}$ and $\sigma_\mathrm{tot,amb}$ in Equation \ref{eq:Pext}, we get a typical value of $P_\mathrm{amb}/k_\mathrm{B} \approx 2.7^{+4.7}_{-1.8}\times 10^5$ K cm$^{-3}$ for the droplets (see Table \ref{table:virial} for the result of each droplet) and $P_\mathrm{amb}/k_\mathrm{B} \approx 1.2 \times 10^5$ K cm$^{-3}$ for the coherent core in B5.  Following Equation \ref{eq:virial_P}, we then estimate the virial energy term corresponding to the ambient pressure confinement of the droplets to be $\left|\Omega_\mathrm{P}\right| \approx 6.8^{+3.0}_{-6.3}\times 10^{41}$ erg and that of the coherent core in B5 to be $\left|\Omega_\mathrm{P}\right| \approx 6.3\times 10^{43}$ erg.  See Table \ref{table:virial} for the estimated $P_\mathrm{amb}/k_\mathrm{B}$ and $\Omega_\mathrm{P}$ of each droplet.

Since the 1980s, there have been efforts to find predominantly pressure confined structures and to estimate the magnitude of such pressure confinement.  The earlier works focused on estimating the magnitude of ``inter-clump'' pressure based on models of pressure-confined clumps \citep{Keto_1986, Bertoldi_1992}.  These models of pressure-confined clumps often presumed an equilibrium between the internal kinetic energy, the gravitational potential energy, and the energy terms representing pressure confinement through various physical processes.  For example, using observations of molecular line emission and extinction to estimate the kinetic energy and the gravitational potential energy of dense clumps, \citet{Keto_1986} estimated that an inter-clump pressure, $P/k_\mathrm{B}$, between $10^{3.5}$ and $10^{4.5}$ K cm$^{-3}$ was needed to keep the dense clumps at virial equilibrium.  In a similar fashion, \citet{Bertoldi_1992} estimated that the ``molecular cloud pressure'' acting on the dense clumps within the molecular cloud ranged from $1.2\times 10^4$ K cm$^{-3}$ in Cepheus to $1.1\times 10^5$ K cm$^{-3}$ in Ophiuchus, in both cases balancing the observed internal pressure.  Because of the relatively coarse resolution available at that time, these works focused on clumps with sizes between $\sim$ 0.5 to 1.0 pc.

At smaller size scales, work has been done to estimate the core confining pressure using direct observations of velocity dispersion in the host molecular clouds \citep[see an incomplete summary in Table \ref{table:virialP_comparison}; for example,][]{Johnstone_2000, Lada_2008, Maruta_2010, Kirk_2017b}.  In these works, observations of molecular line emission were devised to estimate the velocity dispersion.  Then, by assuming that the molecular line emission traces a certain (range of) density, the pressure was estimated by equations similar to Equation \ref{eq:Pext}.  While these works found a large range of gas pressure from $P_\mathrm{amb}/k_\mathrm{B} \approx 5\times 10^4$ K cm$^{-3}$ to $2\times 10^7$ K cm$^{-3}$ for structures with sizes from 0.006 to 0.26 pc, they similarly concluded that a substantial portion of targeted structures was pressure confined.  However, these works were limited by the lack of observations suitable for estimating the variation in the confining pressure from structure to structure.

Notably, previous analyses done by \citet{Pattle_2015} of structures in Ophiuchus with sizes slightly smaller than the droplets gave an estimate of the ambient pressure two orders of magnitude larger than that estimated for the droplets.  However, \citet{Pattle_2015} found $\left|\Omega_\mathrm{P}\right| \approx 9\times 10^{41}$ erg for the same structures, which was comparable to the typical value found for the droplets, $\left|\Omega_\mathrm{P}\right| \approx 7.6\times 10^{41}$ erg.  This is because the estimation of the virial energy term, $\Omega_\mathrm{P}$, representing the confinement provided by the ambient gas pressure, is dominated by the size of the targeted structure, $\Omega_\mathrm{P} \propto R^3$ (Equation \ref{eq:virial_P}), and so a size difference of a factor of 2 amounts to roughly an order of magnitude difference in $\Omega_\mathrm{P}$.  Similarly, \citet{Johnstone_2000} found a larger ambient gas pressure, $P_\mathrm{amb}/k_\mathrm{B} \approx 2\times 10^7$ K cm$^{-3}$, and a comparable energy term, $\left|\Omega_\mathrm{P}\right| \approx 2.2\times 10^{41}$ to $1.3\times 10^{44}$ erg, for even smaller structures with sizes between 0.006 and 0.05 pc.  On the other hand, \citet{Maruta_2010} found both an ambient pressure larger than that estimated for the droplets, $P_\mathrm{amb}/k_\mathrm{B} \approx 3\times 10^6$ K cm$^{-3}$, and a pressure energy term larger than that estimated for the droplets, $\left|\Omega_\mathrm{P}\right| \approx 1.6\times 10^{42}$ to $5.0\times 10^{43}$ erg, for structures in Ophiuchus with sizes of 0.022 to 0.069 pc.  To some extent, the difference between the ambient gas pressure estimated in this work for the droplets and the gas pressure estimated for structures in the same region given by previous works can be attributed to the effects of a large uncertainty in the assumed critical density.  Moreover, in previous works, the tracer used for estimating the gas pressure is usually different from the tracer used to define the structures themselves.  This could result in the estimated gas pressure deviating from the actual local ambient gas pressure that is relevant in confining the structures under discussion.

\begin{deluxetable*}{lcccccc}
\tablecaption{External Pressure of Droplets Compared to Previous Works\tablenotemark{a} \label{table:virialP_comparison}}
\tablehead{\colhead{} & \colhead{Region} & \colhead{$P_\mathrm{amb}/k_\mathrm{B}$\tablenotemark{b}} & \colhead{$\Omega_\mathrm{P}$\tablenotemark{b=c}} & \colhead{Sizes of Targeted Structures} & \colhead{Tracer of $\sigma_\mathrm{amb}$} & \colhead{$n_\mathrm{amb}=\rho_\mathrm{amb}/m_\mathrm{ave}$\tablenotemark{d}} \\ \colhead{} & \colhead{} & \colhead{K cm$^{-3}$} & \colhead{erg} & \colhead{pc} & \colhead{} & \colhead{cm$^{-3}$}}
\startdata
Droplets & Oph/Tau & $2.7^{+4.7}_{-1.8}\times 10^5$ & $6.8^{+3.0}_{-6.3}\times 10^{41}$ & 0.02--0.08 & NH$_3$ (1, 1) & Herschel N$_{\mathrm{H}_2}$ \\
B5 & Per & $1.2\times 10^5$ & $6.3\times 10^{43}$ & 0.2 & NH$_3$ (1, 1) & Herschel N$_{\mathrm{H}_2}$ \\
\hline
\citet{Johnstone_2000} & Oph & $2\times 10^7$ & ($2.2\times 10^{41}$--$1.3\times 10^{44}$) & 0.006--0.05 & CO (1--0) & $3\times 10^{4}$ \\
\citet{Lada_2008} & Pipe & $5\times 10^4$ & ($3.2\times 10^{41}$--$4.5\times 10^{43}$) & 0.05--0.26 & $^{13}$CO (1--0) & $1\times 10^{3}$ \\
\citet{Maruta_2010} & Oph & $3\times 10^6$ & ($1.6\times 10^{42}$--$5.0\times 10^{43}$) & 0.022--0.069& H$^{13}$CO$^+$ (1--0) & (0.5--1.0)$\times 10^{5}$ \\
\citet{Pattle_2015} & Oph & $1.8\times 10^7$ & $9\times 10^{41}$ & 0.01& C$^{18}$O (3--2) & $\leq 1\times 10^{5}$ \\
\citet{Kirk_2017b} & Ori & $9.5\times 10^5$ & ($2.7\times 10^{41}$--$1.1\times 10^{44}$) & 0.017--0.13& C$^{18}$O (1--0) & $5\times 10^{3}$
\enddata
\tablenotetext{a}{This table compares estimates of the ambient pressure and the corresponding virial energy term presented in \S\ref{sec:analysis_virial_P} with previous estimates for other density structures found in molecular clouds.  We only include estimates based on direct observations of the velocity dispersion of the ambient material in this table, and the table is by no means meant to be complete.  Other efforts to estimate the ambient pressure include the work presented by \citet{Seo_2015}, where estimates are made by modeling the surface pressure using measurements at the peripheries of cores, and that presented by \citet{Fischera_2012}, where estimates are made for filamentary structures based on surface brightness models of near-equilibrium cylinders, for example.  See discussion in \S\ref{sec:analysis_virial_P}.}
\tablenotetext{b}{The pressure due to the thermal and non-thermal motions of the gas surrounding the targeted structures.  See \S\ref{sec:analysis_virial_P} for details.}
\tablenotetext{c}{The energy term is calculated according to Equation \ref{eq:virial_P}.  Numbers in parentheses are not reported by the original authors and are instead derived here based on the ambient gas pressures and the radii of corresponding structures.}
\tablenotetext{d}{For each of the droplets and the coherent core in B5, the density of the ambient gas is estimated based on the Herschel column density map.  Other works derived the ambient gas density by assuming a ``critical density'' that the velocity dispersion tracer traces.  The number density assumed to be traced by the ambient gas tracer is listed for reference.}
\end{deluxetable*}

Fig.\ \ref{fig:virialGP}b shows a comparison between the kinetic energy and the energy term representing the ambient pressure confinement.  Before including the gravitational potential energy (due to self-gravity acting as a confining force; see Equation \ref{eq:virial}), it already seems that the ambient gas pressure is substantial in both the droplets and the dense cores compared to the kinetic energy.  Here for the dense cores, due to the lack of molecular line observations of the ambient gas, we follow \citet{Kirk_2017b} and adopt a single value of $P_\mathrm{amb}/k_\mathrm{B} = 9.5\times 10^5$ K cm$^{-3}$ based on observations of C$^{18}$O (1--0) emission in nearby molecular clouds.  The result is consistent with the conclusion drawn by \citet{Johnstone_2000} that the ambient gas pressure is ``instrumental'' in confining the dense structures in the Ophiuchus cloud.

It is worth mentioning that a similar effort to obtain the local turbulent pressure structure-by-structure is done by \citet{Seo_2015} for cores identified in the B218 region in Taurus.  \citet{Seo_2015} used the velocity dispersion and column density measurements at the circumference of the targeted core to estimate the work done by the ambient gas pressure, $W_\mathrm{amb} \approx 5\times 10^{40}$ to $1\times 10^{42}$ erg, and by assuming that the density distribution of the core follows the density profile of a critical Bonnor-Ebert sphere, \citet{Seo_2015} estimated that the pressure at the surface of the core is $P/k_\mathrm{B} \approx 8\times 10^5$ K cm$^{-3}$.  Both numbers are similar to the numbers we get for the droplets, and similarly, \citet{Seo_2015} conclude that some of the cores in the B218 region are pressure confined.  A similar value of the ambient pressure, $P/k_\mathrm{B} \approx 2\times 10^4$ K cm$^{-3}$, is found structure-by-structure for filamentary structures in molecular clouds by \citet{Fischera_2012}, by modeling Herschel surface brightness profiles with near-equilibrium cylinders.  See discussion below in \S\ref{sec:discussion_definition}.

\subsubsection{Full Virial Analysis}
\label{sec:analysis_virial_virial}
\begin{figure}[ht!]
\plotone{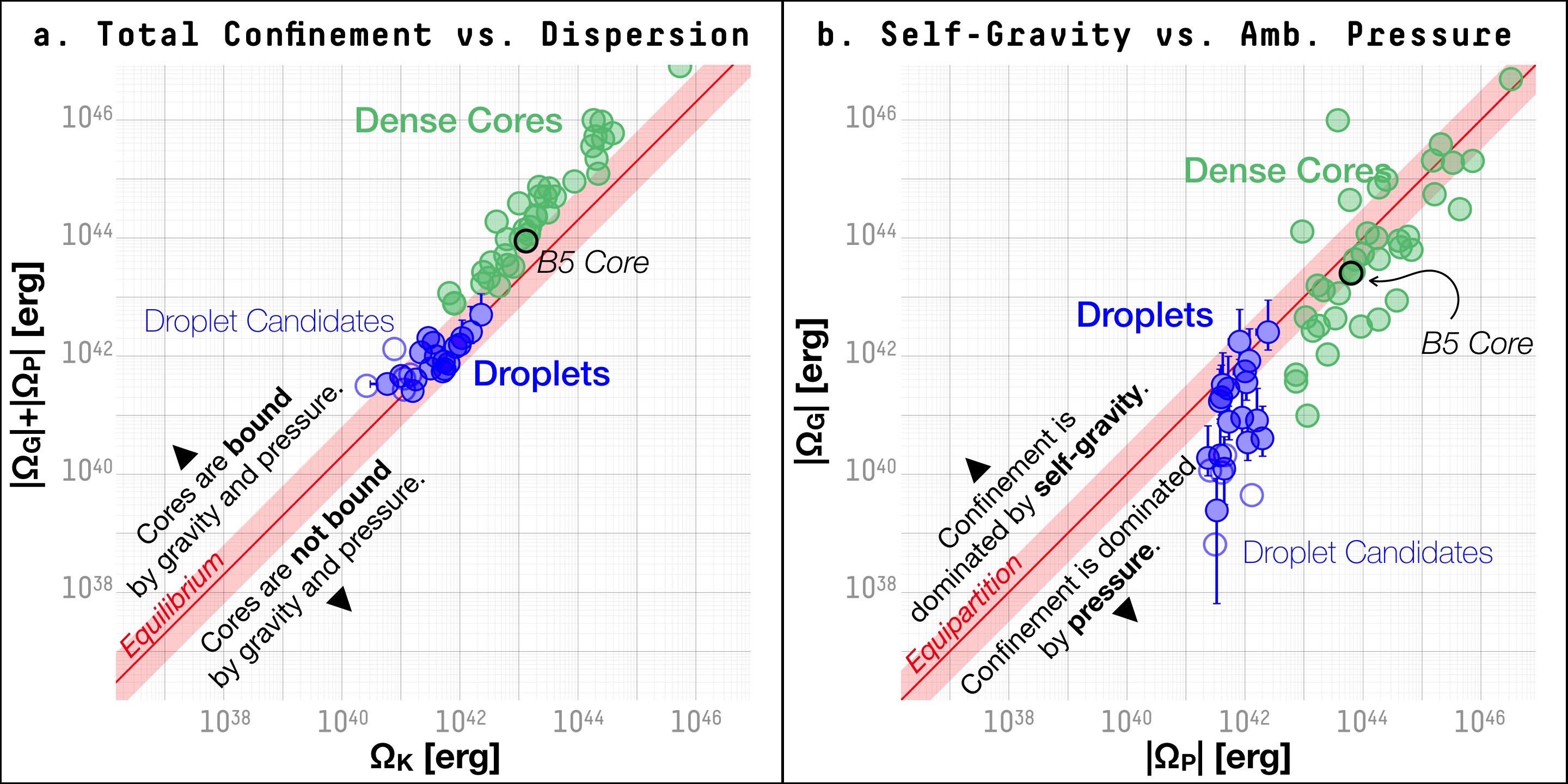}
\caption{\label{fig:virialAll} \textbf{(a)} The sum of gravitational potential energy, $\Omega_\mathrm{G}$, and the energy term representing the confinement provided by the ambient gas, $\Omega_\mathrm{P}$, plotted against the internal kinetic energy, $\Omega_\mathrm{K}$, for dense cores (green circles), the coherent core in B5 (a green circle marked with a black edge), and the newly identified coherent structures: droplets (filled blue circles) and droplet candidates (empty blue circles).  The red band from the lower left to the top right marks the equilibrium between the sum of confining terms and the internal kinetic energy (solid red line) within an order of magnitude (pink red band).  \textbf{(b)} The gravitational potential energy, $\Omega_\mathrm{G}$, plotted against the ambient pressure energy, $\Omega_\mathrm{P}$, for the same structures as in (a).  The red band from the lower left to the top right marks an equipartition between $\Omega_\mathrm{G}$ and $\Omega_\mathrm{P}$.}
\end{figure}

Combining the estimates of $\Omega_\mathrm{K}$, $\Omega_\mathrm{G}$, and $\Omega_\mathrm{P}$, we can assess the balance between the internal kinetic energy and the sum of ``confining forces'' in the form of the gravitational potential energy and the energy term representing the confinement provided by the ambient gas motions (Equation \ref{eq:virial}).  Fig.\ \ref{fig:virialAll}a shows the distribution of the sum of the energy terms on the right-hand side of Equation \ref{eq:virial} ($\Omega_\mathrm{G}$ and $\Omega_\mathrm{P}$) plotted against the internal kinetic energy, $\Omega_\mathrm{K}$.  Both the newly identified droplets and the dense cores appear to be virially bound (by self-gravity and the ambient gas pressure combined) or at least within an order of magnitude around an equilibrium.  The dense cores appear to have the sum of $\Omega_\mathrm{G}$ and $\Omega_\mathrm{P}$ roughly half an order of magnitude larger than $\Omega_\mathrm{K}$.  By contrast, the newly identified droplets and droplet candidates appear to be slightly closer to an equilibrium between the internal kinetic energy and the sum of energy terms representing the confining forces.  That is, Equation \ref{eq:virial} holds for the droplets within an order of magnitude.

In Fig.\ \ref{fig:virialAll}b, we examine the equipartition between the gravitational potential energy, $\Omega_\mathrm{G}$, and the energy term measuring the confinement provided by the ambient gas pressure, $\Omega_\mathrm{P}$.  Most of the coherent cores, including the droplets, have $\left|\Omega_\mathrm{P}\right| \geq \left|\Omega_\mathrm{G}\right|$, showing that even for dense cores which are often gravitationally bound, the ambient gas pressure is substantial.  The full results from the virial analysis are listed in Table \ref{table:virial}, and below in \S\ref{sec:discussion_confinement}, we discuss the nature of the confinement provided by the ambient gas pressure.

\section{Discussion}
\label{sec:discussion}

\subsection{Nature of the Pressure Confinement}
\label{sec:discussion_confinement}
The fact that the newly identified coherent structures, \textit{droplets}, are dominated by the ambient gas pressure but relatively less so by self-gravity (\S\ref{sec:analysis_virial}; see also Fig.\ \ref{fig:virialGP} and Fig.\ \ref{fig:virialAll}) seems to suggest that the confinement of the droplets is primarily provided by the ambient gas pressure.  Understanding the nature of such pressure confinement and the related velocity structures is key to understanding the formation of the droplets and also to understanding the potential role the droplets, as well as the coherent structures, play in star/structure formation in nearby molecular clouds.

\subsubsection{Comparison to the Bonnor-Ebert Sphere}
\label{sec:discussion_confinement_BE}
The droplets are likely confined by the pressure exerted on the surface by the ambient gas (\S\ref{sec:analysis_virial}), and the subsonic velocity dispersion in the droplets indicates that the internal kinetic energy is largely provided by the thermal motions (\S\ref{sec:analysis_basic}).   The interior of each droplet has a virtually uniform distribution of the velocity dispersion dominated by the thermal motions, with the non-thermal component being roughly half of the thermal component (see Fig.\ \ref{fig:sigmas} and Fig.\ \ref{fig:basic}).  These results prompt us to compare the droplets to the Bonnor-Ebert model, which describes an isothermal core embedded in a pressurized medium \citep{Ebert_1955, Bonnor_1956, Spitzer_1968}.

\begin{figure}[ht!]
\plotone{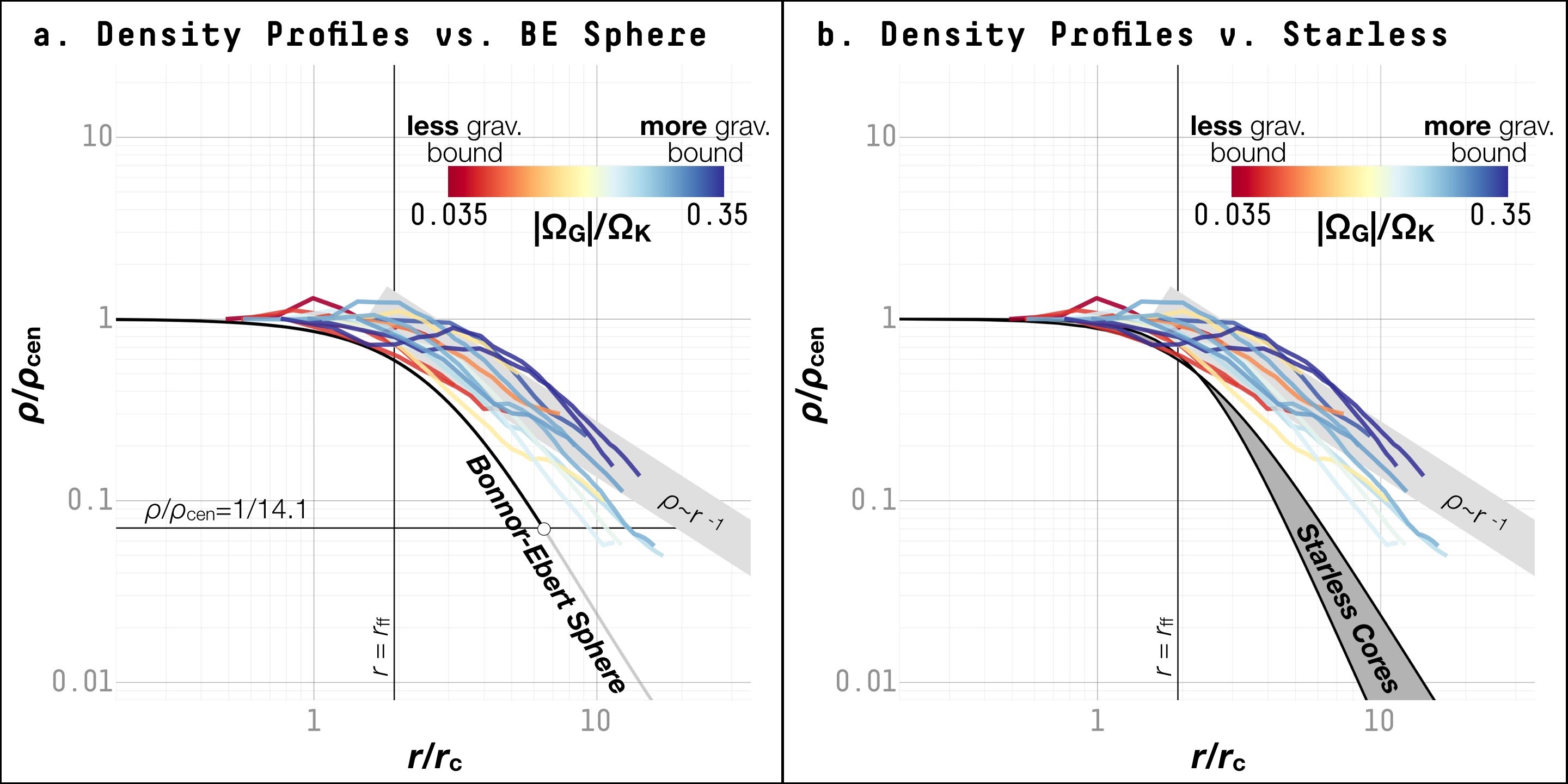}
\caption{\label{fig:profilesDensity} \textbf{(a)} The radial profile of volume density in normalized units of each droplet, compared to a critical Bonner-Ebert density profile.  Each curve is the average radial profile of a droplet, color coded according to the ratio between gravitational potential energy and internal kinetic energy.  The thick black curve plots the density profile of a critical Bonor-Ebert sphere, and the lower horizontal line marks the critical contrast in volume density.  The light gray band shows the slope of a density profile as a power-law function of the radius, $\rho \propto r^{-1}$.  The vertical gray line marks the free-fall length scale, $r_\mathrm{ff}$.  \textbf{(b)} Same as (a), the radial profile of volume density in normalized units of each droplet, this time plotted against previous observations of starless cores \citep[the dark gray band;][]{Tafalla_2004}.  Since the dumbbell shape of L1688-d1 affects this analysis which assumes spherical geometry, L1688-d1 is not included in these plots.  The typical uncertainty for each volume density measurement along a density profile is $\sim$ 25\%.}
\end{figure}

By a similar approach described in \S\ref{sec:analysis_basic}, we derive the radial profiles of volume density, assuming a spherical geometry (see also Appendix \ref{sec:appendix_baseline} and Fig.\ \ref{fig:cartoon}).  In the analysis below, we repeat the procedure for layers of regions at different distances to obtain the radial density profile.  We use one half of the GBT beam FWHM as the bin size in the radial direction.  The resulting radial density profiles are shown in Fig.\ \ref{fig:profilesDensity}.  The typical uncertainty in the density measurement due to the assumption of spherical geometry is $\sim$ 25\%, estimated based on the variation in column density at pixels within each radial distance bin.


We then compare the resulting density profiles of the droplets to the density profile of a Bonnor-Ebert sphere (Fig.\ \ref{fig:profilesDensity}).  A Bonnor-Ebert sphere describes an isothermal sphere of gas in a pressurized medium.  Assuming a pressure distribution satisfying the ideal gas law, $P = \rho c_\mathrm{s}^2$, a Bonnor-Ebert sphere satisfies the Lane-Emden equation:

\begin{equation}
\label{eq:emden_BE}
\frac{1}{r^2}\frac{\mathrm{d}}{\mathrm{d}r}\left(\frac{r^2}{\rho}\frac{\mathrm{d}\rho}{\mathrm{d}r}\right) = -\frac{4 \pi G}{c_\mathrm{s}^2}\rho\ \mathrm{,}
\end{equation}

\noindent where $r$, $\rho$, and $P$ are the radial distance from the center, the density as a function of the radius, and the pressure at $r$, respectively \citep{Ebert_1955, Bonnor_1956}.  A set of non-singular numerical solutions can be found for Equation \ref{eq:emden_BE}.  Following analyses presented by \citet{Ebert_1955}, \citet{Bonnor_1956}, and \citet{Spitzer_1968}, we compare the observed density profiles with the density profile of a critical Bonnor-Ebert sphere in the normalized and dimensionless units of the density, $y = \rho/\rho_\mathrm{cen}$, where $\rho_\mathrm{cen}$ is the density at the center ($r = 0$), and of the distance, $x = r/r_\mathrm{c}$, where $r_\mathrm{c} = c_\mathrm{s}/\sqrt{4 \pi G \rho_\mathrm{cen}}$, corresponding to the y-axis and the x-axis of Fig.\ \ref{fig:profilesDensity}, respectively.  Note that $x$ is proportional to the free-fall length scale, $r_\mathrm{ff} \simeq 1.92 r_\mathrm{c}$.

Fig.\ \ref{fig:profilesDensity}a shows the result of the comparison, with the observed density profiles shown as curves color coded by the ratio between $\Omega_\mathrm{G}$ and $\Omega_\mathrm{K}$ and the density profile of the critical Bonnor-Ebert sphere plotted as the thick black line.  The resulting Bonnor-Ebert sphere has a critical minimum radius for which the sphere is stable, $x_\mathrm{crit} = 6.5$, corresponding to a critical density contrast of $y_\mathrm{crit} = 1/14.1$ (the horizontal dashed line in Fig.\ \ref{fig:profilesDensity}a; see discussions in \citealt{Bonnor_1956} and \citealt{Ebert_1955} for details).  In a \emph{critical} Bonnor-Ebert sphere, the kinetic support and self-gravity is at a critical equilibrium, and the non-critical, stable solutions form a set of density profiles shallower than the critical Bonnor-Ebert sphere.  In this model, a core with a density profile steeper than that of the critical Bonnor-Ebert sphere would collapse under self-gravity.



Fig.\ \ref{fig:profilesDensity}a shows that the density profiles at $r \lesssim r_\mathrm{ff}$ appear to be near-constant, while the density profiles at $r \gtrsim r_\mathrm{ff}$ appear to be shallower than the critical Bonnor-Ebert sphere.  On the outer edge, the density profiles of the droplets approach $\rho \propto r^{-1}$, which can arise from structures having a constant column density and thus following a mass-size relation of $M \propto R^2$.  This mass-size relation has been observed for cloud-scale structures (see examples in \citealt{Larson_1981} and discussions in \citealt{Kauffmann_2010a, Kauffmann_2010b}).  The non-critical, shallow density profiles can be consistent with the virial analysis presented in \S\ref{sec:analysis_virial}, where the droplets are found to be bound by ambient pressure but not self-gravity.  For reference, we also compare the radial density profiles of the droplets to previously observed starless cores \citep{Tafalla_2004}, and we find that the droplets have shallower density profiles than starless cores (Fig.\ \ref{fig:profilesDensity}b; see also Appendix \ref{sec:appendix_profiles} for the radial profiles in physical units).


\begin{figure}[ht!]
\plotone{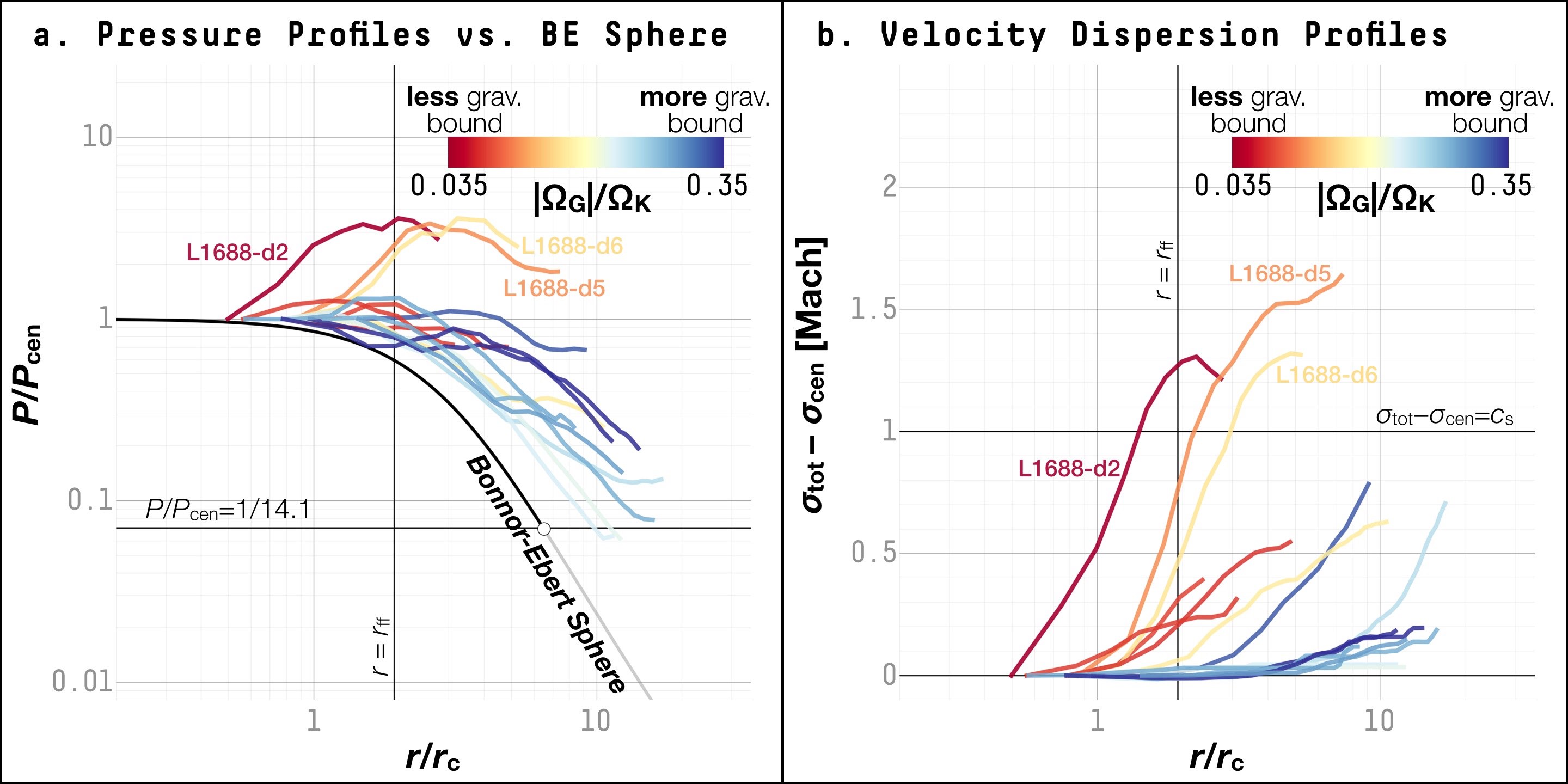}
\caption{\label{fig:profilesKinematics} \textbf{(a)} Like Fig.\ \ref{fig:profilesDensity}a but for the radial profile of pressure in normalized units of each droplet.  Since L1688-d2 and L1688-d5 sit near the edge of the regions with significant detection of NH$_3$ (1, 1) emission, the profiles at larger radii could be dominated by fewer pixels, and thus the corresponding curves are specifically marked.  L1688-d6 is also marked due to its highly elongated shape, for which the measurements using equidistant annuli could be biased.  \textbf{(b)} The radial profile of total velocity dispersion, $\sigma_\mathrm{tot}$, relative to the value at the center of the droplet, $\sigma_\mathrm{tot,cen}$, in Mach numbers (ratios to the sonic velocity).  The horizontal line marks when the change in $\sigma_\mathrm{tot}$ with respect to $\sigma_\mathrm{tot,cen}$ is equal to the sonic speed.  Since the dumbbell shape of L1688-d1 affects this analysis, which assumes spherical geometry, L1688-d1 is not included in these plots.}
\end{figure}

%
Since the Bonnor-Ebert sphere describes a thermal (no turbulent motions) and isothermal (uniform temperature) sphere, the radial profile of the gas pressure, derived from the ideal gas law, $P = \rho c_\mathrm{s}^2$, in the Bonnor-Ebert model, is the same as the density profile of a Bonnor-Ebert sphere in dimensionless units.  In Fig.\ \ref{fig:profilesKinematics}a, we compare the observed radial profiles of the gas pressure (due to the turbulent and thermal motions of the gas) in droplets to the pressure profile of a critical Bonnor-Ebert sphere.  Intriguingly, L1688-d2, L1688-d5, and L1688-d6 have pressure profiles increasing \textit{outwards}, and these droplets also appear to be less gravitationally bound (redder curves in Fig.\ \ref{fig:profilesKinematics}).  However, note that L1688-d2 and L1688-d5 sit near the edge of the region where NH$_3$ emission is detected, such that the profiles at larger radii are dominated by fewer pixels.  Also note that the assumption of spherical geometry could break down due to the elongated shape of L1688-d6.  Fig.\ \ref{fig:profilesKinematics}b shows that the increases in velocity dispersion across the edges of the droplets are usually more abrupt than the change in the density profiles (Fig.\ \ref{fig:profilesDensity}).  See Appendix \ref{sec:appendix_profiles} for the radial profiles of density and pressure in physical units.

\begin{figure}[ht!]
\plotone{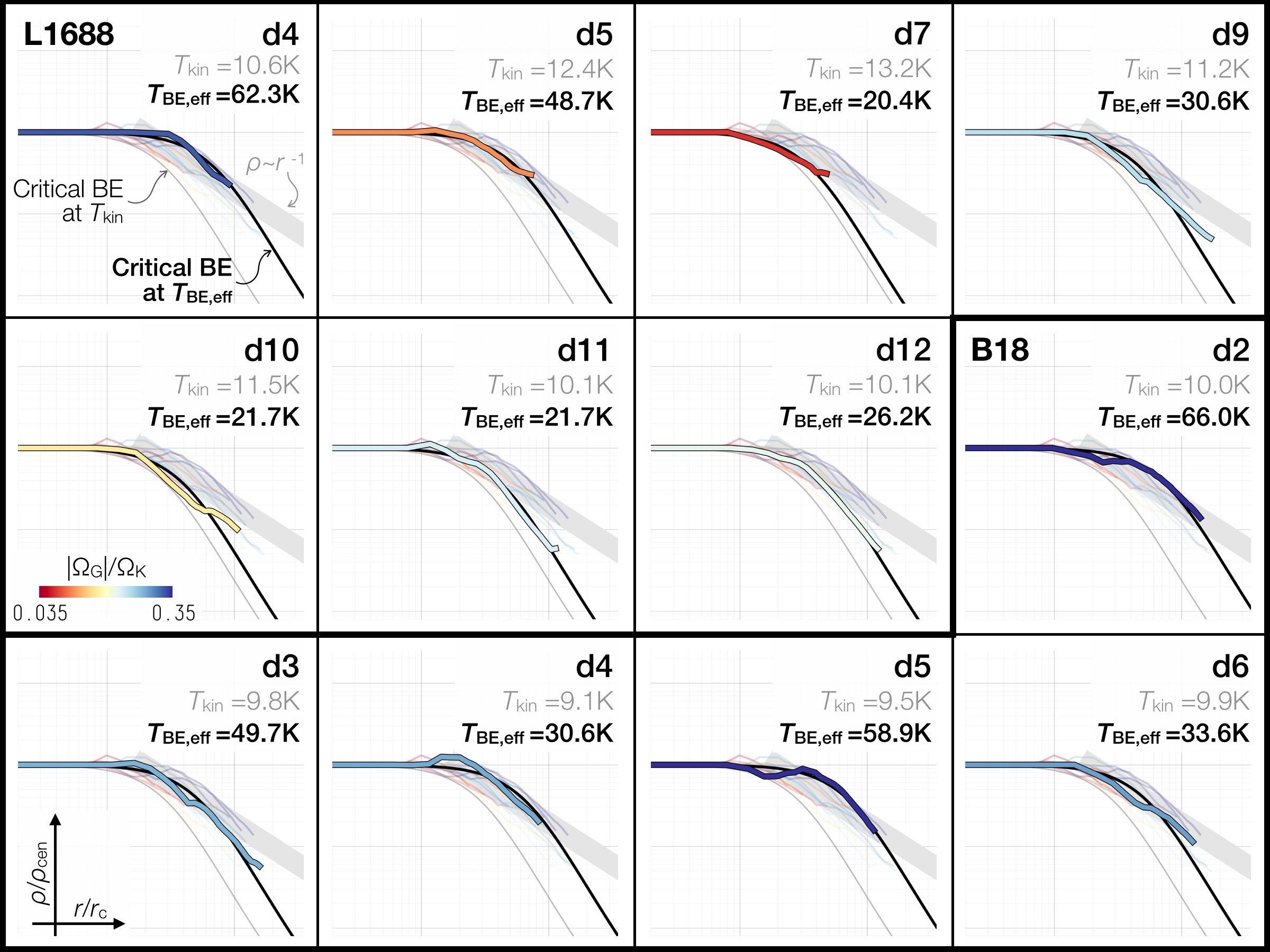}
\caption{\label{fig:profilesEffBE} Individual radial profiles of normalized volume density, compared to critical Bonnor-Ebert spheres at best-fit effective temperatures.  Each panel shows the radial density profile of a droplet, with the ID labeled at the top right of the panel.  The observed radial density profile in each panel is plotted as thick curves, color coded according to $\Omega_\mathrm{G}/\Omega_\mathrm{K}$.  The radial density profile of a Bonnor-Ebert sphere at the best-fit effective temperature are shown as black curves (see surrounding text for details).  The radial density profile of a Bonnor-Ebert sphere at the observed $T_\mathrm{kin}$, corresponding to the radial density profile of the critical Bonnor-Ebert sphere shown in Fig.\ \ref{fig:profilesDensity}, is plotted as a light gray curve in each panel.  The gray band corresponds to a power-law density profile, $\rho \propto r^{-1}$.  Density profiles of droplets other than the one highlighted in each panel are plotted as transparent curves.}
\end{figure}

Using a free parameter---the ``effective temperature,'' $T_\mathrm{BE,eff}$---instead of the observed kinetic temperature, $T_\mathrm{kin}$, to derive $c_\mathrm{s}$ in the ideal gas law, we can fit the critical Bonnor-Ebert profile to the observed density profiles of the droplets.  Fig.\ \ref{fig:profilesEffBE} shows the resulting critical Bonnor-Ebert spheres at best-fit effective temperatures for droplets where we have reliable measurements of radial density profiles beyond the characteristic size scale.  As Fig.\ \ref{fig:profilesEffBE} shows, most of the droplets have an excess in density compared to the best-fit critical Bonnor-Ebert profile at larger distances, approaching a power-law like density profile.  And, for most droplets, the best-fit effective temperature, $T_\mathrm{BE,eff}$, is unreasonably higher than the kinetic temperature measured from NH$_3$ line fitting.  Again, the results suggest that density and pressure profiles of the droplets cannot be well modeled with a critical Bonnor-Ebert sphere.


\subsubsection{Comparison to the Logotropic Sphere}
\label{sec:discussion_confinement_logo}
Based on the observational results obtained in the 1990s that 1) the density distribution at large radial distances from the center of a core is close to a power-law expression, $\rho \propto r^{-1}$ \citep[instead of the singular isothermal solution, $\rho \propto r^{-2}$;][]{Shu_1977}, 2) the core is supported by both thermal and non-thermal (turbulent) velocity distributions, and 3) the total velocity dispersion is close to being purely thermal at the center and increases outwards, \citet{McLaughlin_1996, McLaughlin_1997} proposed that a dense core has a velocity dispersion distribution with a constant (isothermal) thermal component and a purely logotropic non-thermal component, i.e., $P_\mathrm{T} \propto \rho$ and $P_\mathrm{NT} \propto \ln{\rho/\rho_\mathrm{cen}}$ in terms of pressure distribution, respectively.  The resulting solution, known as the logotropic sphere, has an equation of state

\begin{equation}
\label{eq:eos_logo}
P = \rho_\mathrm{cen} c_\mathrm{s}^2 \left[1 + A\ln{\left(\frac{\rho}{\rho_\mathrm{cen}}\right)}\right]\ \mathrm{,}
\end{equation}

\noindent
where $A > 0$ is an adjustable parameter of the logotropic component.  Replacing the pressure term, derived from the ideal gas law, in the Bonnor-Ebert model with Equation \ref{eq:eos_logo}, we can find a non-singular numerical solution of pressure distribution for the logotropic sphere.







\begin{figure}[ht!]
\plotone{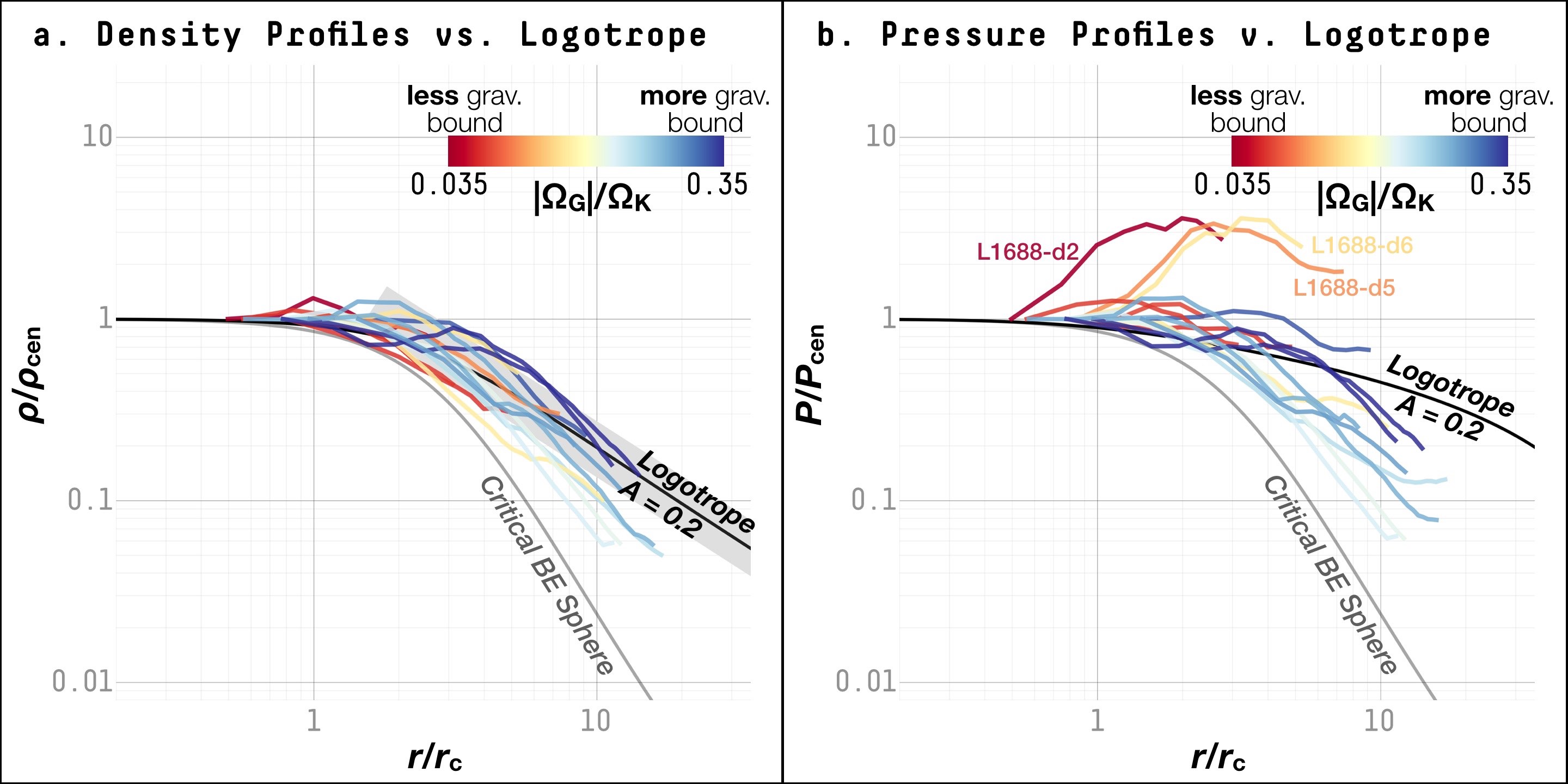}
\caption{\label{fig:profilesLogotrope} \textbf{(a)} The radial profile of volume density in normalized units of each droplet, compared to a logotropic density profile.  Each curve is the radial profile of a droplet, color coded according to $\Omega_\mathrm{G}/\Omega_\mathrm{K}$.  The thick black curve plots the density profile of a logotropic sphere, and the gray curve shows the density profiles of a critical Bonnor-Ebert sphere.  The light gray band shows the slope of the power-law density profile, $\rho \propto r^{-1}$.  The typical uncertainty for each volume density measurement along a density profile is $\sim$ 25\%.  \textbf{(b)} The radial profile of pressure in normalized units of each droplet, same as the color-coded curves shown in Fig.\ \ref{fig:profilesKinematics}a.  The black curve plots the radial profile of normalized pressure for a logotropic sphere, and the gray curve plots the radial pressure profiles of the critical Bonnor-Ebert sphere.  As in Fig.\ \ref{fig:profilesKinematics}, L1688-d2, L1688-d5, and L1688-d6 are marked either because the droplet sits near the edge of the regions with significant detection of NH$_3$ (1, 1) emission or because of the highly elongated shape.}
\end{figure}

Following the analysis presented by \citet{McLaughlin_1996, McLaughlin_1997} and similar to the comparison with the Bonnor-Ebert model, we compare the observed radial profiles of density and pressure to a logotropic sphere with $A = 0.2$ \citep[Equation \ref{eq:eos_logo}, also used by][]{McLaughlin_1996} in dimensionless units (Fig.\ \ref{fig:profilesLogotrope}).  While Fig.\ \ref{fig:profilesLogotrope}a shows that a logotropic sphere has a density profile generally matching the droplet density profiles, the observed pressure profiles of the droplets decrease faster at increasing distances than the pressure profile of a logotropic sphere (see Fig.\ \ref{fig:profilesLogotrope}b).  The result suggests that the logotropic solution cannot describe the droplets, either.

In summary, we find that neither a critical Bonnor-Ebert sphere or a logotropic sphere describes the density and pressure profiles of the droplets well.  Instead, the shallow radial density and pressure profiles of the droplets can be approximated by a uniform density at smaller radii and a power-law density distribution approaching $\rho \propto r^{-1}$ at larger radii, the latter of which has also been observed for cloud-scale structures.


\subsubsection{Velocity Distribution of the Droplet Ensemble}
\label{sec:discussion_confinement_PPV}
The virial analysis presented in \S\ref{sec:analysis_virial} suggests that the confinement of the droplets is primarily provided by the ambient gas pressure.  Consistently, we find that the droplets have non-critical and relatively shallow density profiles approaching $\rho \propto r^{-1}$ at the outer edges.  Both results point to a close relation between the droplets and the local cloud environment.  Below, to investigate this relationship between the droplets and the surrounding cloud, we examine the distribution of emission in the position-position-velocity (PPV) space.

\begin{figure}[ht!]
\plotone{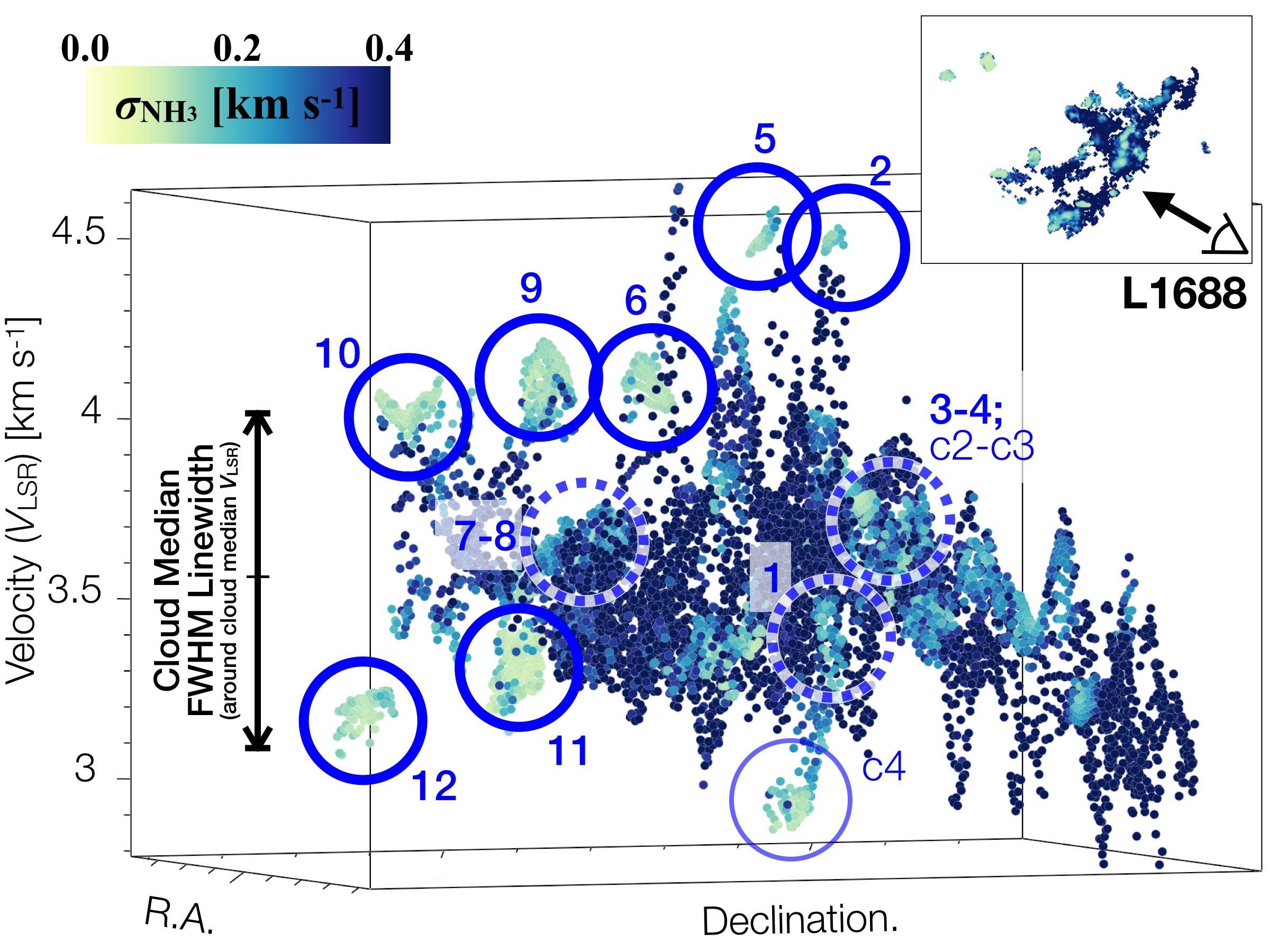}
\caption{\label{fig:PPV} The position-position-velocity (PPV) distribution of the best Gaussian fits to the NH$_3$ hyperfine line profiles, color-coded by the NH$_3$ velocity dispersion, $\sigma_{\mathrm{NH}_3}$.  Each dot corresponds to a pixel in the plane of the sky.  The map at the top right corner shows the projected point of view on the plane of the sky.  The droplets that are distinguishable in PPV space from the distribution of the bulk material in the cloud (as traced by NH$_3$ hyperfine line emission; usually the darker points) are marked by solid circles, while the approximate positions of the droplets that are more embedded in the cloud in PPV space are marked by dashed circles.  The numbers correspond to the droplet IDs in Table \ref{table:basic}, with the header ``L1688-'' removed for better visualization.  The visualization is made with the aid of Glue.}
\end{figure}

Fig.\ \ref{fig:PPV} shows the PPV distribution of the best fits to the NH$_3$ hyperfine line profiles observed at the pixels shown in Fig.\ \ref{fig:L1688_VlsrSigma}b, with the locations along the velocity axis equal to the velocity centroids of the best fits.  With each data point (the location of the Gaussian peak) color-coded by $\sigma_{\mathrm{NH}_3}$, several low linewidth features stand out having different line-of-sight velocities from the system velocity of the cloud, by $\sim$ 0.5 km s$^{-1}$.  Overall, we find that roughly half of the total 12 droplets in L1688 sit at the local extremes in $V_\mathrm{LSR}$, while the other half of the 12 droplets appear more embedded in the main cloud component in the PPV space.  Note that the distribution of emission in the PPV space does not correspond to the distribution of material in the position-position-position (PPP) space \citep{Beaumont_2013}, and the deviation in $V_\mathrm{LSR}$ from the main cloud component does not necessarily suggest that the droplet is separated from the cloud in the PPP space.



\begin{figure}[ht!]
\epsscale{0.6}
\plotone{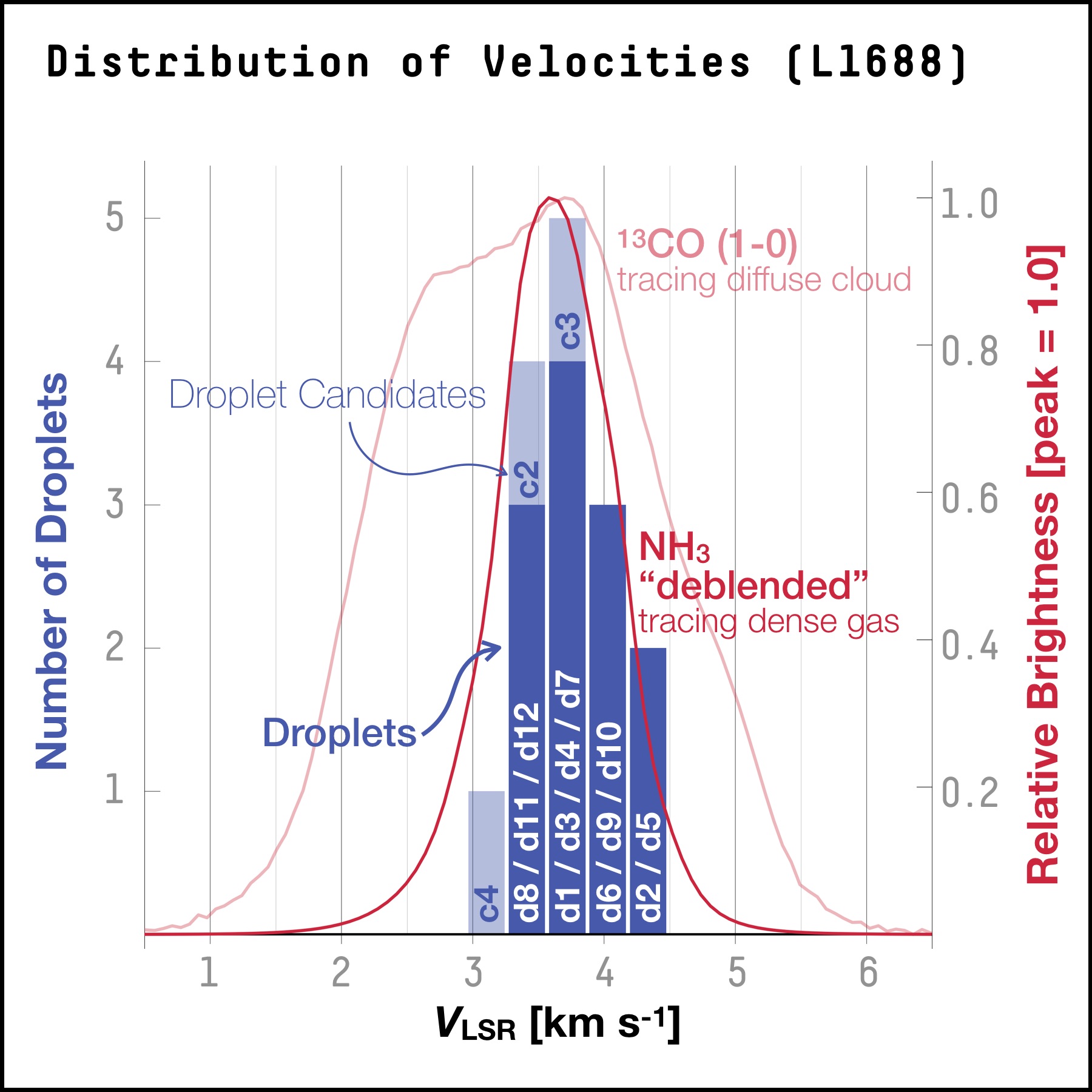}
\caption{\label{fig:Vlsr} Distribution of velocity centroids of the droplets and droplet candidates (blue histogram; lighter parts correspond to droplet candidates), plotted against the average spectra of NH$_3$ (dark red curve) and $^{13}$CO (1-0) emission (light red curve).  The NH$_3$ average spectrum is calculated from a ``deblended'' data cube created based on the results of NH$_3$ line fitting, such that each spectrum is a Gaussian with a center the same as the velocity centroid and a spread ($\sigma$) the same as the velocity dispersion.  The structure IDs of the structures included in each bin of the histogram are noted, with the leading ``L1688-'' removed for better visualization.  The spectra are shown in relative units wherein the peak has a value of 1.}
\end{figure}

Notably, the typical $V_\mathrm{LSR}$ difference of $\sim$ 0.5 km s$^{-1}$ between the $V_\mathrm{LSR}$ of droplets found at local velocity extremes and the system velocity of the cloud component traced by the NH$_3$ emission is comparable to half of the median FWHM linewidth of the NH$_3$ (1, 1) emission, $\sim$ 0.46 km s$^{-1}$ (shown as a vertical line along the velocity axis in Fig.\ \ref{fig:PPV}; FWHM$_{\mathrm{NH}_3}$ $\approx$ 0.92 km s$^{-1}$, measured for pixels outside the droplet boundaries---dark blue regions in Fig.\ \ref{fig:L1688_VlsrSigma}b).  A more detailed comparison shows that the dispersion in the velocity centroids of the droplets \citep[analogous to the ``core-to-core velocity'' examined by][]{Kirk_2010} agrees well with the median NH$_3$ velocity dispersion measured at pixels outside the droplet boundaries (see Table \ref{table:Vlsr}).  In Fig.\ \ref{fig:Vlsr}, we compare the distribution of droplet $V_\mathrm{LSR}$ to the average ``deblended'' spectrum of the entire L1688 region\footnote{\citet{GAS_DR1} constructed ``deblended'' spectral cubes from the results of the NH$_3$ hyperfine line fitting, assuming a single Gaussian line profile with the mean and the dispersion equal to $V_\mathrm{LSR}$ and $\sigma_{\mathrm{NH}_3}$ from the best fit for each pixel.  The deblending removes the NH$_3$ hyperfine line components and allows direct comparison with other spectra and velocity distributions.} and show that the distribution of droplet $V_\mathrm{LSR}$ has a shape similar to the deblended NH$_3$ line profile.  Given that the NH$_3$ velocity dispersion, $\sigma_{\mathrm{NH}_3}$, is associated with the thermal and turbulent motions of the dense gas, the results suggest that the droplets are traveling in the dense component of the cloud at velocities on par with the thermal and turbulent motions of the dense gas traced by NH$_3$ emission.  The result further suggest that the velocities of the droplets are inherited from the velocity dispersion of materials in the environment.


For reference, we also compare the distribution of droplet $V_\mathrm{LSR}$ to the average $^{13}$CO (1-0) spectrum\footnote{The $^{13}$CO spectrum is from the COMPLETE Survey of the molecular cloud in Ophiuchus \citep{Ridge_2006}.} and find that $^{13}$CO (1-0) has a line profile 2 to 3 times as broad as the droplet-to-droplet velocity distribution.  The result is consistent with what \citet{Kirk_2010} observed in Perseus.  Using the N$_2$H$^+$ emission to trace the dense core motions in the molecular cloud, \citet{Kirk_2010} found that the core-to-core velocity dispersion is about half of the total $^{13}$CO velocity dispersion in the region.

In the analyses presented in \S\ref{sec:analysis_virial} and \S\ref{sec:discussion_confinement}, we find that 1) the droplets generally appear not to be bound by self-gravity and predominantly confined by the ambient gas pressure and that 2) there is a close relation between the droplets and the local cloud component traced by the NH$_3$ emission.  Together, the results point to the possibility that the droplets, primarily defined by their subsonic and uniform interiors, are the result of compression due to the relatively more turbulent motions in the dense gas component of the cloud.  Below in \S\ref{sec:discussion_simulation}, we look for similar structures in a magnetohydrodynamic (MHD) simulation and speculate on the potential formation mechanism of the droplets.


\begin{deluxetable*}{lcc}
\tablecaption{Velocity Distribution of the Droplets and the Entire Cloud in the L1688 Region\label{table:Vlsr}}
\tablehead{\colhead{} & \colhead{Median Velocity} & \colhead{Velocity Dispersion}}
\startdata
Entire Cloud\tablenotemark{a} & $3.54^{+0.23}_{-0.22}$ & $0.39^{+0.16}_{-0.14}$ \\
Droplets\tablenotemark{b} & $3.68^{+0.42}_{-0.23}$ & $0.39$\tablenotemark{c}
\enddata
\tablenotetext{a}{Measured from pixel-by-pixel distributions on the maps of $V_\mathrm{LSR}$ and $\sigma_{\mathrm{NH}_3}$, excluding pixels within the droplet boundaries.}
\tablenotetext{b}{Measured from the droplet samples, as listed in Table \ref{table:basic}.}
\tablenotetext{c}{Measured by taking the standard deviation of the $V_\mathrm{LSR}$ distribution (see Table \ref{table:basic}); the velocity resolution of the observations is $\sim$ 0.07 km s$^{-1}$.}
\end{deluxetable*}

\subsection{Comparison with Hydrodynamic Models}
\label{sec:discussion_simulation}


\begin{deluxetable*}{lcccc}
\tablecaption{Comparison between the Droplets and the Structure Found in the MHD Simulation\label{table:models}}
\tablehead{\colhead{} & \colhead{Mass} & \colhead{Effective Radius} & \colhead{$\sigma_\mathrm{tot}$} & \colhead{Difference in LOS Velocity from the Cloud} \\ \colhead{} & \colhead{M$_\odot$} & \colhead{pc} & \colhead{km s$^{-1}$} & \colhead{km s$^{-1}$}}
\startdata
Droplets (Observation)\tablenotemark{a} & $0.4^{+0.4}_{-0.3}$ & $0.04\pm 0.01$ & $0.22\pm 0.02$ & 0.39/$\sim$0.5\tablenotemark{b} \\
Droplets (MHD Simulation)\tablenotemark{c} & $0.2\pm 0.1$ & $0.04\pm 0.01$ & $0.24\pm 0.02$ & 0.37 \\
\hline
Sim-d1 (MHD Simulation)\tablenotemark{c} & $0.96$ & $0.036$ & $0.24$ & $0.63$ \\
Sim-c1 (MHD Simulation)\tablenotemark{d} & $0.44$ & $0.031$ & $0.23$ & $0.14$
\enddata
\tablenotetext{a}{Median values with the lower and the upper bounds correspond to the 16th and 84th percentiles, respectively.}
\tablenotetext{b}{The standard deviation of the droplet $V_\mathrm{LSR}$ distribution is 0.39 km s$^{-1}$ (see Table \ref{table:Vlsr}).  For droplets that sit at local velocity extremes, the typical $V_\mathrm{LSR}$ difference is $\sim$ 0.5 km s$^{-1}$.}
\tablenotetext{c}{The droplets in the MHD simulation, including Sim-d1, are identified in the synthesized NH$_3$ spectral cube following the same procedure described in \S\ref{sec:analysis_id} (Fig.\ \ref{fig:MHD}; Smullen et al.\ \textit{in prep}).  See \S\ref{sec:discussion_simulation}.}
\tablenotetext{d}{Sim-c1 is found to associate with a shock-induced structure not unlike the one associated with Sim-d1.  While Sim-c1 also has a subsonic velocity dispersion, it is less clear whether a transition to coherence happens at its periphery (see Fig.\ \ref{fig:MHD}).  Thus, it is categorized as a ``droplet candidate.''}
\end{deluxetable*}


Simple analytical models could hint at the formation mechanism of droplets.  For example, by extending the Jeans model \citep{Jeans_1902}, \citet{Myers_1998} proposed a ``kernel'' model, in which a condensation with a mass of 1 M$_\sun$ and a size of 0.03 pc can exist within a dense core under ambient pressure provided by the thermal and turbulent motions.  Below, we demonstrate that formation of droplets is also possible in an MHD simulation of a turbulent cloud with self-gravity and sink particles, representing protostars.

We analyze an MHD simulation of a star-forming turbulent molecular cloud (Smullen et al.\ \textit{in prep}).  The simulation is carried out with the ORION2 adaptive mesh refinement (AMR) code \citep{Li_2012}.  The domain represents a piece of a molecular cloud 5 pc on a side with physical parameters and initialization identical to those of the W2T2 simulation in \citet{Offner_2015}.  The mean gas density is $\rho$ = 440 cm$^{-3}$ ($2.04\times10^{-21}$ g cm$^{-3}$).  The initial gas temperature is 10 K.  The ratio of thermal to magnetic pressure is $\beta = 8\pi c_\mathrm{s}^2/B^2 = 0.1$ and becomes 0.02 after 2 crossing times of driving.  The gas has a velocity dispersion of 1.98 km s$^{-1}$, which is set such that the cloud falls on the observed linewidth-size relation.  The calculation has 5 AMR levels with a maximum resolution of 125 AU.  We analyze a snapshot at 0.52 Myr or 0.35 t$_\mathrm{ff}$ as measured from when the initial driving phase ends and self-gravity is turned on.  At this time 1.3\% of the gas is in stars.

We use RADMC-3D\footnote{See \url{http://www.ita.uni-heidelberg.de/~dullemond/software/radmc-3d/index.html} for documentation.} to calculate the NH$_3$ emission given the simulated gas density and temperature distribution.  We adopt a uniform NH$_3$ abundance of $2\times10^{-9}$ n$_\mathrm{H}$.  We adopt the collisional parameters from the Leiden atomic and molecular database \citep{Schoeier_2005} and compute the radiative transfer using the non-local thermodynamic equilibrium large velocity gradient approximation \citep{Shetty_2011}.  To look for structures that show 1) a sharp change in velocity dispersion, and 2) locally concentrated emission, we derive the moment maps using the synthesized NH$_3$ spectral cube.  We then follow the same identification procedure described in \S\ref{sec:analysis_id} and identify a total of 8 droplets that show clear signs of a change in velocity dispersion and coincide with concentrated synthetic NH$_3$ emission, as well as another 4 droplet candidates.

The identified droplets in the MHD simulations have a typical effective radius of $0.04\pm 0.01$ pc, a typical mass of $0.2\pm 0.1$ M$_\sun$, and a typical total velocity dispersion of $0.24\pm 0.02$ km s$^{-1}$.  The droplets found in the simulation also have a typical difference in $V_\mathrm{LSR}$ of 0.37 km s$^{-1}$.  These values span a range similar to those found for the droplets identified in the observations within uncertainty (Table \ref{table:models}; see also Fig.\ \ref{fig:MHDproperties}).

Following the virial analysis presented in \S\ref{sec:analysis_virial}, we find that, similar to the droplets found in L1688 and B18, the droplets identified in the MHD simulation are generally not bound by self-gravity and are instead confined by the ambient pressure.  The ambient pressure of the droplets in the simulation is $P_\mathrm{amb}/k_\mathrm{B} \approx 1.4^{+1.7}_{-1.0} \times 10^5$ K cm$^{-3}$, comparable to the typical value of $P_\mathrm{amb}/k_\mathrm{B} \approx 2.7^{+4.7}_{-1.8} \times 10^5$ K cm$^{-3}$ for the droplets in L1688 and B18.



\begin{figure}[ht!]
\plotone{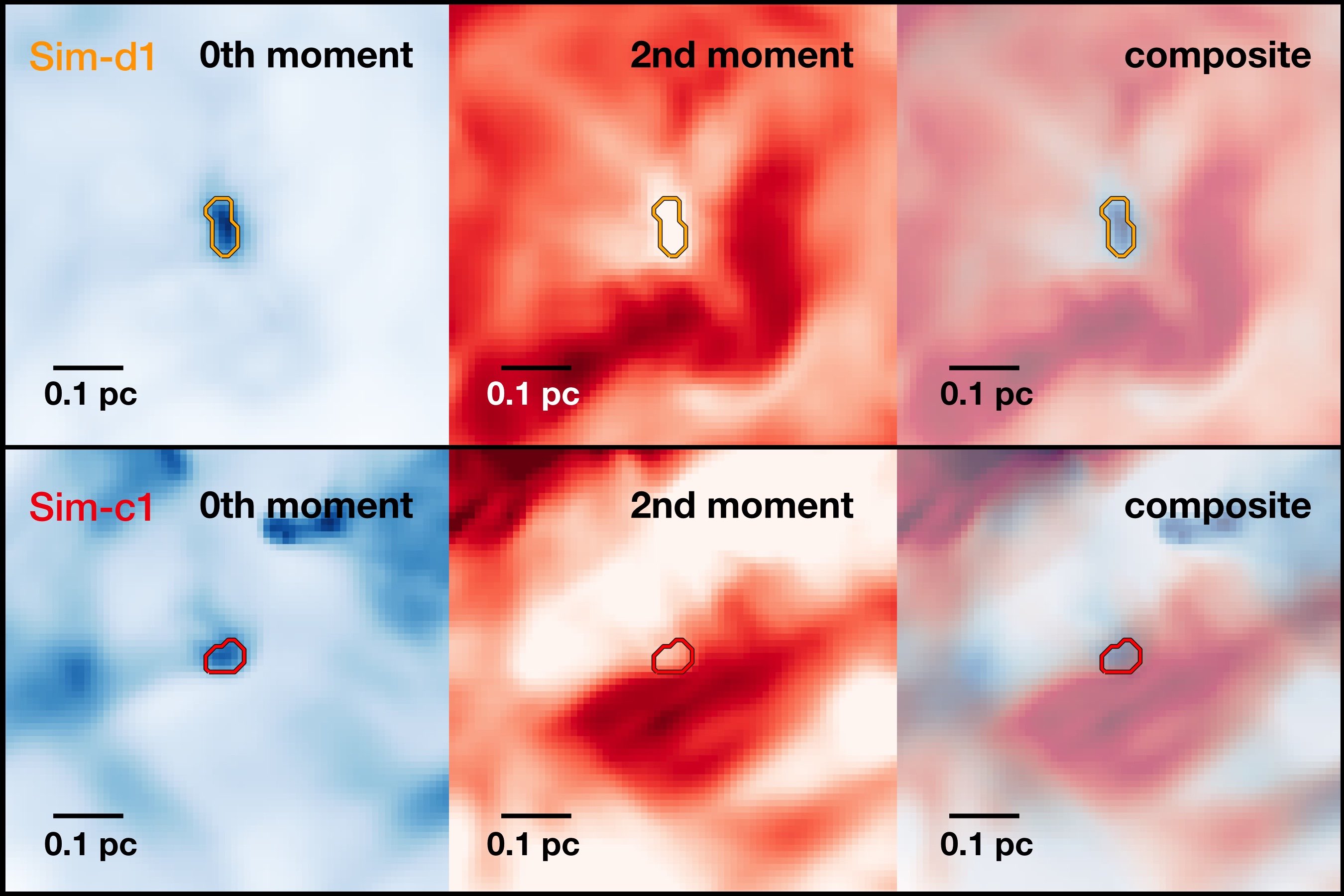}
\caption{\label{fig:MHD} \textbf{Top Row.} Sim-d1, a droplet identified in the MHD simulation following the procedure described in \S\ref{sec:analysis_id} (Smullen et al.\ \textit{in prep}), with identifiable rise in integrated emission (0th moment; left panel) and a sharp drop in velocity dispersion (2nd moment; middle panel) near its edge.  The rightmost panel shows an overlay of the 0th and 2nd moment maps.  \textbf{Bottom Row.} Same as the top row but showing Sim-c1, another structure associated with an isolated shock-induced feature which has a subsonic velocity dispersion but where signs of a transition to coherence are less clear.   Note that each panel in this figure shows only a 0.7 pc by 0.7 pc region near the dense structures.}
\end{figure}


\begin{figure}[ht!]
\plotone{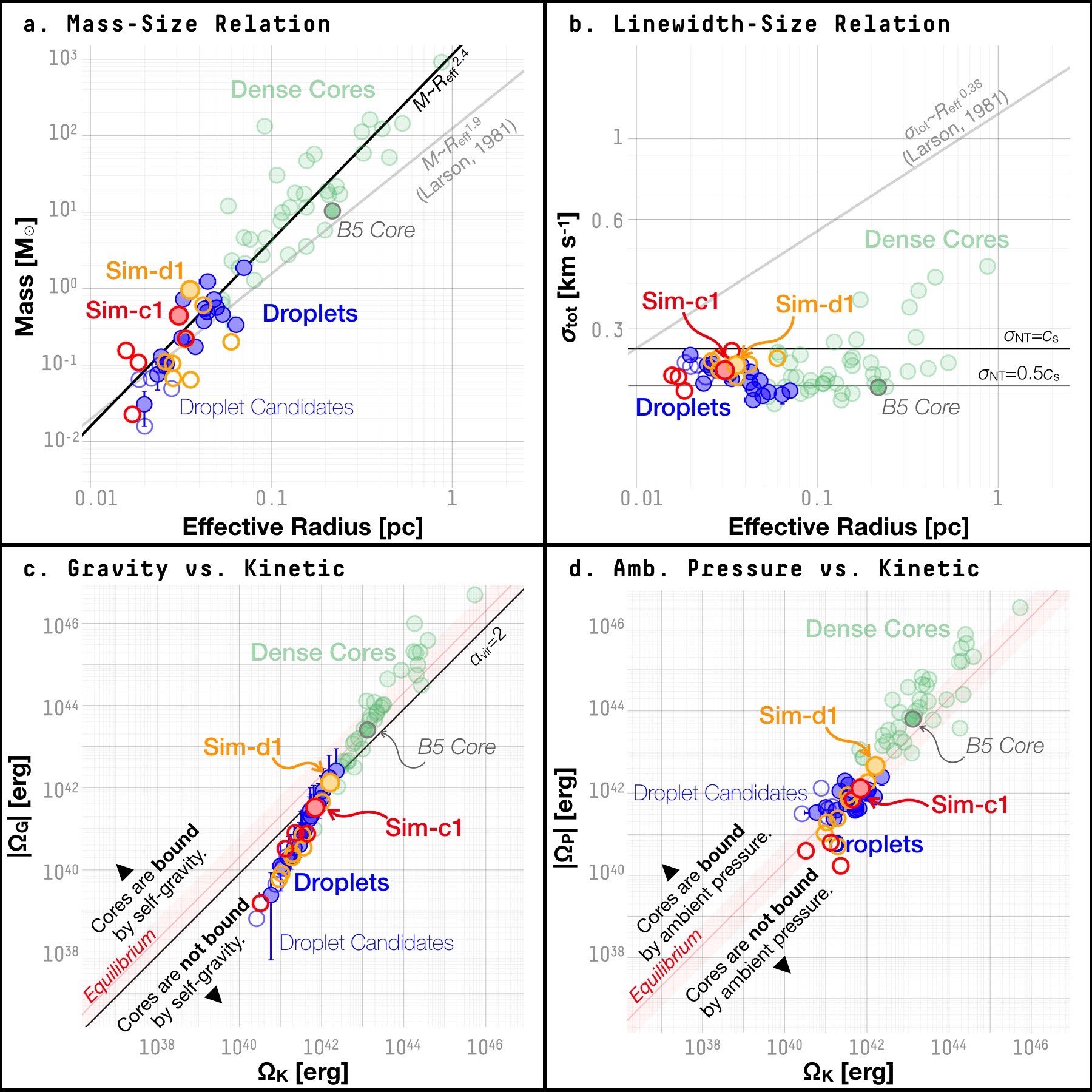}
\caption{\label{fig:MHDproperties} Figs.\ \ref{fig:basic} and \ref{fig:virialGP} but for droplets (yellow circles) and droplet candidates (red circles) in the MHD simulation.  Sim-d1 and Sim-c1 shown in Fig.\ \ref{fig:MHD} are plotted as a filled yellow circle and a filled red circle, respectively, against the dense cores (green circles) and the droplets (blue circles).  \textbf{(a)} The mass-size distribution (see Fig.\ \ref{fig:basic}a).  \textbf{(b)} The $\sigma_\mathrm{tot}$-size distribution (see Fig.\ \ref{fig:basic}b).  \textbf{(c)} Gravitational potential energy, $\Omega_\mathrm{G}$, plotted against internal kinetic energy, $\Omega_\mathrm{K}$ (see Fig.\ \ref{fig:virialGP}a).  \textbf{(d)} The energy term representing the confinement provided by the ambient gas pressure, $\Omega_\mathrm{P}$, plotted agains the internal kinetic energy, $\Omega_\mathrm{K}$ (see Fig.\ \ref{fig:virialGP}b).}
\end{figure}

\begin{figure}[ht!]
\plotone{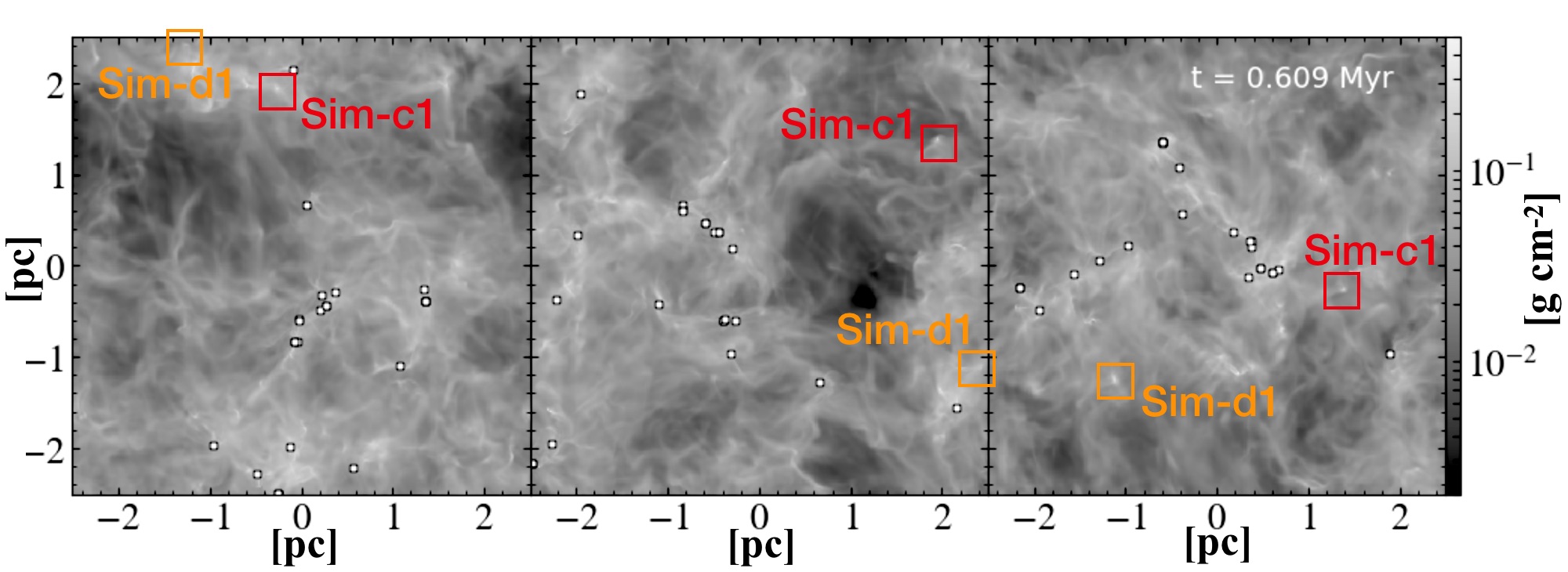}
\caption{\label{fig:MHD_fullcube} The integrated density along three different viewing directions of the MHD simulation examined in \S\ref{sec:discussion_simulation}.  The rightmost panel shows the view adopted for the synthesized NH$_3$ cube is derived and used to calculate the moment maps shown in Fig.\ \ref{fig:MHD}.  The yellow square marks the droplet-like structure identified following the same procedure described in \S\ref{sec:analysis_id} and shown in Fig.\ \ref{fig:MHD}.  The red square marks another shock-induced feature, around which signatures of a subsonic velocity dispersion and a concentrated density distribution are found.  The white dots mark the positions of the sink particles.  \textbf{In an 40-second animated version of this plot available in the HTML version of this article, we show the evolution of the simulated density cube in the same integrated density units (g cm$^{-2}$) from $1.46\times 10^5$ yr to $6.09\times 10^5$ yr in the evolutionary time.  In the animation, multiple filamentary structures as well as more isolated shock-induced features are found moving across the star forming medium.  The static version shown here corresponds to the last frame of the animation at $6.09\times 10^5$ yr in the evolutionary time, at which time step we identified Sim-d1 and Sim-c1 using synthetic observations of NH$_3$ emission.  We use a monotonic color map in the static version to better show the integrated density in contrast to the orange and red boxes used to mark the identified structures.  The animation is also available at \url{https://goo.gl/PEd9Pd}.}}
\end{figure}

To gain insight into droplet formation, we select an isolated droplet, Sim-d1, and follow its evolution in the MHD simulation (Figs.\ \ref{fig:MHD} and \ref{fig:MHD_fullcube}).  We find that Sim-d1 corresponds to a relatively isolated shock-induced feature in the MHD simulation, moving generally toward the viewer along the line of sight on which we ``observe'' the synthesized NH$_3$ cube (see Fig.\ \ref{fig:MHD_fullcube} and the video showing the evolution of the MHD cube linked in the caption).  Meanwhile, material seems to accumulate at the converging point of the shock-induced feature as the simulation evolves.  The general movement of Sim-d1 toward the viewer is consistent with the relatively high line-of-sight velocity difference observed in the synthesized NH$_3$ cube, $\sim$ 0.63 km s$^{-1}$.  The association between a droplet and a shock-induced feature, viewed from different angles, might explain why the observed droplets are sometimes found at local line-of-sight velocity extremes.


A significant portion of droplets and droplet candidates identified in the simulation are found to be associated with similar shock-induced features (e.g., Sim-c1 marked by the red square in Fig.\ \ref{fig:MHD_fullcube}).  These structures generally have subsonic internal velocity dispersions but do not necessarily satisfy all criteria used for identification in \S\ref{sec:analysis_id}.  A typical example is a droplet candidate, Sim-c1 (Fig.\ \ref{fig:MHD}).  We find that Sim-c1 has properties consistent with the physical properties of the droplets identified in observations (see Fig.\ \ref{fig:MHDproperties} Table \ref{table:models}).


Notably, the most active star-forming regions are found near points where shocks are colliding and gas is converging.  This creates local density enhancements, which are conducive to core formation if matter continues to accumulate (see Fig.\ \ref{fig:MHD_fullcube}).  It seems possible that the droplet-like features associated with isolated shock features might evolve into star-forming cores through continuing accumulation of material and/or through converging with other shock-induced features.  See more discussion below in \S\ref{sec:discussion_definition}.


\subsection{Cores \& Droplets}
\label{sec:discussion_definition}



In this work, we examine the physical properties of two closely related populations of structures: the droplets and the dense cores (among which many were found to be coherent cores).  In the analyses presented above, we essentially use the two terms, the droplets and the dense cores, to indicate structures identified in this work and those examined by \citet{Goodman_1993}, respectively.  Although the two populations indeed have different physical properties, we are not satisfied with this rather arbitrary use of terminology.  Thus, we provide a more physical set of definitions for different groups of structures discussed in this paper below.

We define \textit{droplets} to be \emph{gravitationally unbound and pressure confined coherent structures}, wherein the \textit{coherent structures} include any structures that have subsonic velocity dispersions and show transitions to coherence.  In comparison, a \textit{coherent core} is a gravitationally bound coherent structure, and a \textit{dense core} is a centrally concentrated density feature with a velocity dispersion approaching a transonic or subsonic value (not necessarily showing a transition to coherence).  Just like a coherent core may correspond to the densest region of a dense core, a droplet may be the innermost region of a larger structure.  The definitions are summarized in Table \ref{table:definition}.

\begin{deluxetable*}{lccccc}
\tablecaption{Cores and Droplets\label{table:definition}\tablenotemark{a}}
\tablehead{\colhead{} & \colhead{Therm. Dominated Linewidths\tablenotemark{b}} & \colhead{Transition to Coherence\tablenotemark{c}} & \colhead{Assoc. with YSOs} & \colhead{Gravitationally Bound} & \colhead{Pressure Bound}}
\startdata
Droplets & Yes & Yes & Neutral & No & Yes \\
\hline
Dense Core\tablenotemark{d} & Yes & Neutral & Neutral & (Yes) & Neutral \\
Coherent Core\tablenotemark{e} & Yes & Yes & Neutral & Yes & Neutral \\
Starless Cores\tablenotemark{f} & Neutral & Neutral & No & No & Neutral \\
Prestellar Cores\tablenotemark{f} & Neutral & Neutral & No & Yes & Neutral \\
Protostellar Cores\tablenotemark{g} & Neutral & Neutral & Yes & Yes & Neutral
\enddata
\tablenotetext{a}{This table lists the definitions of names commonly given to subsets of cores, as logical combinations of several criteria.  A value of ``Yes'' means that a core needs to meet the criterion in order to be assigned to a certain category.  A value of ``No'' means that a core needs to satisfy the \emph{negation} of the criterion.  A value of ``Neutral'' means that the definition of a certain category does not concern the criterion.}
\tablenotetext{b}{The thermal component of the velocity dispersion is larger than the non-thermal (turbulent) component.}
\tablenotetext{c}{Observation of ``transition to coherence,'' as described by \citet{Goodman_1998}.  The observation of ``transition to coherence'' may be done by observing the same core with multiple tracers and focusing on the change in the linewidth-size relation going from one tracer to the next \citep[Type 4 in Fig.\ 9 of][later used by \citealt{Caselli_2002} in observations of coherent cores]{Goodman_1998}.  Another way to observe the ``transition to coherence'' is to spatially resolve the transition with a single tracer.  Examples include observations of NH$_3$ emission in B5 \citep{Pineda_2010} and the droplets in this work.  Note that the criterion of ``transition to coherence'' is stricter than ``thermally dominated linewidths.''  The thermally dominated linewidths concern the overall measurement of velocity dispersion in the core but not the spatial change in velocity dispersion.}
\tablenotetext{d}{The canonical example of dense cores \citep[e.g.\ ][]{Myers_1983a} is simply defined by a centrally concentrated density distribution.  Based on observations of NH$_3$ emission and emission from other higher density molecular line tracers, \citet{Myers_1983c} found that most of the dense cores examined by \citet{Myers_1983a} had velocity dispersions approaching transonic or subsonic values.  Gravitational boundedness was less clear, oftentimes because of a lack of necessary observations to accurately estimate the boundedness of these structures.  The dense cores analyzed by \citet{Goodman_1993} and included in this paper are mostly gravitationally bound, as shown in Fig.\ \ref{fig:virialGP}.}
\tablenotetext{e}{As described by \citet{Goodman_1998} and later observed by \citet{Caselli_2002} and \citet{Pineda_2010}.}
\tablenotetext{f}{In the literature, the starless cores and the prestellar cores are both not associated with any YSOs.  A criterion often used to distinguish between the two categories is the gravitational boundedness.  Prestellar cores are cores that are gravitationally bound, and starless cores are those that are not \citep[e.g.,][]{Tafalla_2004}.  In some cases, density features at smaller scales within the cores are used to further investigate the ``starlessness'' of the starless cores \citep[e.g.,][]{Kirk_2017a}.}
\tablenotetext{g}{By definition, protostellar cores are cores associated with YSOs.}
\end{deluxetable*}

Due to the lack of observations needed to determine whether the cores examined by \citet{Goodman_1993} show any signs of a transition to coherence, we simplify the criteria and recategorize the structures identified in this work and those examined by \citet{Goodman_1993} into two categories: 1) ``droplets'': structures not virially bound by self-gravity and with subsonic velocity dispersions, and 2) ``dense cores'': structures virially bound by self-gravity and with subsonic velocity dispersions.  In the discussion below, we retain the quotation marks around these terms to differentiate them from the terms used throughout the analyses above.

\begin{figure}[ht!]
\plotone{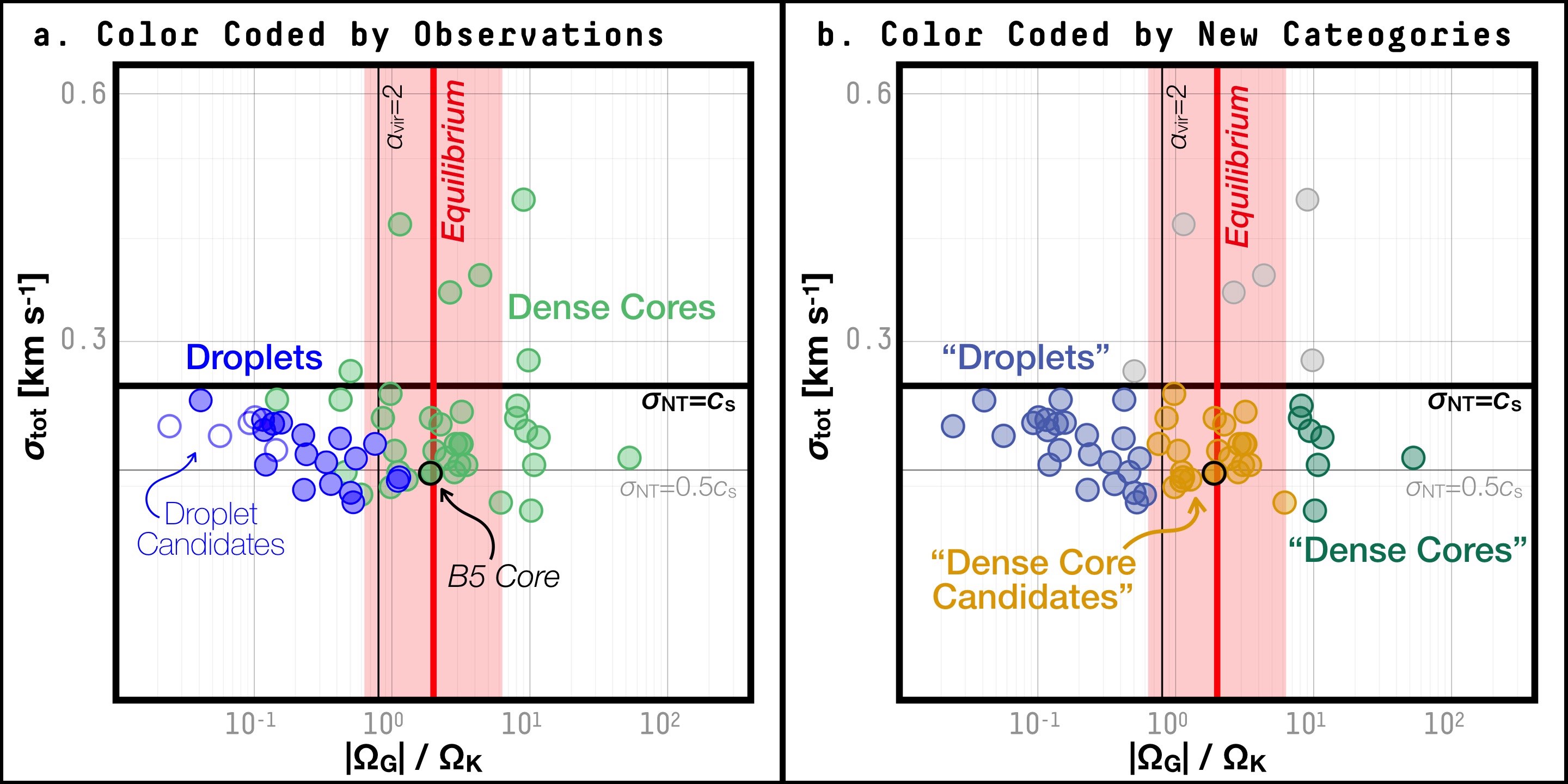}
\caption{\label{fig:definition} \textbf{(a)} Total velocity dispersion, $\sigma_\mathrm{tot}$, plotted against the ratio between $\Omega_\mathrm{G}$ and $\Omega_\mathrm{K}$.  The droplets and droplet candidates are plotted as solid and empty blue circles, and the dense cores are plotted as green circles.  The horizontal lines show the total velocity dispersions expected for structures where the non-thermal component is equal to the sonic speed (thicker, black line) and half the sonic speed (thinner, gray line) of a medium with a mean molecular weight of 2.37 u at a temperature of 10 K.  The vertical red band marks an equilibrium between $\Omega_\mathrm{G}$ and $\Omega_\mathrm{K}$ (solid red line) within an order of magnitude (pink band).  The vertical black line marks $\alpha_\mathrm{vir} = 2$.  \textbf{(b)} Same as (a), but with the data points color coded according to the proposed recategorization of structures into the ``dense cores'' (dark green circles), the ``dense core candidates'' (dark yellow circles), and the ``droplets'' (dark blue circles; see \S\ref{sec:discussion_definition}).  Structures that have supersonic velocity dispersions are omitted in this recategorization and plotted as gray circles.}
\end{figure}

Fig.\ \ref{fig:definition} shows how this recategorization would change the groupings of structures examined in the analyses in this paper.  To proceed with caution and to avoid uncertainty in using a virial analysis to determine the equilibrium state of a structure, we categorize those structures with subsonic velocity dispersions and within an order of magnitude of a virial equilibrium between the gravitational potential energy and the kinetic energy (see details in \S\ref{sec:analysis_virial}) as ``dense core candidates.''  Notice in Fig.\ \ref{fig:definition}, most of these ``dense core candidates'' have virial parameters $\leq$ 2 and would conventionally be considered virially bound by self-gravity.  In this recategorization, we temporarily omit structures with supersonic velocity dispersions, although the largest turbulent Mach number (the ratio between the turbulent component of velocity dispersion and the sonic speed) found in these structures is $\lesssim$ 1.5, i.e., not anywhere close to the turbulence measured for the entire molecular cloud.  For example, using observations of the $^{13}$CO (1--0) emission, we measure a turbulent Mach number of $\sim$ 10 for Ophiuchus.  In total, three out of 43 dense cores are recategorized as ``droplets,'' and three out of 18 droplets are recategorized as ``dense core candidates.''

\begin{figure}[ht!]
\plotone{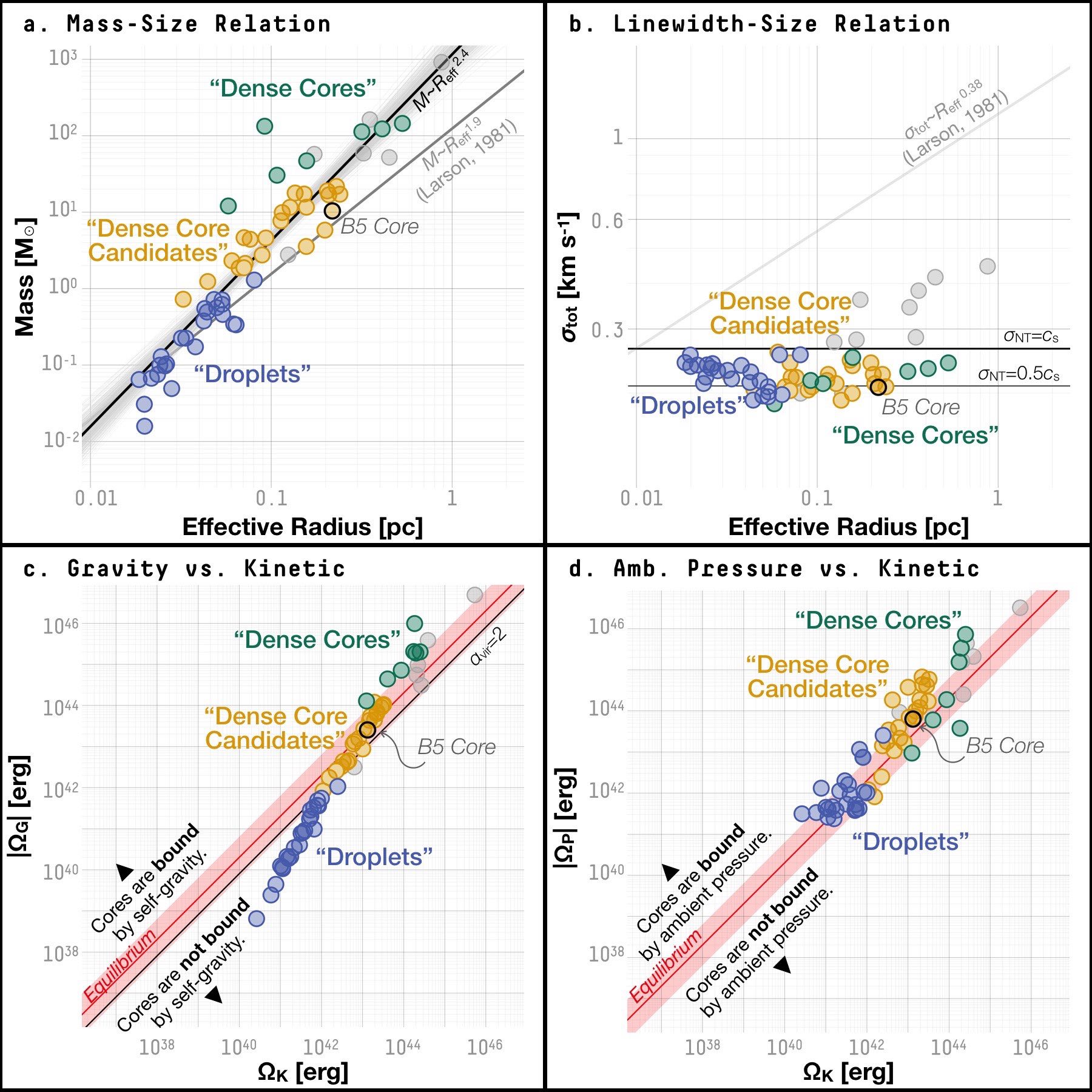}
\caption{\label{fig:DefinitionProperties} Like Figs.\ \ref{fig:basic} and \ref{fig:virialGP} but color-coded for recategorized structures: the ``dense cores'' (dark green circles), the ``dense core candidates'' (dark yellow circles), and the ``droplets'' (dark blue circles). Structures that have supersonic velocity dispersions are omitted in this recategorization and plotted as gray circles.  \textbf{(a)} The mass-size distribution (see Fig.\ \ref{fig:basic}a).  \textbf{(b)} The $\sigma_\mathrm{tot}$-size distribution (see Fig.\ \ref{fig:basic}b).  \textbf{(c)} Gravitational potential energy, $\Omega_\mathrm{G}$, plotted against internal kinetic energy, $\Omega_\mathrm{K}$ (see Fig.\ \ref{fig:virialGP}a).  \textbf{(d)} The energy term representing the confinement provided by the ambient gas pressure, $\Omega_\mathrm{P}$, plotted agains the internal kinetic energy, $\Omega_\mathrm{K}$ (see Fig.\ \ref{fig:virialGP}b).}
\end{figure}

Fig.\ \ref{fig:DefinitionProperties} shows the distributions of the recategorized structures in various parameter spaces.  It is evident that different populations---``droplets'' and ``dense cores''---categorized according to their physical properties are mingled and form a continuous distribution.  The continuity might suggest that different populations of coherent structures emerge from the same set of physical processes, although they differ in gravitational boundedness.  Based on the MHD simulation examined in \S\ref{sec:discussion_simulation}, one possible scenario is that both the gravitationally bound coherent structures---coherent cores---and the gravitationally unbound and pressure confined coherent structures---droplets---arise from shock-induced overdensities.  In this scenario, we would likely find star-forming coherent cores at the converging points of multiple shocks, and we would find droplets around more isolated shock-induced structures and/or isolated pairs of colliding shocks.

%
Based on an examination of the MHD simulation (\S\ref{sec:discussion_simulation}), a shock induced formation mechanism of coherent structures might point to an evolutionary sequence connecting droplets and coherent cores.  If droplets are formed as isolated shock-induced structures, they might still evolve into star-forming and gravity dominated coherent cores in the future if the isolated shocks converge/collide with other shock-induced structures.  Similarly, based on observations of core structures in the B218 region in Taurus, \citet{Seo_2015} find that it is more likely to find Class 0/I YSOs associated with gravitationally bound cores than with pressure confined structures.  Thus, \citet{Seo_2015} suggest that there exists an evolutionary sequence connecting the pressure confined structures to the gravitationally bound cores.  The evolutionary sequence suggested by \citet{Seo_2015} might conform with the conventional evolutionary sequence observed toward the prestellar and the protostellar cores (see Table \ref{table:definition}).

\begin{figure}[ht!]
\plotone{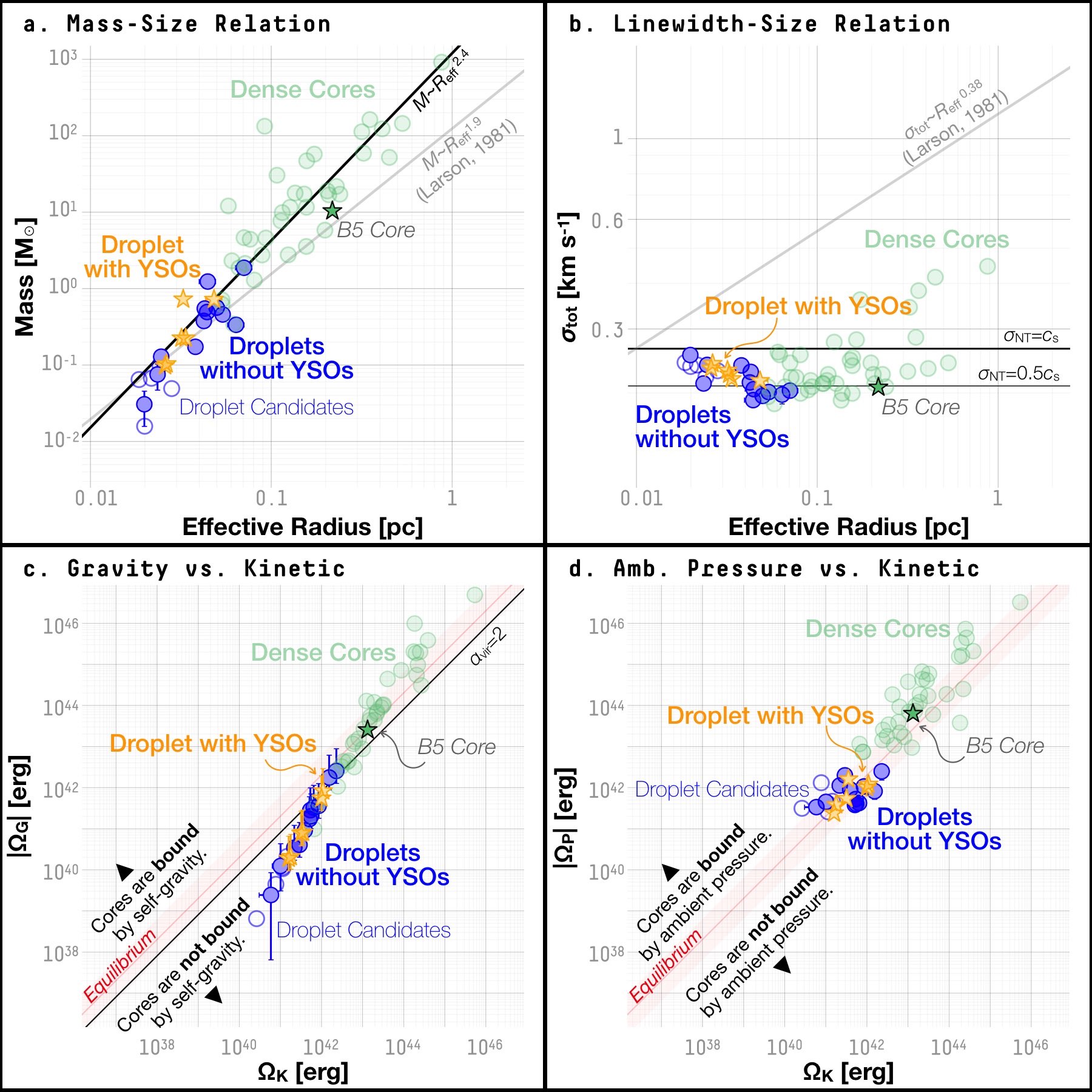}
\caption{\label{fig:YSOproperties} Like Figs.\ \ref{fig:basic} and \ref{fig:virialGP} but marked for the droplets with YSOs within the boundaries (yellow star marks) and the droplets without YSOs within boundaries (blue circles), in comparison to the dense cores (green circles) and the coherent core in B5 (the green star mark; the coherent core in B5 has at least one YSO within its boundary).  \textbf{(a)} The mass-size distribution (see Fig.\ \ref{fig:basic}a).  \textbf{(b)} The $\sigma_\mathrm{tot}$-size distribution (see Fig.\ \ref{fig:basic}b).  \textbf{(c)} Gravitational potential energy, $\Omega_\mathrm{G}$, plotted against internal kinetic energy, $\Omega_\mathrm{K}$ (see Fig.\ \ref{fig:virialGP}a).  \textbf{(d)} The energy term representing the confinement provided by the ambient gas pressure, $\Omega_\mathrm{P}$, plotted agains the internal kinetic energy, $\Omega_\mathrm{K}$ (see Fig.\ \ref{fig:virialGP}b).}
\end{figure}

Since droplets are found within active star-forming regions, the projection effect makes it difficult to determine whether or not the droplets are associated with any YSOs.  If we simply look at the existence of YSO(s) within the droplet boundary projected on the plane of the sky, we find that five out of 12 droplets in L1688 and one out of 6 droplets in B18 coincide with at least one YSO within each of their boundaries.  This gives a $\lesssim$ 40\% chance of finding at least one YSO within the droplet boundary.  As described in \S\ref{sec:analysis_id}, we do not find significant rises in $T_\mathrm{kin}$ in any of the droplets where we find YSOs, and as shown above in \S\ref{sec:analysis} and in Fig.\ \ref{fig:YSOproperties}, we do not find a statistically significant difference in physical properties between the droplets with YSOs within their boundaries and those without.  Compared to the coherent core in B5, where a YSO and at least three star-forming substructures are found, we do not find signs of star-forming substructures in the droplets with YSOs within their boundaries \citep[see \S\ref{sec:analysis_virial};][]{Pineda_2015}.  These facts indicate that the innermost part not resolved by our observations may be gravitationally bound in these droplets, and if this is proved true, it is likely that there exists an evolutionary sequence connecting the pressure-confined droplets to the star-forming gravitationally bound coherent cores \citep[such as the one in B5; see also discussions in][]{Seo_2015}.  Follow-up higher-resolution observations are needed to establish the association between the droplets and the YSOs found within their boundaries and the effects of the YSOs on the evolution of droplets.

In conclusion, droplets are a previously omitted sub-population of coherent structures.  Although imminent star formation within droplets is unlikely because of the gravitational unboundedness,  droplets might form from the same set of physical processes that lead to the formation of star-forming coherent cores.  Since the subsonic velocity dispersion within a coherent core is expected to be disturbed by ongoing formation of stars, droplets may provide a precious chance to examine the internal kinematics and the formation of coherent structures.  Furthermore, there could exist an evolutionary sequence connecting the pressure dominated droplets to the star-forming coherent cores, but this cannot be confirmed with present data.  More works to systematically examine droplets in simulations and to compare them with droplets and other cores identified in observations are needed to answer the following questions: How do droplets form?  Do droplets evolve into star-forming cores, and if so, how?  What is the relation between coherent structures, including both star-forming coherent cores and pressure dominated droplets, and other populations of cores (e.g., starless and protostellar cores) and structures \citep[e.g., filaments and bundles][]{Hacar_2013}?  Are there observable velocity gradients and potentially associated rotational and/or shear motions in the interiors of coherent structures?  Would the coherent structures fragment into smaller features in the future?  We will address some of these questions in subsequent papers of this series.

\section{Conclusion}
\label{sec:conclusion}
In search of coherent structures defined by a change in velocity dispersion from supersonic to nearly constant subsonic values (\S\ref{sec:analysis_id}; Figs.\ \ref{fig:L1688_TpeakTkin}, \ref{fig:L1688_VlsrSigma}, and \ref{fig:L1688_Herschel} for L1688, and Figs.\ \ref{fig:B18_TpeakTkin}, \ref{fig:B18_VlsrSigma}, and \ref{fig:B18_Herschel} for B18), we identify a total of 18 coherent structures in the L1688 region of Ophiuchus and the B18 region of Taurus, using data from the first data release of the Green Bank Ammonia Survey \citep[see \S\ref{sec:data_GAS};][]{GAS_DR1}.  The 18 coherent structures newly identified within a total projected area of $\sim$ 0.6 pc$^2$ suggest that the coherent structures are ubiquitous in nearby molecular clouds and allow statistical analyses of coherent structures for the first time.

The newly identified coherent structures have a typical radius of 0.04 pc and a typical mass of 0.4 M$_\sun$ (\S\ref{sec:analysis_basic}; Table \ref{table:basic}) and appear to follow the same mass-linewidth-size relation as the dense cores previously examined by \citet{Goodman_1993}, many of which are later found to be coherent cores \citep[see Fig.\ \ref{fig:basic};][]{Goodman_1998, Caselli_2002, Pineda_2010}.  In a virial analysis, we find that the newly identified coherent structures are not virially bound by self-gravity and are instead confined by the pressure provided by the ambient gas motions (see \S\ref{sec:analysis_virial}).  This clearly differentiates the newly identified coherent structures from previously known coherent cores, which have been found to be gravitationally bound and sometimes hosting ongoing star formation \citep{Pineda_2010, Pineda_2015}.  We term this newly discovered population of \emph{gravitationally unbound and pressure confined coherent structures} the \textit{droplets}.

The radial density and pressure profiles of the droplets cannot be well described by either a critical Bonnor-Ebert sphere or a logotropic sphere (see \S\ref{sec:discussion_confinement_BE} and \S\ref{sec:discussion_confinement_logo}).  The droplets have relatively shallow density profiles (e.g., compared to previously observed starless cores; see Fig.\ \ref{fig:profilesDensity}b), and their density profiles can generally be approximated by a constant density at smaller radial distances and a power-law density distribution approaching $\rho \propto r^{-1}$ at larger distances, the latter of which has been observed toward cloud-scale structures (Fig.\ \ref{fig:profilesDensity}).  While the droplets are sometimes found at local extremes of the line-of-sight velocity, the $V_\mathrm{LSR}$ distribution of the droplets has a shape similar to that of the average NH$_3$ line profile (see \S\ref{sec:discussion_confinement_PPV}; see also Fig.\ \ref{fig:PPV} and Fig.\ \ref{fig:Vlsr}).  Both the power-law density profiles ($\rho \propto r^{-1}$) and the distribution of $V_\mathrm{LSR}$ suggest a close relation between the droplets and the natal cloud environment.

By identifying droplet-like structures in the synthesized NH$_3$ cube, we demonstrate that the formation of droplets is possible in an MHD simulation of a star-forming cloud.  The droplet-like structures examined in \S\ref{sec:discussion_simulation} appear to correspond to shock-induced features in the simulation, and throughout the evolution, material accumulates at shock-induced converging points.  Given the active star formation in the same simulation emerges in regions near the converging points of multiple shocks, we speculate that a droplet might evolve into a star-forming core if accumulation continues and/or if the associated shock-induced feature converges/collides with other shocks.

More work is needed to understand the formation and evolution of droplets and coherent structures in general.  With the GAS data, we hope to extend our analyses on coherent structures to other nearby molecular clouds.  The GAS observations of NH$_3$ hyperfine line emission also allows an analysis of the internal velocity structures, which would shed light on the potential rotational and shear motions in the droplets/coherent cores.  On the other hand, more targeted modeling and a statistical approach are needed to further understand the physical processes involved in the formation of droplets and the role droplets might play in star formation.

\acknowledgments
The National Radio Astronomy Observatory is a facility of the National Science Foundation operated under cooperative agreement by Associated Universities, Inc.  This work was supported by a Cottrell Scholar Award.  A.P. acknowledges the support of the Russian Science Foundation project 18-12-00351.  

\vspace{5mm}
\facilities{GBT (KFPA+VEGAS), Herschel Space Observatory (PACS+SPIRE)}

\software{astropy \citep{astropy}, Glue \citep{glue, glue_2017}, PySpecKit \citep{pyspeckit}, RADMC-3D \citep{radmc3d}}

\appendix

\section{Summary of Updated Distances for Cores Presented by Goodman et al.\ (1993)}
\label{sec:appendix_distances}
To compare the droplets identified in this work to previously known dense cores, we correct the physical properties summarized in \citet{Goodman_1993} by more recent distance measurements.  The updated distances are summarized below:

\begin{enumerate}
\item Regions associated with the molecular cloud in Perseus: \textbf{PER3}, \textbf{PER6}, and \textbf{B5}.  We adopt distances measured by \citet{Schlafly_2014} using \textit{PanSTARRS-1} photometry, which are $260\pm 26$ pc for the western part of the Perseus molecular cloud (including PER3 and PER6) and $315\pm 32$ pc for the eastern part (including B5).

\item Regions associated with the molecular cloud in Taurus: \textbf{L1489}, \textbf{L1498}, \textbf{L1495}, \textbf{L1495NW}, \textbf{L1495SE}, \textbf{TAU11}, \textbf{TAU16}, \textbf{B217}, \textbf{L1524}, \textbf{TMC-2A}, \textbf{L1534} (TMC-1A), \textbf{L1527}, \textbf{TMC-1C}, and \textbf{L1517B}.  We adopt a distance of $126.6\pm 1.7$ pc, measured by \citet{Galli_2018}.

\item Regions associated with $\lambda$ Orionis: \textbf{L1582A} and \textbf{B35A}.  We adopt a distance of $420\pm 42$ pc, measured by \citet{Schlafly_2014}.

\item A region associated with the molecular cloud and the YSO cluster in Ophiuchus (sometimes referred to as ``$\rho$ Oph''): \textbf{L1696A}.  We adopt a distance of $137.3\pm 6$ pc, measured by \citet{OrtizLeon_2017} using parallax.

\item Regions associated with clouds and clumps in Oph N: \textbf{L43/RNO90}, \textbf{L43}, \textbf{L260} (a.k.a.\ L255), \textbf{L158}, \textbf{L234E}, \textbf{L234A}, and \textbf{L63}.  These regions are usually associated with the Ophiuchus complex or, on a larger scope, the Upper Sco-Oph-Cen complex. \citet{Goodman_1993} adopted the same distance for these regions as for L1696A.  Here we use an updated distance measurement of $125\pm 18$ pc to the Ophiuchus complex by \citet{Schlafly_2014}.  This is in good agreement with the widely used $125\pm 45$ pc, measured by \citet{deGeus_1989}.

\item Regions associated with Cepheus Flare: The Cepheus Flare spans more than 10 degrees from North to South on the plane of the sky, and is known to have a complicated structure with multiple concentrations of material at different distances.  Here we adopt different distance measurements for different regions in Cepheus Flare, and note that these distances were used by \citet{Kauffmann_2008} side-by-side.  Note that \citet{Schlafly_2014} measured $360\pm 35$ pc for the southern part of Cepheus Flare and $900\pm 90$ pc for the northern part of Cepheus Flare.  See discussions in \citet{Schlafly_2014}.

\begin{itemize}
\item \textbf{L1152}: $325\pm 13$ pc, measured by \citet{Straizys_1992} using photometry.

\item \textbf{L1082C}, \textbf{L1082A}, and \textbf{L1082B}: $400\pm 50$ pc, measured by \citet{Bourke_1995} using photometry.

\item \textbf{L1174} and \textbf{L1172A}: $288\pm 25$ pc, measured by \citet{Straizys_1992} using photometry.

\item \textbf{L1251A}, \textbf{L1251E}, and \textbf{L1262A}: we update the distance used by \citet{Kauffmann_2008} based on \citet[][$300^{+50}_{-10}$ pc]{Kun_1998} with a more recent measurement of $286\pm 20$ pc made by \citet{Zdanavicius_2011} using photometry.
\end{itemize}

\item Regions with distances measured from masers:
\begin{itemize}
\item \textbf{L1400G} and \textbf{L1400K}: $170\pm 50$ pc, measured by \citet{Montillaud_2015}.

\item\textbf{L134A}: $110\pm 10$ pc, measured by \citet{Montillaud_2015}.
\end{itemize}

\item Other regions of which the distances have not updated since the 1990s but are cited recently.  Here we provide a list of the original references and the most recent year when each reference was cited.
\begin{itemize}
\item \textbf{L483}: 200 pc \citep[][with citations as recent as 2017]{Dame_1985}.

\item \textbf{L778}: 200 pc \citep[][with citations as recent as 2017]{Schneider_1979}.

\item \textbf{B361}: 350 pc \citep[][with citations as recent as 2010]{Schmidt_1975}.

\item \textbf{L1031B}: 900 pc \citep[][with citations as recent as 2017]{Hilton_1995}.
\end{itemize}
\end{enumerate}

The resulting change in distance, $D$, affects the measured radius, $R$, of each core listed in Table 1 in \citet{Goodman_1993} according to a linear relation, $R \propto D$.  Since \citet{Goodman_1993} calculated the mass based on volume density derived from NH$_3$ hyperfine line fitting, the change in distance affects the mass by $M \propto D^3$.  See \S\ref{sec:data_catalogs_Goodman93} for details.

\section{A Gallery of Close-Up Views of the Droplets and the Droplet Candidates}
\label{sec:appendix_gallery}
In \S\ref{sec:analysis_id}, we explain the steps we take to identify the droplets and the droplet candidates.  The resulting droplets and droplet candidates are shown in Figs.\ \ref{fig:L1688_TpeakTkin}, \ref{fig:L1688_VlsrSigma}, and \ref{fig:L1688_Herschel} for L1688, and Fig.\ \ref{fig:B18_TpeakTkin}, \ref{fig:B18_VlsrSigma}, and \ref{fig:B18_Herschel} for B18.  Here we provide a gallery of close-up views of these droplets and droplet candidates.  The gallery can be found at \url{https://github.com/hopehhchen/Droplets/tree/master/Droplets/plots/droplets}.

The quantity shown in each panel of the figure is denoted in the top left corner, where $N_{\mathrm{H}_2}$ is the Herschel column density, $T_\mathrm{dust}$ is the Herschel dust temperature, $T_\mathrm{peak}$ is the NH$_3$ brightness, $\sigma_{\mathrm{NH}_3}$ is the observed NH$_3$ velocity dispersion, $V_\mathrm{LSR}$ is the velocity centroid from fitting the NH$_3$ (1, 1) hyperfine line profile, and $T_\mathrm{kin}$ is the kinetic temperature from fitting the NH$_3$ (1, 1) and (2, 2) profiles (thus its smaller footprint due to the lack of detection of the NH$_3$ (2, 2) emission at some pixels).

The thick contour (black or white) in each panel marks the outline of the mask used to define the boundary of the droplet.  The crosshair and the circle (red or blue) show the position centroid and the effective radius ($R_\mathrm{eff}$) of the droplet.  The red contour in the panel that shows the observed NH$_3$ velocity dispersion ($\sigma_{\mathrm{NH}_3}$) corresponds to the outline of the regions where velocity dispersion is found to be subsonic.

Due to the varying contrast, we adjust the color scale used in each panel from droplet to droplet, but the span of the color scale between the two extreme colors remains the same for each quantity across different droplets.  The grayscale used to plot $N_{\mathrm{H}_2}$ ranges from lower column density in lighter gray to higher column density in darker gray and spans a total of one order of magnitude in column density (from white---the lowest column density, to black---the highest column density).  The color scale used to plot $T_\mathrm{dust}$ ranges from lower dust temperature in darker orange to higher dust temperature in lighter yellow and spans a total of six degrees in dust temperature (from dark red---the lowest dust temperature, to light yellow---the highest dust temperature).  Similar to the grayscale used to plot $N_{\mathrm{H}_2}$, the grayscale used to plot $T_\mathrm{peak}$ spans a total of an order of magnitude in NH$_3$ brightness (from white---the lowest $T_\mathrm{peak}$, to black---the $T_\mathrm{peak}$).  The color scale used to plot $\sigma_{\mathrm{NH}_3}$ is fixed and shows observed NH$_3$ velocity dispersion between 0.05 km s$^{-1}$ (light yellow) to 0.40 km s$^{-1}$ (dark blue).  The color scale used to plot $V_\mathrm{LSR}$ ranges from more redshifted $V_\mathrm{LSR}$ in red to more blueshifted $V_\mathrm{LSR}$ in blue.  Similar to the color scale used to plot $T_\mathrm{dust}$, the color scale used to plot $T_\mathrm{kin}$ spans a total of six degrees in dust temperature (from dark red---the lowest kinetic temperature temperature, to light yellow---the highest kinetic temperature temperature).

The physical scale on the plane of the sky is noted by the horizontal line in the top right corner.  The black circular area in the lower left corner corresponds to the GAS beam at 23 GHz.


\section{Droplets at Positions of Dense Cores and Other Known Structures}
\label{sec:appendix_overlap}
\begin{figure}[ht!]
\plotone{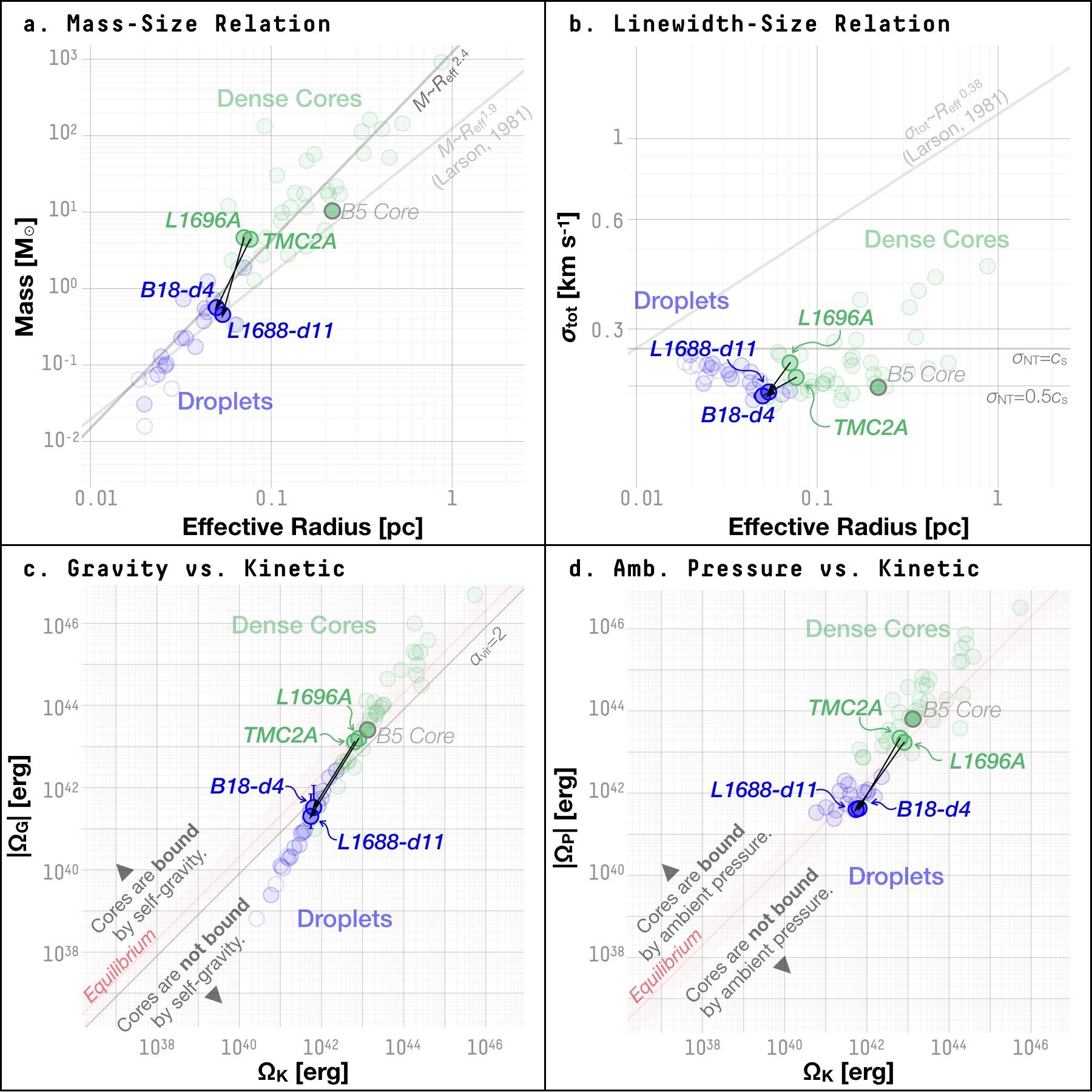}
\caption{\label{fig:OverlapPropertiesNH3corr} \textbf{(a)} The mass-size distribution of the droplets identified in this work (blue circles) and the dense cores examined by \citet{Goodman_1993} (green circles).  The droplets found at positions of known dense cores and the corresponding dense cores are highlighted and connected by black lines.  These are L1688-d11, found at the position of L1696A, and B18-d4, found at the position of TMC-2A.  As in Fig.\ \ref{fig:basic}a, the black line shows a power-law relation between the mass and the effective radius, and randomly selected 10\% of the accepted parameters in the MCMC chain used to find the power-law fit are plotted as transparent lines for reference.  The solid gray line shows the empirical relation based on observations of larger-scale structures examined by \citet{Larson_1981}.  \textbf{(b)} The $\sigma_\mathrm{tot}$-size distribution of the same structures shown in (a).  As in Fig.\ \ref{fig:basic}b, the horizontal lines show $\sigma_\mathrm{tot}$ expected for structures where the non-thermal component is equal to the sonic speed ($c_\mathrm{s}$; thicker line) and half the sonic speed (thinner line) of a medium with a mean molecular weight of 2.37 u at a temperature of 10 K.  \textbf{(c)} Gravitational potential energy, $\Omega_\mathrm{G}$, plotted against internal kinetic energy, $\Omega_\mathrm{K}$ (Equation \ref{eq:virial}), for the same structures shown in (a).  As in Fig.\ \ref{fig:virialGP}a, the red band from the lower left to the top right marks the equilibrium between $\Omega_\mathrm{G}$ and $\Omega_\mathrm{K}$ (solid red line) within an order of magnitude (pink band), according to the virial equation (Equation \ref{eq:virial}; omitting the pressure term).  The black line marks where the conventional virial parameter, $\alpha_\mathrm{vir}$, has a value of 2.  \textbf{(d)} The energy term representing the confinement provided by the ambient gas pressure, $\Omega_\mathrm{P}$, plotted agains the internal kinetic energy, $\Omega_\mathrm{K}$ (Equation \ref{eq:virial}), for the same structures shown in (a).  As in Fig.\ \ref{fig:virialGP}b, the red band from the lower left to the top right marks an equilibrium between $\Omega_\mathrm{P}$ and $\Omega_\mathrm{K}$ (solid red line) within an order of magnitude (pink red band), according to the virial equation (Equation \ref{eq:virial}; omitting the gravitational term).  Structures in the parameter space above the red line (equilibrium) are expected to be dominated by the ambient gas pressure.}
\end{figure}

Two of the 18 droplets defined in \S\ref{sec:analysis_id} are found near the positions of two dense cores observed and analyzed by \citet{Benson_1989}, \citet{Goodman_1993}, and \citet{Ladd_1994}.  These are L1688-d11 and B18-d4, with centroid positions found within one GBT FWHM beam size (32\arcsec) of the centers of L1696A and TMC-2A, respectively.  Fig.\ \ref{fig:OverlapPropertiesNH3corr} shows how the basic properties measured in this work using data from the Green Bank Ammonia Survey compare to properties measured by \citet{Benson_1989}, \citet{Goodman_1993}, and \citet{Ladd_1994}.  We note that the observations done by \citet{Benson_1989} and \citet{Ladd_1994} did not spatially resolve the ``transition to coherence'' \citep{Goodman_1998}, as was done by \citet{Pineda_2010} for B5.  For reference, the spatial resolution of the observations done by \citet{Benson_1989} and \citet{Ladd_1994} is a factor of $\sim$ 2.5 coarser than that of modern GBT observations.  The velocity resolution (at 23 GHz) of the observations done by \citet{Benson_1989} and \citet{Ladd_1994} ranges from 0.07 to 0.20 km s$^{-1}$, compared to 0.07 km s$^{-1}$ of the GBT observations done by the Green Bank Ammonia Survey \citep{GAS_DR1}.  See \S\ref{sec:data} for details.

Here we also list previously known cores and density features potentially associated with droplets in Table \ref{table:association32} and Table \ref{table:association96}, based on a thorough search of the SIMBAD Astronomical Database\footnote{The database can be accessed via \url{http://simbad.u-strasbg.fr/simbad/}.  As pointed out on the SIMBAD Astronomical Database page, while the database includes most of the published catalogs, it is by no means complete.}.

\begin{longrotatetable}
\begin{deluxetable*}{ll}
\tablecaption{Droplets and Previously Known Objects Potentially Assocaited with Each Droplet\tablenotemark{a}\label{table:association32}}
\tablehead{\colhead{ID} & \colhead{Features located within 32\arcsec\tablenotemark{b}}}
\startdata
L1688-d1 & Core 18 (M10\tablenotemark{c}) / MMS055 (S06\tablenotemark{d}) \\
L1688-d2 &  Core 23, 24, 25 (M10) \\
L1688-d3 &  [Oph C-N]\tablenotemark{e} / Core 27 (M10) \\
L1688-d4 &  Bolo 12 (Y06\tablenotemark{f}) / Oph C-MM4, MM5, MM6 (M98\tablenotemark{g}) / Oph C-HC$_5$N, C$_2$S, A1, A3 (F09\tablenotemark{h}) / MMS022 (S06) \\
L1688-d5 &  Oph E-MM2c, MM8 (M98) / Core 31 (M10) \\
L1688-d6 &  Oph F-MM2b, MM3 (M98) / Core 46 (M10) / Oph F-A2, A3 (F09) / [Oph F-1] \\
L1688-d7 &  [Oph F] / Oph F-MM8 (M98) \\
L1688-d8 &  Core 59 (M10) / [Oph E-1] / MMS081 (S06) \\
L1688-d9 &  Bolo 24 (Y06) \\
L1688-d10 &  MMS126 (S06) / Core 67 (M10) \\
L1688-d11 &  [L1696A] / [Oph D] \\
L1688-d12 &  [L1696B] / MMS041, MMS075 (S06) \\
\hline
L1688-c2\tablenotemark{i} &  \nodata \\
L1688-c3 &  Oph C-A2 (F09) / Oph C-MM12 (M98) \\
L1688-c4 &  Oph B3-N1 (F10\tablenotemark{j}) / Oph B3-A1 (F09) \\
\hline
\hline
B18-d1 &  \nodata \\
B18-d2 &  \nodata \\
B18-d3 &  L1524-4 (L99\tablenotemark{k}) / L1524-4 C1 (K08\tablenotemark{l}) \\
B18-d4 &  B18-1 C1 (K08) / MC 31 (O02\tablenotemark{m}) \\
B18-d5 &  [TMC-2] \\
B18-d6 & \nodata
\enddata
\tablenotetext{a}{This table is based on the SIMBAD Astronomical Database and lists only molecular clouds (or a subpart of it), cores, and features identified in millimeter/submillimeter emission.  Other types of objects, including YSOs and Herbig-Haro objects, are excluded from this table.  For many droplets, multiple objects identified using observations made with the Submillimeter Common-User Bolometer Array (SCUBA) on the James Clerk Maxwell Telescope (JCMT) are found within a GBT beam FWHM from the droplet centroid.  To be concise, we omit these objects and point to catalogs presented by \citet{Johnstone_2000}, \citet{Jorgensen_2008}, and \citet{Gurney_2008}.}
\tablenotetext{b}{Angular distances from the droplet centroids listed in Table \ref{table:basic}.  The angular distance of 32\arcsec corresponds to the GBT beam FWHM at 23 GHz and is $\sim$ 0.02 pc at the distances of Ophiuchus and Taurus.  The angular distance of 96\arcsec is three times of the GBT beam FWHM and is $\sim$ 0.06 pc at the distances of Ophiuchus and Taurus.  The third column (``features located within 96\arcsec'') does not repeat features listed in the second column (``features located within 32\arcsec'').  See \S\ref{sec:data_GAS}.}
\tablenotetext{c}{M10: \citet{Maruta_2010}.  The objects are grouped by catalogs and ordered by the distance of the the object closest to each droplet on the plane of the sky in each catalog.}
\tablenotetext{d}{S06: \citet{Stanke_2006}.}
\tablenotetext{e}{Object names enclosed in square brackets are names of molecular clouds or subparts of a molecular cloud.  These often point to relatively loosely defined regions within a molecular cloud, and the names have often been in use in literatures for the past decades.}
\tablenotetext{f}{Y06: \citet{Young_2006}.}
\tablenotetext{g}{M98: \citet{Motte_1998}.}
\tablenotetext{h}{F09: \citet{Friesen_2009}.}
\tablenotetext{i}{Since L1688-c1E and L1688-c1W overlap with L1688-d1, they are not listed here where the associations are purely based on the locations of the objects.}
\tablenotetext{j}{F10: \citet{Friesen_2010}.}
\tablenotetext{k}{L99: \citet{Lee_1999}.}
\tablenotetext{l}{K08: \citet{Kauffmann_2008}.}
\tablenotetext{m}{O02: \citet{Onishi_2002}.  MC stands for ``molecular condensation.''}
\end{deluxetable*}
\end{longrotatetable}

\begin{longrotatetable}
\begin{deluxetable*}{ll}
\tablecaption{Droplets and Previously Known Objects Potentially Assocaited with Each Droplet (3 $\times$ GBT FWHM)\tablenotemark{a}\label{table:association96}}
\tablehead{\colhead{ID} & \colhead{Features located within 96\arcsec (but outside 32\arcsec)}}
\startdata
L1688-d1  & [Oph C-W] / Oph C-MM9 (M98) \\
L1688-d2 &  Oph E-MM1 (M98) / Core 28 (M10) / MMS068 (S06) \\
L1688-d3 &  MMS044 (S06) / Oph C-MM10, MM12 (M98) / Oph C-A2 (F09) \\
L1688-d4 &  Core 29 (M10) / MMS039, MMS059 (S06) / Oph C-MM1, MM2, MM7, MM12 (M98) / Oph C-A2 (F09)  \\
L1688-d5 &  Oph E-MM2b, MM9 (M98) / [Oph E] / Core 34 (M10) \\
L1688-d6 &  Oph F-MM2, MM2a (M98) / Core 44, 51, 52 (M10) / Oph F-A1 (F09) / MMS040 (S06) \\
L1688-d7 &  Oph F-MM6, MM7, MM9 (M98) / [L1681B] / SMM073 (S06) / Core 57, 58 (M10) \\
L1688-d8 &  \nodata \\
L1688-d9 &  Core 61, 62, 63, 64 (M10) / MMS051 (S06) \\
L1688-d10 &  Core 66, 68 (M10) / MMS060 (S06) \\
L1688-d11 &  R26 (L89\tablenotemark{b}) / Oph D-MM1, MM3, MM4, MM5 (M98) / MMS047, MMS052 (S06) / Bolo 27 (Y06) \\
L1688-d12 & Oph I-MM1 (M98) \\
\hline
L1688-c2 & Core 27 (M10) / MMS044 (S06) / [Oph C-N] \\
L1688-c3 & Oph C-MM2, MM4, MM5, MM10 (M98) / Oph C-C$_2$S, A1, HC$_5$N (F09) / [Oph C] / Bolo 12 (Y06) / MMS044 (S06) \\
L1688-c4 & [Oph B-3] / Core 42, 47 (M10) / MMS108, MMS 143 (S06) \\
\hline
\hline
B18-d1 &  \nodata\\
B18-d2 &  MC 29 (O02) / [L1524] \\
B18-d3 &  MC 29 (O02) \\
B18-d4 &  B18-1 (L99) / [TMC-2A] \\
B18-d5 &  TMC-2 C1 (K08) / TMC-2 (L99) / MC 33b (O02) / Tau E2 (W94\tablenotemark{c}) \\
B18-d6 & B18-4 C1 (K08) / MC 35 (O02) / [TMC-3B, 3A \& B18-I] / B18-4 (L99)
\enddata
\tablenotetext{a}{This table is like Table \ref{table:association32} but lists objects that are within 96\arcsec (but outside 32\arcsec) from the droplet centroids on the plane of the sky.  The angular distance of 96\arcsec is three times of the GBT beam FWHM and is $\sim$ 0.06 pc at the distances of Ophiuchus and Taurus.  The notations remain the same.}
\tablenotetext{b}{L89: \citet{Loren_1989}.}
\tablenotetext{c}{W94: \citet{Wood_1994}.}
\end{deluxetable*}
\end{longrotatetable}

\begin{figure}[ht!]
\plotone{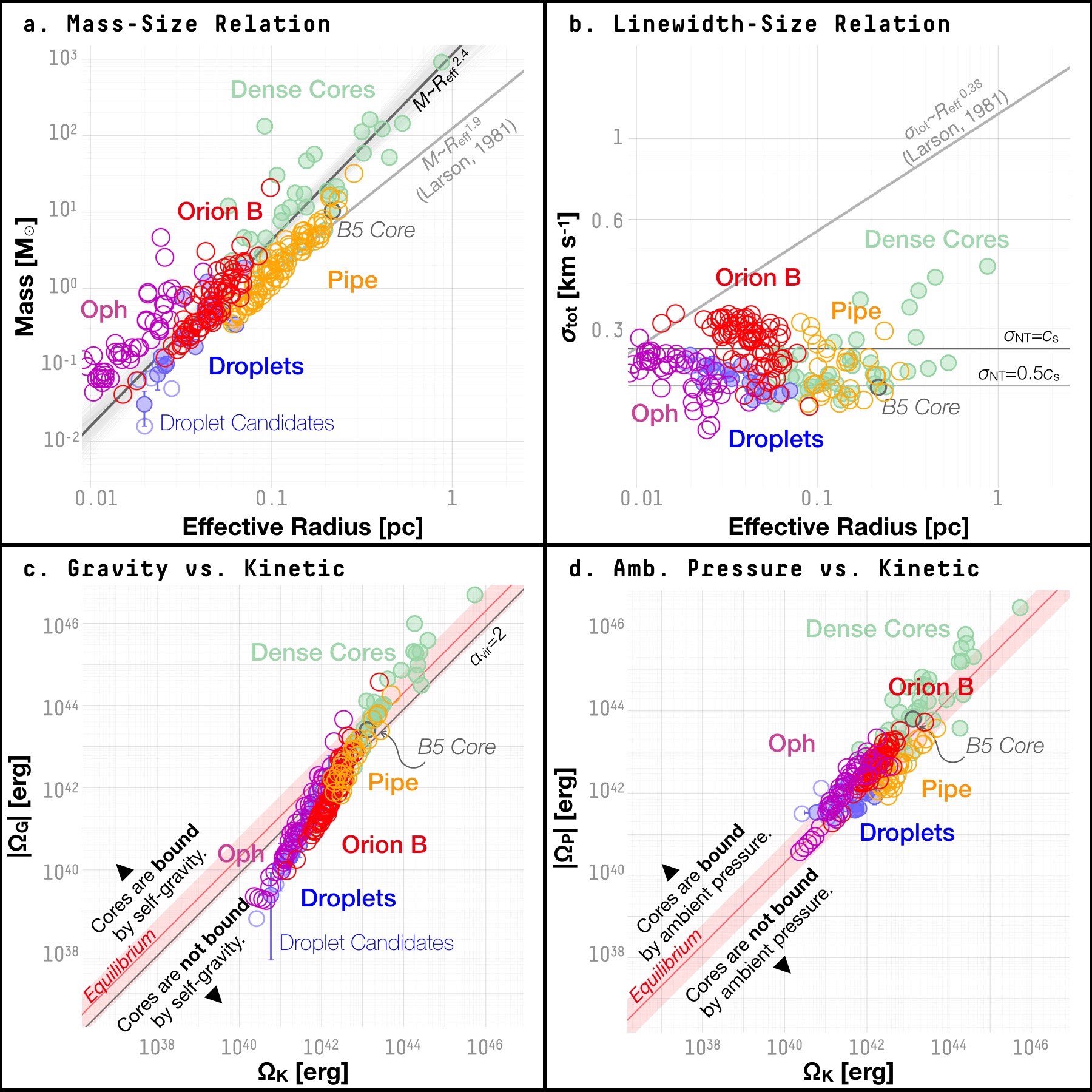}
\caption{\label{fig:OtherCoreProperties} Like Figs.\ \ref{fig:basic} and \ref{fig:virialGP} but plotted with previous observations of cores in Ophiuchus (purple circles), Orion B (red circles), and the Pipe Nebula (yellow circles). \textbf{(a)} The mass-size distribution (see Fig.\ \ref{fig:basic}a).  \textbf{(b)} The $\sigma_\mathrm{tot}$-size distribution (see Fig.\ \ref{fig:basic}b).  \textbf{(c)} Gravitational potential energy, $\Omega_\mathrm{G}$, plotted against internal kinetic energy, $\Omega_\mathrm{K}$ (see Fig.\ \ref{fig:virialGP}a).  \textbf{(d)} The energy term representing the confinement provided by the ambient gas pressure, $\Omega_\mathrm{P}$, plotted agains the internal kinetic energy, $\Omega_\mathrm{K}$ (see Fig.\ \ref{fig:virialGP}b).}
\end{figure}

We can also compare the physical properties of droplets to cores found similarly in nearby molecular clouds.  Risking comparing measurements that are biased by the difference in observation setups and the methods used in analyses, Fig.\ \ref{fig:OtherCoreProperties} shows a comparison of physical properties between the droplets and the core populations found in Ophiuchus, Orion B, and the Pipe Nebula, respectively observed and analyzed by \citet{Johnstone_2000}, \citet{Johnstone_2001}, and \citet{Lada_2008} (see also discussions in \S\ref{sec:analysis_virial_P}; distances have been updated with modern measurements).  Like droplets, many of these cores are also found to be unbound by self-gravity, and the ambient pressure likely contributes to the confinement of these cores \citep{Johnstone_2000, Lada_2008}.  In general, these cores have physical properties comparable to that of the droplets and some of the dense cores.  A more careful treatment of differences in observational setups and the methods used to define the structures is needed to fully understand the relation between the droplets and these cores.  We leave a comprehensive comparison between droplets and previously observed cores to a paper in the future.

\section{Uncertainty in the Radius Measurement}
\label{sec:appendix_radius}
The uncertainty in the radius measurement lies in two aspects.  First, the radius measurement is limited by the intrinsic resolution of the observations.  In the case of this paper, since the droplet boundary is defined by the change in linewidth based on GBT observations of NH$_3$ emission, the uncertainty in the radius measurement scales with the pixel size of the linewidth map.  For the Nyquist-sampled linewidth map produced by \citet{GAS_DR1}, the pixel size equals to $\sim$ 0.007 pc at the distance of L1688 and B18.

\begin{figure}[ht!]
\epsscale{0.6}
\plotone{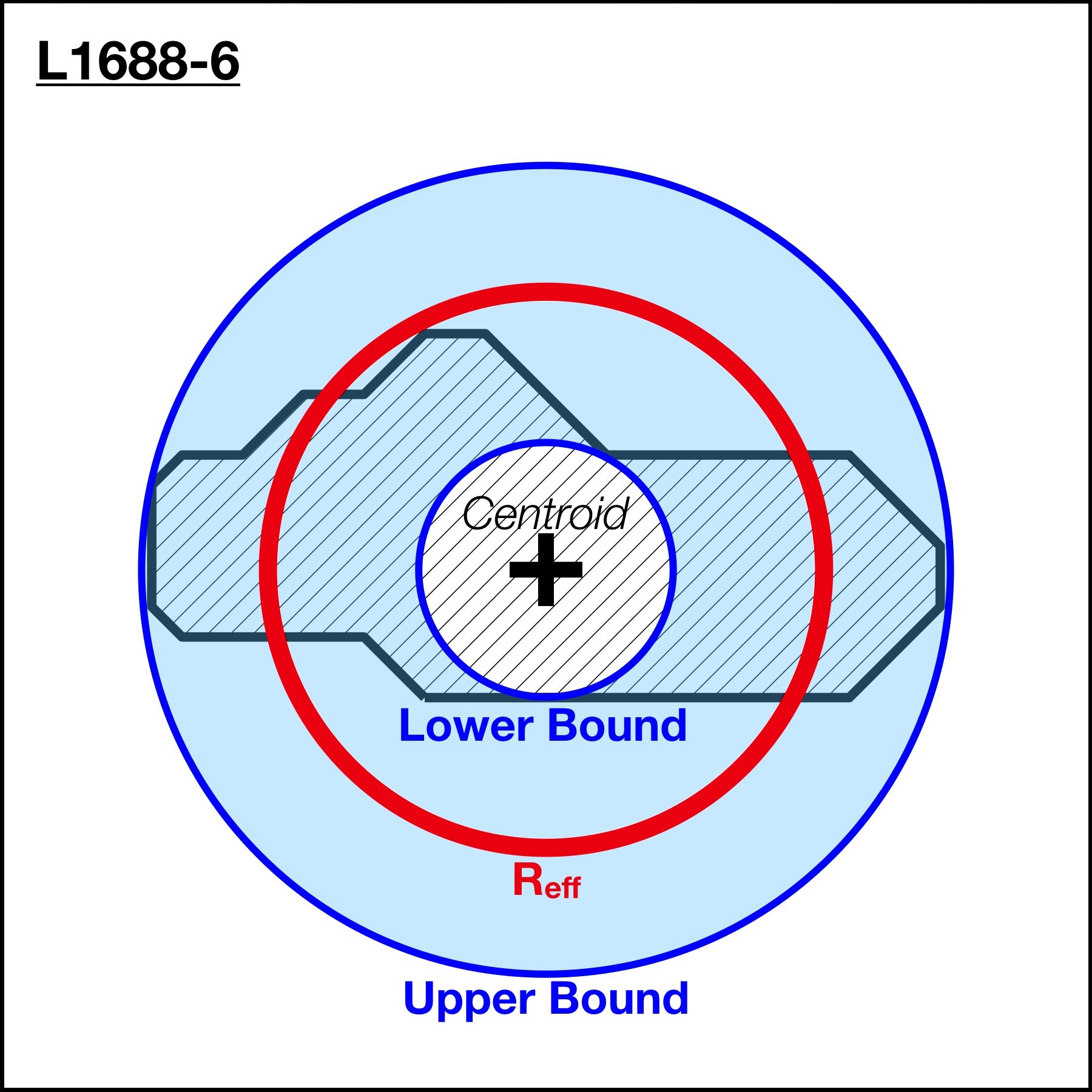}
\caption{\label{fig:radius} Droplet L1688-d6 as an example of how the lower and the upper bounds of the radius measurement are defined.  The droplet is shown as an irregular hatched area.  The red circle shows the effective radius, $R_\mathrm{eff}$, derived from the principal component analysis (PCA) weighted by the peak NH$_3$ brightness, $T_\mathrm{peak}$ (see \S\ref{sec:analysis_basic}).  The inner blue circle marks the largest circle that can \emph{be enclosed} by the droplet boundary, which we use as the lower bound in the radius measurement.  The outer blue circle marks the smallest circle that can \emph{enclose} the droplet boundary, which we use as the upper bound in the radius measurement.  Notice that the circles used in determining the lower and the upper bounds are required to center at the position centroid of the droplet (the positions listed in Table \ref{table:basic}).}
\end{figure}

Second, since the droplet boundary is not perfectly circular, assigning a single number to describe the size (radius) of the droplet boundary is subject to the uncertainty due to the non-circular shape of the boundary.  In this paper, we estimate the lower and the upper bounds of the radius by measuring the radius of the largest circle that can be enclosed by the droplet boundary and the smallest circle that can enclose the droplet boundary, respectively.  See Fig.\ \ref{fig:radius}.

The difference between $R_\mathrm{eff}$ and the lower or the upper bound of the radius is then required to be larger than the uncertainty due to the finite resolution of the GBT observations (i.e., $\gtrsim$ 0.02 pc at the distances of L1688 and B18).  The resulting uncertainty is listed in Table \ref{table:basic}.

Besides using the principal component analysis (PCA; see \S\ref{sec:analysis_basic}) to find the radius, another common way to determine ``effective radius'' for a non-circular shape is to measure the projected area, $A$, and find the radius of the circle that has the same area \citep[e.g.,][]{Rosolowsky_2006}.  The effective radius found through the projected area is then:

\begin{equation}
R_\mathrm{eff,A} = \sqrt{\frac{A}{\pi}}\ \mathrm{.}
\end{equation}

For each droplet, $R_\mathrm{eff,A}$ lies within the range between the lower and the upper bounds determined using the enclosed circles (see above and Fig.\ \ref{fig:radius}), and deviates by less than 10\% from $R_\mathrm{eff}$, determined from the PCA.  For example, this translates to $\lesssim$ 10\% of difference in the gravitational potential energy, $\Omega_\mathrm{G}$.  In the analyses presented in this paper, the uncertainty in $R_\mathrm{eff}$ determined using the lower and the upper bounds (Fig.\ \ref{fig:radius}) is propagated to the uncertainties of other quantities, which are shown in corresponding plots and tables.  Since $R_\mathrm{eff,A}$ lies between the lower and the upper bounds determined for the radius measurement, using $R_\mathrm{eff,A}$, instead of $R_\mathrm{eff}$, in these analyses will only have an effect within the uncertainty reported throughout this paper.

\section{Baseline Subtraction}
\label{sec:appendix_baseline}
\begin{figure}[ht!]
\plotone{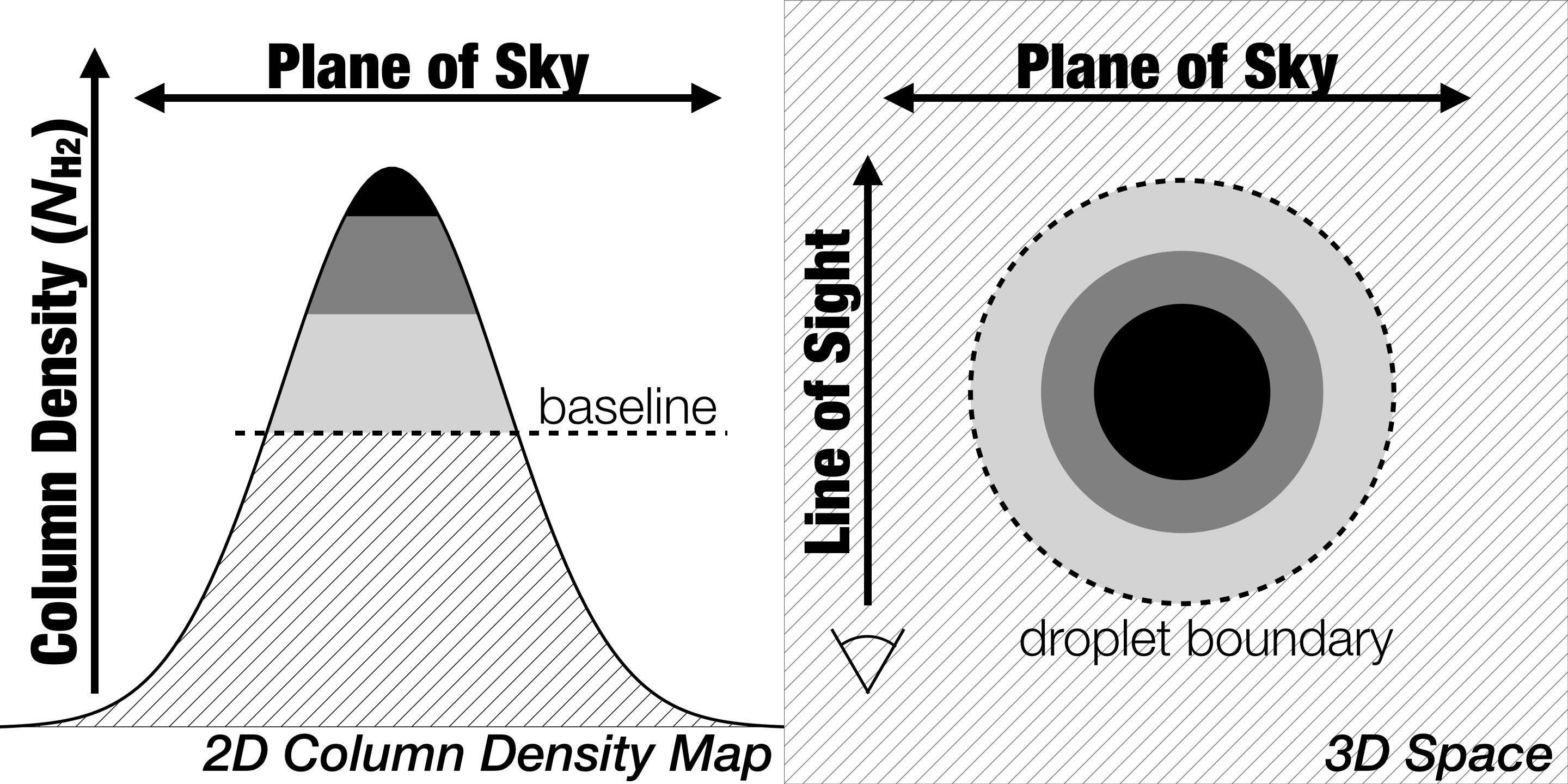}
\caption{\label{fig:cartoon} This cartoon shows the corresponding ``layers'' of material along a cut on the \emph{2D column density map} (\textbf{left}; with the vertical axis corresponding to the column density) and in a top-down view in the \emph{3D space} (\textbf{right}; with the line of sight along the vertical axis).  The solid shaded area (in black/dark gray/gray) corresponds to materials inside a schematic spherical ``droplet,'' while the hatched area corresponds to the material outside the droplet.  On the right hand side, the dashed line marks the boundary of the droplet in the 3D space.  On the left hand side, the dashed line shows how a constant column density baseline, corresponding to the minimum value of the solid shaded area, can be a reasonable estimate of the contribution from material outside the droplet (hatched area).  The baseline subtraction method is similar to the ``clipping paradigm'' analyzed by \citet{Rosolowsky_2008b} and applied to mass measurements of sub-0.1 pc density features by \citet{Pineda_2015}.}
\end{figure}

In the analyses presented in this paper, the mass and related quantities such as the density are estimated after a baseline correction for the line-of-sight material outside the targeted volume.  The method we apply in this paper is similar to the ``clipping paradigm'' examined by \citet{Rosolowsky_2008b} to estimate the physical properties of a compact structure.  The clipping method produces a mass estimate that would correspond better to the mass calculated from fitting the NH$_3$ emission, which traces only the material within the targeted structure and was used by \citet{Goodman_1993} to estimate the masses of the dense cores (see \S\ref{sec:data_catalogs_Goodman93}).  Fig.\ \ref{fig:cartoon} schematically demonstrates how the clipping paradigm can be a reasonable way to remove the contribution to column density measurements from the material along the same line of sight but outside the targeted structure.  In the virial analysis presented in \S\ref{sec:analysis_virial} and in the analyses of the radial density profiles presented in \S\ref{sec:discussion_confinement_BE} and \S\ref{sec:discussion_confinement_logo}, we use the same method to estimate the density of the gas surrounding the volume under discussion.

\begin{figure}[ht!]
\plotone{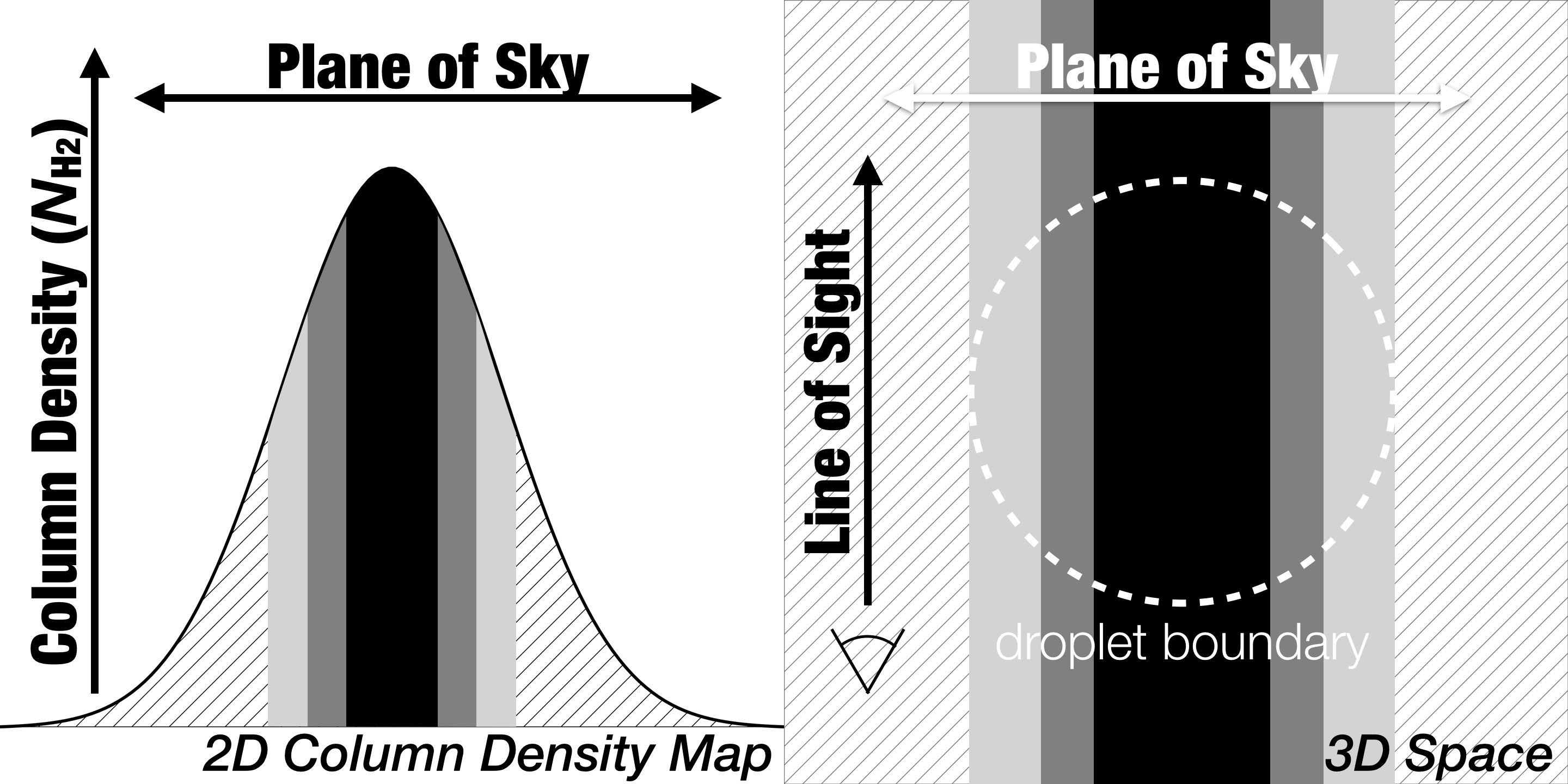}
\caption{\label{fig:cartoon_fullcolumn} This cartoon shows the mass (integrated column density) on a column density distribution derived from a \emph{2D column density map} (\textbf{left}; with the vertical axis corresponding to the column density) and the corresponding volumes in the \emph{3D space} (\textbf{right}; with the line of sight along the vertical axis), if no baseline removal is applied.  The solid shaded area (in black/dark gray/gray) corresponds to the mass (integrated column density) estimated from the 2D column density map (left) and the material occupying the corresponding volume in the 3D space (right).}
\end{figure}

In comparison, a simple sum of column densities measured within a certain projected area on the plane of the sky overestimates the mass by including the contribution from the material along the entire line of sight (see Fig.\ \ref{fig:cartoon_fullcolumn}).  Similarly, when estimating the average density within a shell-shaped volume surrounding the targeted structure (as done in \S\ref{sec:analysis_virial_P}), summing the column densities measured within a ring-shaped area on the plane of the sky overestimates the mass and thus, the density (different shades of gray in Fig.\ \ref{fig:cartoon_fullcolumn}).  The typical difference between the mass estimated after applying the clipping and the mass estimated without any clipping (a simple sum) is $\sim$ 25\%.  In the virial analysis presented in \S\ref{sec:analysis_virial}, this amounts to a $\sim$ 50\% of uncertainty in the estimate of the gravitational potential energy and a $\sim$ 25\% of uncertainty in either of the kinetic energy and the energy term representing the ambient gas pressure confinement.  These uncertainties are included in the uncertainties listed in Table \ref{table:virial} and do not qualitatively change the results presented in subsequent discussions.  

\begin{figure}[ht!]
\plotone{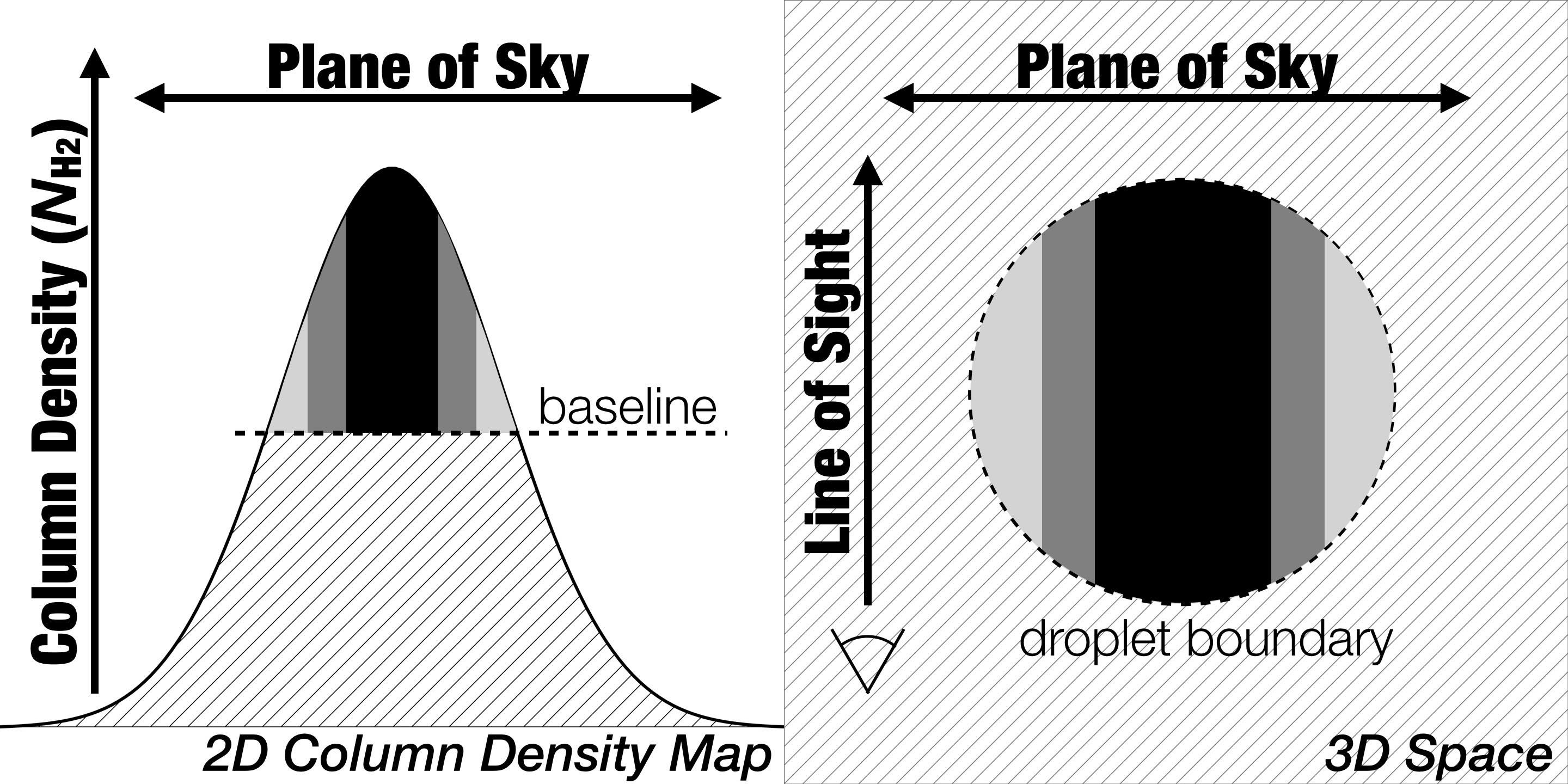}
\caption{\label{fig:cartoon_singlebaseline} This cartoon shows the mass (integrated column density) on a column density distribution derived from a \emph{2D column density map} (\textbf{left}; with the vertical axis corresponding to the column density) and the corresponding volumes in the \emph{3D space} (\textbf{right}; with the line of sight along the vertical axis), if a single baseline removal is applied.  The solid shaded area (in black/dark gray/gray) corresponds to the mass (integrated column density) estimated from the 2D column density map (left) and the material occupying the corresponding volume in the 3D space (right).  Schematically, a single baseline removal corresponds to the ``subtraction'' of column density below the dashed line in the panel on the left.  For a structure with a spherical shape, such subtraction corresponds to the removal of the contribution from material outside the dashed line in the panel on the right.  Depending on the algorithm used for the baseline removal, the ``subtraction'' might not exactly correspond to the removal of a flat-top function as shown in this cartoon.  The cartoon is used only to demonstrate how a single baseline removal does not fit the purpose of simultaneously estimating the masses within different layers at different radial distances.}
\end{figure}

More sophisticated ways to remove contributions from material in the foreground and background may involve removing contributions from column density structures larger than a certain size scale, for example, using a transform algorithm like the wavelet decomposition.  While such algorithms perform well in analyses of compact structures, the uncertainty becomes unclear if we are interested in both the mass within the targeted structure (as we are in \S\ref{sec:analysis_basic}) and the density of the surrounding material (as we are in \S\ref{sec:analysis_virial_P}).  A single background removal can result in an overestimated mass of the structure at the center (the black areas in Fig.\ \ref{fig:cartoon_singlebaseline}, compared to Fig.\ \ref{fig:cartoon}), and when estimating the density of the surrounding material, a single background removal would give an estimate for a hollow cylindrical volume instead of the shell-shaped volume (different shades of gray in Fig.\ \ref{fig:cartoon_singlebaseline}, compared to Fig.\ \ref{fig:cartoon}).  While theoretically, we can resolve the issue by optimizing the transform algorithm to perform differently for different purposes, we adopt the clipping method in this paper 1) to fully avoid double counting the contribution from the material in the same volume when estimating the mass in \emph{and} outside a structure and 2) for its simplicity and ease of error estimation.  As demonstrated in Fig.\ \ref{fig:cartoon}, if the structure is spherical, the clipping method would give the exact masses for the layers of materials at different radial distances.

The main uncertainty resulted from using the clipping method with circular annuli (as done in \S\ref{sec:analysis_virial_P}, \S\ref{sec:discussion_confinement_BE}, and \S\ref{sec:discussion_confinement_logo}) is then the deviation in the shape of the targeted structure from a sphere.  As mentioned in \S\ref{sec:analysis_id}, most of the droplets have aspect ratios between 1 and 2, with the exceptions of L1688-d1 with an aspect ratio of $\sim$ 2.50, L1688-d6 with an aspect ratio of $\sim$ 2.52, and B18-d5 with an aspect ratio of $\sim$ 2.03.  Thus, we estimate that the uncertainty resulted from using circular annuli is no larger than a factor of 2.  See discussions in Appendix \ref{sec:appendix_radius}.

\section{Uncertainty in Mass Due to the Potential Bias in SED Fitting of Herschel Observations}
\label{sec:appendix_bias}
Lastly, we examine the uncertainty in the potentially biased SED fitting of the Herschel observations.  As presented above, we use the column density map obtained via SED fitting of Herschel observations to estimate the mass of the droplets (\S\ref{sec:data_Herschel}).  Using the Herschel column density to estimate the mass and the fits to the NH$_3$ line profiles to estimate the velocity dispersion allows mutually independent measurements of the mass-size and the mass-linewidth relations (Fig.\ \ref{fig:basic}).  However, it is a known issue that SED fitting of emissions in Herschel bands might be biased, especially towards cold and dense regions \citep{Shetty_2009a, Shetty_2009b, Kelly_2012}.  In the cold and dense regions, there can be a certain degree of redundancy between a high dust temperature and a high column density.  As a result, the SED fitting can overestimate the dust temperature and underestimate the column density, which would result in underestimated masses for the droplets in this case.

\begin{figure}[ht!]
\plotone{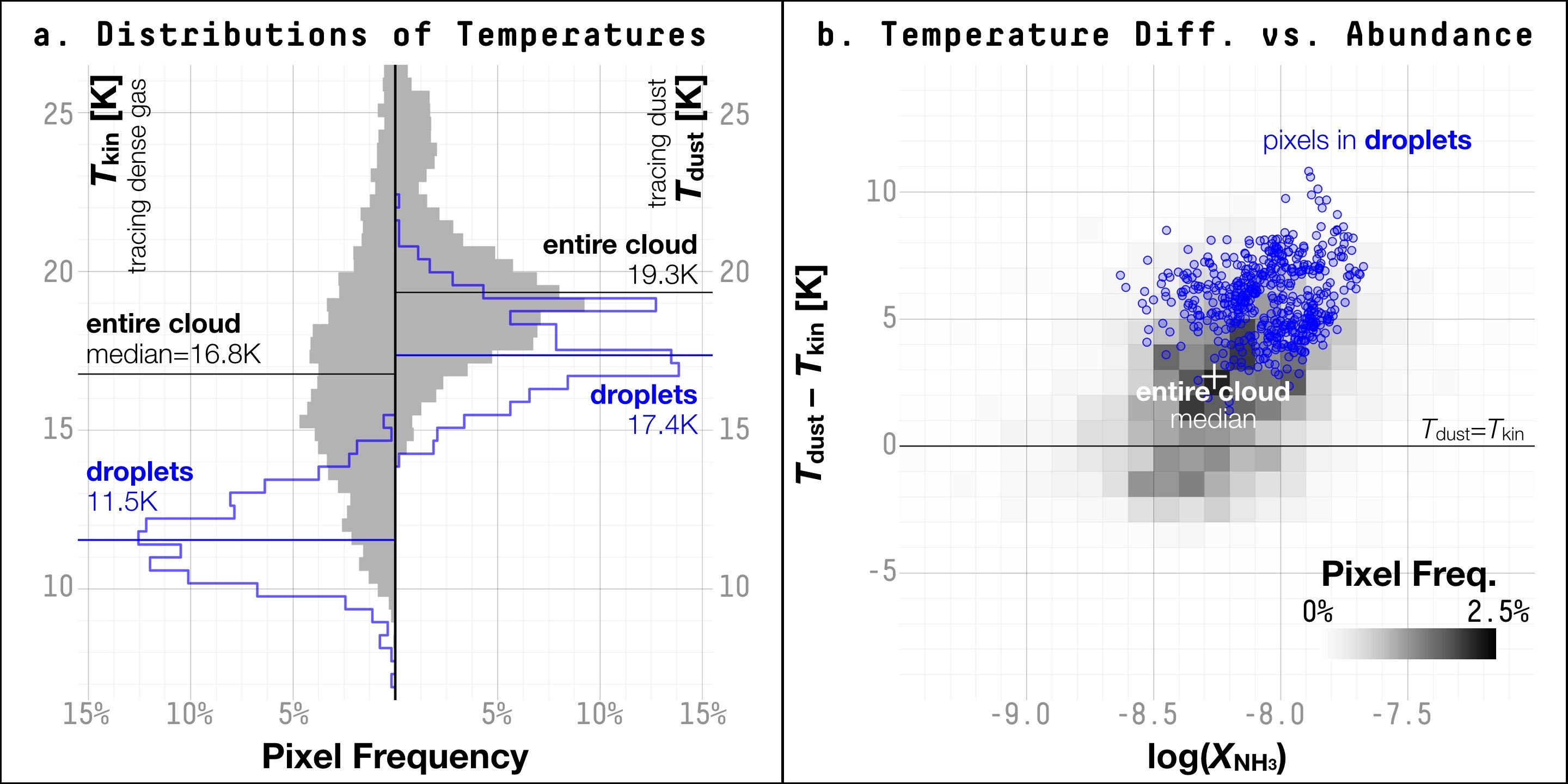}
\caption{\label{fig:TdiffAbundance} A comparison between properties based on fitting NH$_3$ hyperfine line profiles and from the SED fitting of the Herschel observations, using the L1688 region as an example.  \textbf{(a)} Pixel-by-pixel distributions of temperatures measured from fitting the NH$_3$ hyperfine line profiles (kinetic temperature, $T_\mathrm{kin}$; \textit{left}) and from SED fitting of Herschel observations (the dust temperature, $T_\mathrm{dust}$; \textit{right}).  Only pixels with significant detection of NH$_3$ (1, 1) emission are included in this plot (i.e., the total number of pixels included on the left half of the plot is the same as that on the right).  The distributions are normalized by the total number of pixels in each group (the cloud or the droplets), so the height of each bin in the histograms correspond to the frequency of a certain range of values occurring in each group (shown along the horizontal axis in percentage).  The gray histograms show the distributions of all the pixels outside the boundaries of the droplets, and the blue histograms show the distributions of all the pixels inside the droplet boundaries.  The median of each distribution is shown as a horizontal line.  \textbf{(b)} 2D histogram showing the distribution between the difference between the dust temperature and the kinetic temperature ($T_\mathrm{dust}-T_\mathrm{kin}$), as a function of the NH$_3$ abundance ($X_{\mathrm{NH}_3} = N_{\mathrm{NH}_3}/N_{\mathrm{H}_2}$, where $N_{\mathrm{H}_2}$ is derived from the SED fitting of the Herschel observations).  The 2D histogram in each panel shows the distribution of pixels with significant detection of NH$_3$ (1, 1) emission in the entire map, with the pixel frequency defined as the percentage of pixels on the map falling in each 2D bin in the 2D histogram.  The blue dots show the distribution of individual pixels within the droplet boundaries.  The horizontal line shows where $T_\mathrm{dust} = T_\mathrm{kin}$, and the white cross marks the median values of the abundance and the difference in temperature of the entire cloud.}
\end{figure}

Consistently with what \citet{GAS_DR1} pointed out, Fig.\ \ref{fig:TdiffAbundance}a shows that the dust temperature, $T_\mathrm{dust}$, is systematically 2 to 3 K higher than the kinetic temperature of the dense gas traced by NH$_3$ emission, $T_\mathrm{kin}$.  Fig.\ \ref{fig:TdiffAbundance}a also shows that, for pixels within the boundaries of the droplets, the difference between $T_\mathrm{dust}$ and $T_\mathrm{kin}$ could be even larger, up to 6 K.  Fig.\ \ref{fig:TdiffAbundance}b further shows that, for pixels within the boundaries of the droplets, not only is the difference between $T_\mathrm{dust}$ and $T_\mathrm{kin}$ larger than the median value of the entire cloud, the pixel-by-pixel NH$_3$ abundance obtained by dividing the $N_{\mathrm{NH}_3}$ \citep[from fitting the NH$_3$ hyperfine line profiles; see][]{GAS_DR1} by $N_{\mathrm{H}_2}$ (from SED fitting of Herschel observations) is higher than the cloud median.  The distribution is consistent with overestimated temperature and underestimated column density in the SED fitting of Herschel observations.

In this section, we try to estimate the effects of underestimated column density in the SED fitting of Herschel observations.  In particular, we examine the effects of the underestimated column density on the virial analysis presented in \S\ref{sec:analysis_virial}.  We compare $N_{\mathrm{H}_2}$, obtained via the SED fitting of Herschel observations, to $N_{\mathrm{NH}_3}$, obtained via fitting the NH$_3$ hyperfine line profiles \citep[see][for details]{GAS_DR1}.  Unfortunately, for the model used in the NH$_3$ hyperfine line fitting, we need detections of emission from both the NH$_3$ (1, 1) and (2, 2) lines to determine the population ratios between the two states, in order to estimate $N_{\mathrm{NH}_3}$ and $T_\mathrm{kin}$.  While all pixels within the droplets are detected in NH$_3$ (1, 1) emission, not all pixels within the droplets are detected in NH$_3$ (2, 2) emission, so we can only obtain estimates of $N_{\mathrm{NH}_3}$ in the densest regions within the droplets.  Thus, estimating the mass solely from $N_{\mathrm{NH}_3}$ is difficult, especially given that we expect the column density to decrease toward the outer edge of a droplet.

Thus, in order to assess the potential bias in the column density obtained via SED fitting of Herschel observations, we used the pixels within the droplet boundaries where we have measurements of both $N_{\mathrm{NH}_3}$ and $N_{\mathrm{H}_2}$ (i.e., the pixels where we have significant detection of NH$_3$ (2, 2) emission).  We compare the abundance of the droplets (obtained by dividing $N_{\mathrm{NH}_3}$ by $N_{\mathrm{H}_2}$) to the median value of the cloud and assume that the difference in abundance between the droplet values and the cloud median is fully due to the underestimated $N_{\mathrm{H}_2}$ in the droplets.  Assuming the underlying, ``real'' NH$_3$ abundance, $X_\mathrm{real} \equiv N_{\mathrm{NH}_3\mathrm{,droplet}}/N_{\mathrm{H}_2\mathrm{,real}}$, is equal to the median NH$_3$ abundance of the cloud, $X_\mathrm{cloud} = N_{\mathrm{NH}_3\mathrm{,cloud}}/N_{\mathrm{H}_2\mathrm{,cloud}}$, we calculate a correction factor, $\epsilon$:

\begin{eqnarray}
\label{eq:correction}
\epsilon &= &\frac{N_{\mathrm{H}_2\mathrm{,real}}}{N_{\mathrm{H}_2\mathrm{,droplet}}} \\
 &= &\frac{N_{\mathrm{NH}_3\mathrm{,droplet}}}{N_{\mathrm{H}_2\mathrm{,droplet}}} \frac{N_{\mathrm{H}_2\mathrm{,cloud}}}{N_{\mathrm{NH}_3\mathrm{,cloud}}} \nonumber \\
 &= &\frac{X_\mathrm{droplet}}{X_\mathrm{cloud}}\ \mathrm{,} \nonumber
\end{eqnarray}

\noindent where $N_{\mathrm{H}_2\mathrm{,real}}$ is the underlying, ``real'' column density, and $N_{\mathrm{H}_2\mathrm{,droplet}}$ is the column density measured from the SED fitting for pixels within the droplet boundaries.  We can then estimate the ``real'' mass using this correction factor:

\begin{equation}
\label{eq:mass_correction}
M_\mathrm{real} = \epsilon M_\mathrm{droplet}\ \mathrm{,}
\end{equation}

\noindent where $M_\mathrm{real}$ is the underlying, ``real'' mass, and $M_\mathrm{droplet}$ is the measured mass of the droplet (from Herschel column density; \S\ref{sec:data_Herschel}).

\begin{figure}[ht!]
\epsscale{0.6}
\plotone{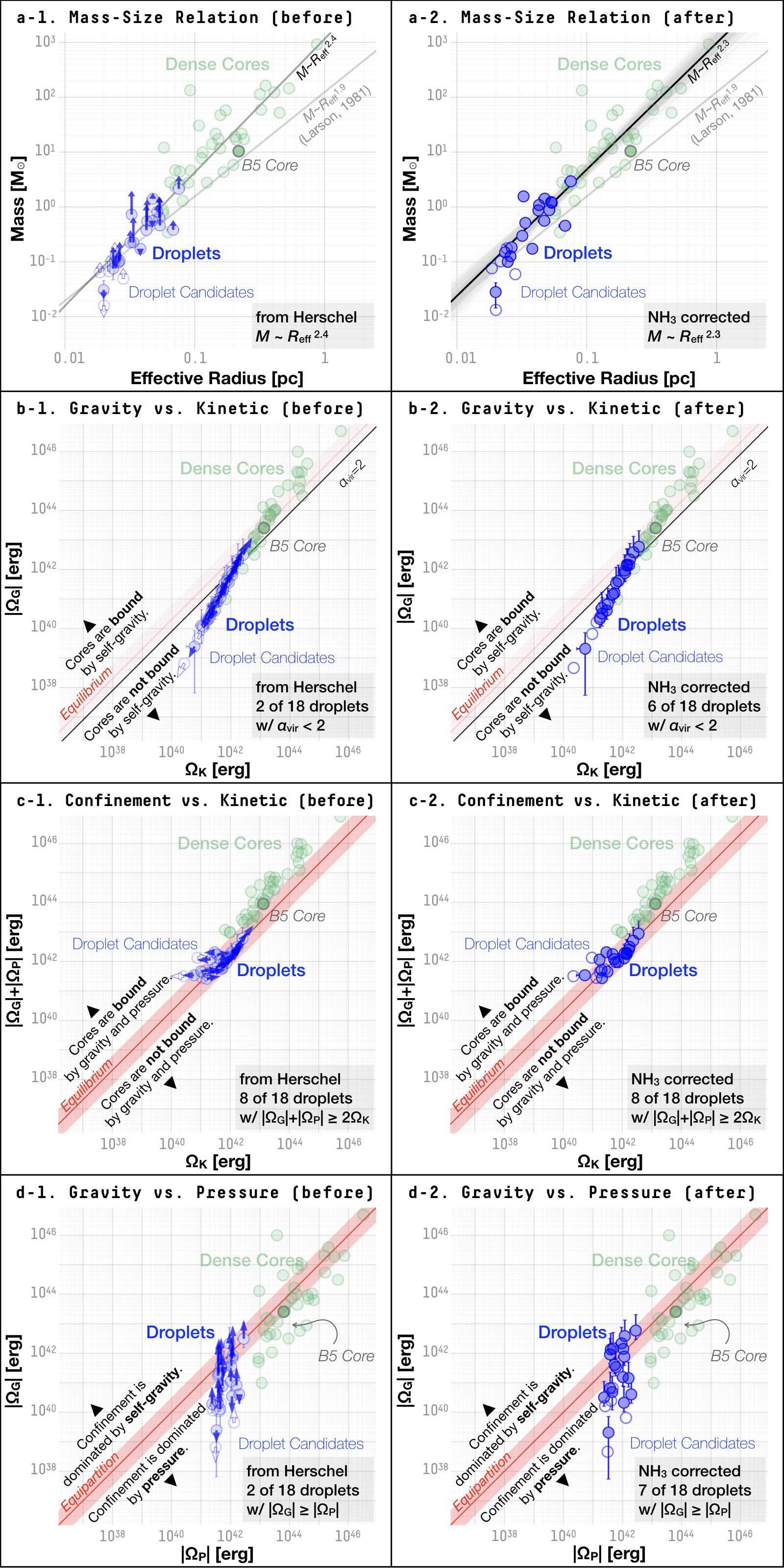}
\caption{\label{fig:change} The left four plots show the change in the physical properties corresponding to the correction using $N_{\mathrm{NH}_3}$ from the fits to the NH$_3$ hyperfine line profiles, plotted on top of the original scatter plots.  The right four plots show the data points after applying the correction, along the same axes.  \textbf{(a-1)} The mass-size distribution of the original data points before the correction, as shown in Fig.\ \ref{fig:basic}a, with the change in mass due to the correction shown as arrows.  \textbf{(a-2)} The mass-size distribution after applying the correction, plotted along the same axes and with the same data points for dense cores (not affected by the correction).  \textbf{(b-1)} The gravitational potential energy plotted against the internal kinetic energy, as shown in Fig.\ \ref{fig:virialGP}a.  The change in both quantities is plotted as arrows.  \textbf{(b-2)} The gravitational potential energy plotted against the internal kinetic energy, after applying the correction, plotted along the same axes and with the same data points for dense cores (not affected by the correction).  \textbf{(c-1)} The sum of the gravitational potential energy and the energy due to the pressure exerted on the cores by the thermal and non-thermal motions of the ambient gas, plotted against the internal kinetic energy, as shown in Fig.\ \ref{fig:virialAll}a.  The change due to the correction in both quantities are plotted as arrows.  \textbf{(c-2)} The sum of gravitational potential energy and the ambient pressure energy term, plotted against the internal kinetic energy, after applying the correction, plotted along the same axes and with the data points for dense cores (not affected by the correction).  \textbf{(d-1)} The gravitational potential energy, plotted against the energy due to the pressure exerted on the core surfaces by the thermal and non-thermal motions of the ambient gas, as shown in Fig.\ \ref{fig:virialAll}.  The change in the gravitational potential energy due to the correction is plotted as arrows.  \textbf{(d-2)} The gravitational potential energy, plotted against the ambient pressure energy, after applying the correction, plotted along the same axes and with the data points for the dense cores (not affected by the correction).}
\end{figure}

After applying the correction factor to the mass, $\Omega_\mathrm{G}$ and $\Omega_\mathrm{K}$ in the virial analysis (\S\ref{sec:analysis_virial}) are changed by ratios $\propto \epsilon^2$ and $\propto \epsilon$, respectively.  The lefthand panels of Fig.\ \ref{fig:change} show the change as arrows, on top of the original mass-size relation (Fig.\ \ref{fig:basic}a) and the original comparisons between various terms in the virial analysis presented in \S\ref{sec:analysis_virial} (cf. Fig.\ \ref{fig:virialGP} and Fig.\ \ref{fig:virialAll}).  The righthand panels of Fig.\ \ref{fig:change} show the resulting plots after applying the correction.  Fig.\ \ref{fig:change}a-2 shows that the mass-size relation after the correction is very slightly less steep and closer to what is found for cloud-scale structures $M \propto R^2$.  Fig.\ \ref{fig:change}b-2 shows that, after the correction, a total of 6 droplets (out of 18 droplets identified in \S\ref{sec:analysis_id}) are now gravitationally ``bound'' based on the conventional criterion of $\alpha_\mathrm{vir} = 2$, as opposed to only 2 droplets that were bound before applying the correction.  The correction on the gravitational potential energy also makes the gravitational term, $\Omega_\mathrm{G}$, in the virial analysis appear more comparable to the ambient pressure term, $\Omega_\mathrm{P}$ (Fig.\ \ref{fig:change}d-2).

\begin{figure}[ht!]
\plotone{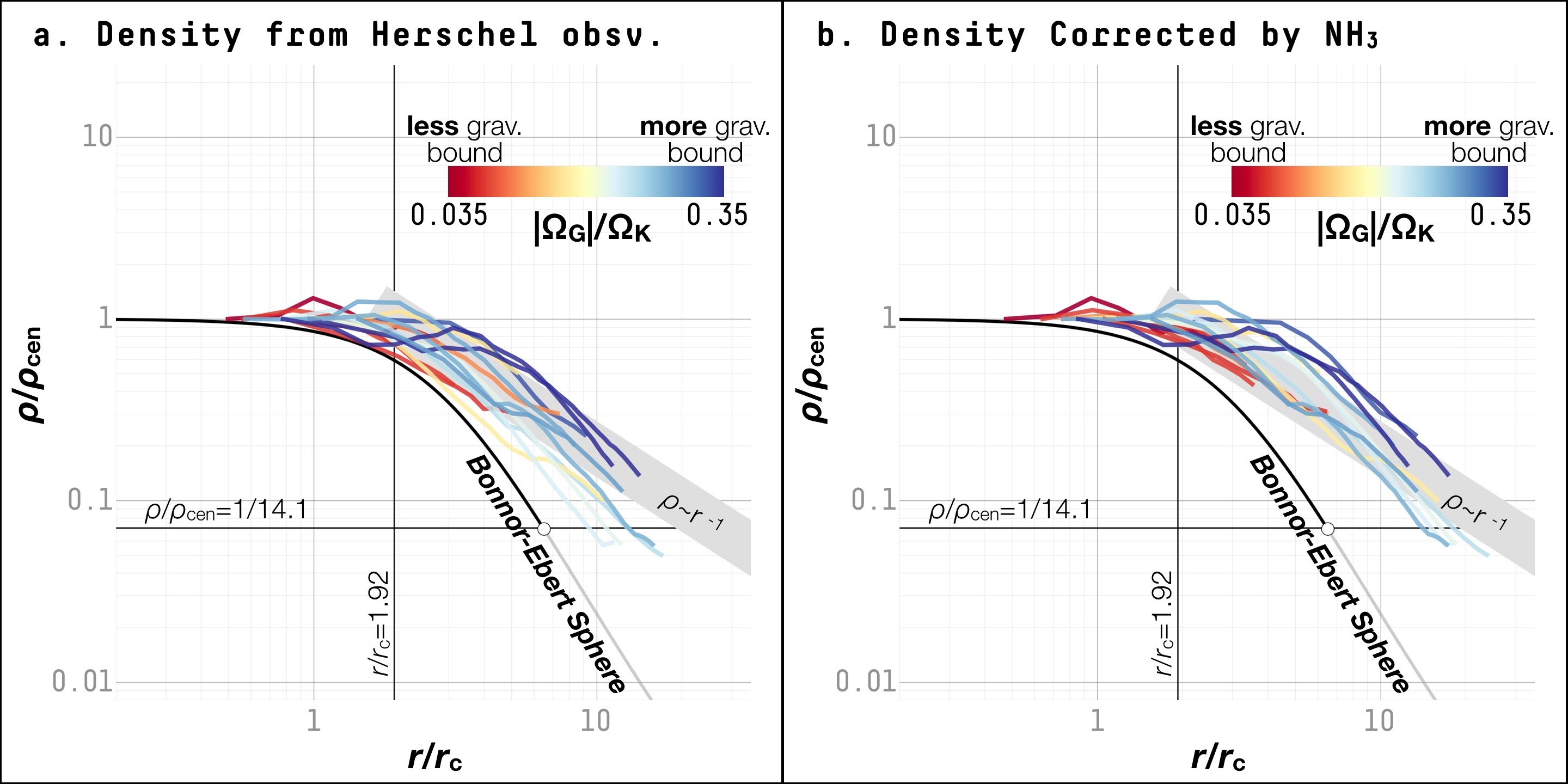}
\caption{\label{fig:profilesChanged} The radial density profiles of the droplets before and after applying the correction using Equation \ref{eq:mass_correction}.  \textbf{(a-1)} The original radial density profiles in dimensionless units, same as Fig.\ \ref{fig:profilesDensity}a.  \textbf{(a-2)} The radial density profiles in dimensionless units, after applying the correction on the characteristic size scale, $r_\mathrm{c}$, which is dependent on $\rho_\mathrm{cen}$.  No correction is applied on $y \equiv \rho / \rho_\mathrm{cen}$, assuming that the same correction factor is applicable on the volume density measured at different distances from the centroid position of a droplet.}
\end{figure}

The correction also affects the normalized radial profiles of density.  The characteristic radius, $r_\mathrm{c}$, which is used to normalize the radial distance from the center is dependent on the density measured at the center of a droplet and is changed by a ratio of $\propto 1/\sqrt{\epsilon}$, making $r_\mathrm{c}$ smaller and consequently $x$ larger.  On the other hand, $y \equiv \rho/\rho_\mathrm{cen}$ is dimensionless and is thus not affected by the correction on the mass (or equivalently, on the density).  The resulting change makes the normalized density profiles look even shallower and closer to $\rho \propto r^{-1}$ at larger distances (Fig.\ \ref{fig:profilesChanged}).

Overall, the correction does not qualitatively alter the results of the analyses presented in \S\ref{sec:analysis_basic} and \S\ref{sec:analysis_virial}.  With the corrected mass, the droplets still appear to follow the same power-law mass-size relation found for dense cores (Fig.\ \ref{fig:change}a-2).  Most of the droplets are still gravitationally unbound (Fig.\ \ref{fig:change}b-2), and the ambient pressure remains important in confining the droplets (Fig.\ \ref{fig:change}c-2).  And, the density profiles of the droplets appear even shallower and seem to remain continuous from the typical density profile found for cloud-scale structures (Fig.\ \ref{fig:profilesChanged}a-2).

Again, since we did not detect NH$_3$ (2, 2) emission everywhere within the droplet boundaries with the GAS observations, and since we hope to examine the mass-size relation with each term independently measured, we base the analyses presented in this paper on the mass estimated from the Herschel column density.  The column density obtained from the SED fitting of Herschel observations also gives better estimates of the radial density profiles outside regions where we find the dense gas (as traced by NH$_3$ emission).  The ``correction'' examined in this section serves only as an estimate of the uncertainty in the SED fitting of Herschel observations, and we emphasize that a more sophisticated approach is needed to further determine the effects of varying NH$_3$ abundances due to changes in the astrochemical environments.

\section{Radial Profiles in Physical Units}
\label{sec:appendix_profiles}
In \S\ref{sec:discussion_confinement_BE}, we examine the radial profiles and compare them to the Bonnor-Ebert sphere \citep{Ebert_1955, Bonnor_1956, Spitzer_1968}.  Following the dimensionless analysis in \citet{Spitzer_1968}, we show that the radial density profiles of the droplets are generally shallower than the Bonnor-Ebert sphere (Fig.\ \ref{fig:profilesDensity}).  The analysis in dimensionless units allows comparisons between droplets of different sizes (\S\ref{sec:discussion_confinement_BE}).

\begin{figure}[ht!]
\plotone{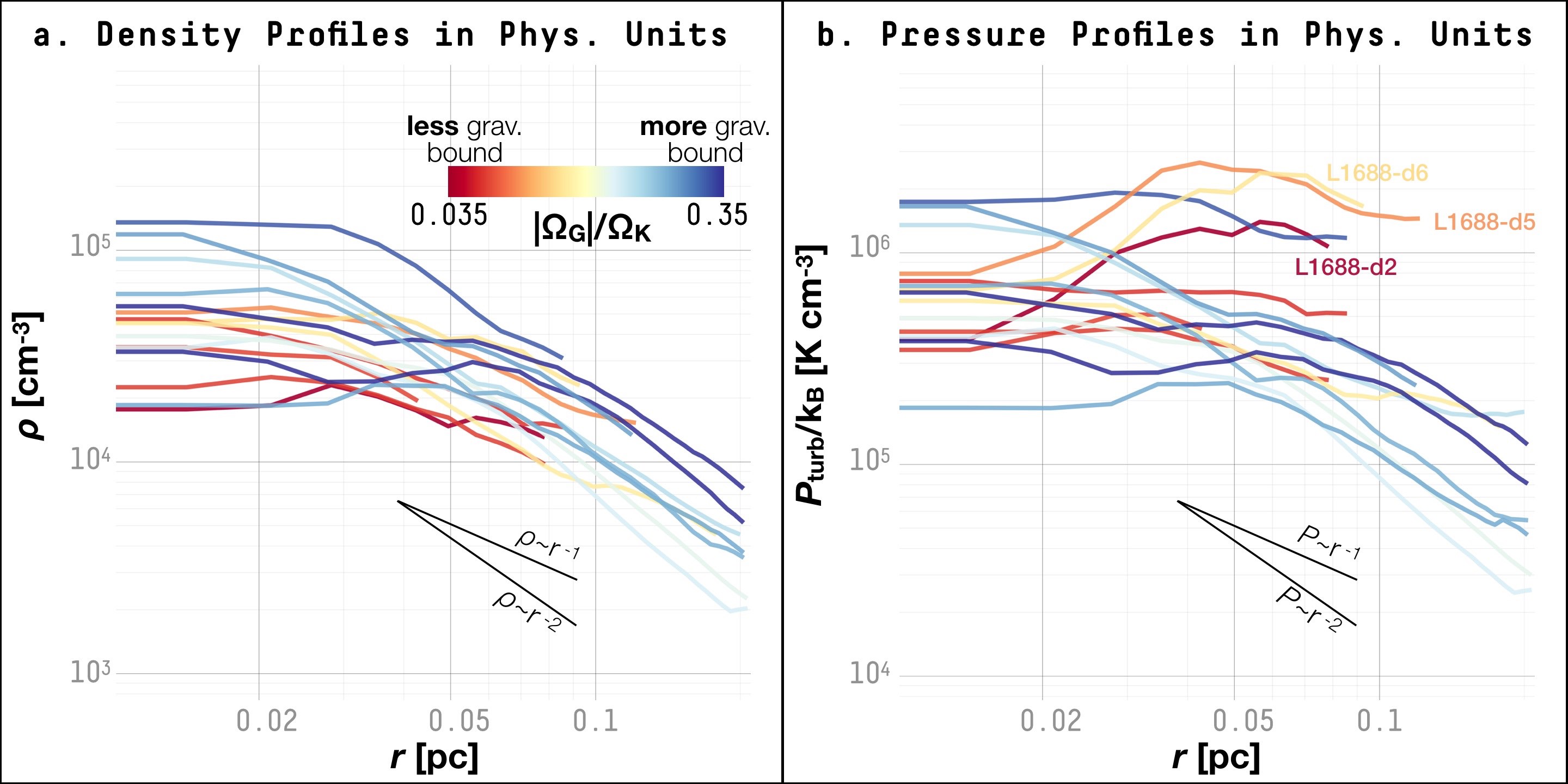}
\caption{\label{fig:profilesPhysical} Radial profiles in physical units.  \textbf{(a)} The radial density profiles of the droplets, color coded according to the ratio between the gravitational potential energy, $\Omega_\mathrm{G}$, and the internal kinetic energy, $\Omega_\mathrm{K}$, similar to Fig.\ \ref{fig:profilesDensity} and Fig.\ \ref{fig:profilesKinematics}.  The unit of the volume density is the number of molecules per cubic centimeter.  \textbf{(b)} The radial pressure profiles of the droplets, similarly color coded by the ratio between the gravitational potential energy, $\Omega_\mathrm{G}$, and the internal kinetic energy, $\Omega_\mathrm{K}$.  Here the pressure measurement is expressed in the unit of K cm$^{-3}$, as a ratio between the measured pressure and the Boltzmann constant.  L1688-d2, L1688-d5, and L1688-d6 are specifically marked, either because the droplets are elongated or because they sit near the edge of the regions where NH$_3$ (1, 1) is detected, both of which render the profiles potentially biased.  Since the dumbbell shape of L1688-d1 affects this analysis which assumes spherical geometry, L1688-d1 is not included in this plot.}
\end{figure}

Fig.\ \ref{fig:profilesPhysical} shows the radial profiles of the volume density and the pressure in physical units.  Again, it demonstrates that the comparison between droplets of different sizes is difficult in the physical units.  See also Fig.\ \ref{fig:sigmas} for the radial profile of linewidths in physical units.

\bibliography{main.bib}

\begin{thebibliography}{}
\expandafter\ifx\csname natexlab\endcsname\relax\def\natexlab#1{#1}\fi
\providecommand{\url}[1]{\href{#1}{#1}}
\providecommand{\dodoi}[1]{doi:~\href{http://doi.org/#1}{\nolinkurl{#1}}}
\providecommand{\doeprint}[1]{\href{http://ascl.net/#1}{\nolinkurl{http://ascl.net/#1}}}
\providecommand{\doarXiv}[1]{\href{https://arxiv.org/abs/#1}{\nolinkurl{https://arxiv.org/abs/#1}}}

\bibitem[{{Andr{\'e}} {et~al.}(2014){Andr{\'e}}, {Di Francesco},
  {Ward-Thompson}, {Inutsuka}, {Pudritz}, \& {Pineda}}]{Andre_2014}
{Andr{\'e}}, P., {Di Francesco}, J., {Ward-Thompson}, D., {et~al.} 2014,
  Protostars and Planets VI, 27,
  \dodoi{10.2458/azu_uapress_9780816531240-ch002}

\bibitem[{Andr{\'e} {et~al.}(2010)Andr{\'e}, Men'shchikov, Bontemps, Konyves,
  Motte, Schneider, Didelon, Minier, Saraceno, Ward-Thompson, Di~Francesco,
  White, Molinari, Testi, Abergel, Griffin, Henning, Royer, Mer{\'\i}n, Vavrek,
  Attard, Arzoumanian, Wilson, Ade, Aussel, Baluteau, Benedettini, Bernard,
  Blommaert, Cambr{\'e}sy, Cox, di~Giorgio, Hargrave, Hennemann, Huang, Kirk,
  Krause, Launhardt, Leeks, Le~Pennec, Li, Martin, Maury, Olofsson, Omont,
  Peretto, Pezzuto, Prusti, Roussel, Russeil, Sauvage, Sibthorpe,
  Sicilia-Aguilar, Spinoglio, Waelkens, Woodcraft, \& Zavagno}]{Andre_2010}
Andr{\'e}, P., Men'shchikov, A., Bontemps, S., {et~al.} 2010, \aap, 518, L102

\bibitem[{{Astropy Collaboration} {et~al.}(2018){Astropy Collaboration},
  {Price-Whelan}, {Sip{\H{o}}cz}, {G{\"u}nther}, {Lim}, {Crawford}, {Conseil},
  {Shupe}, {Craig}, {Dencheva}, {Ginsburg}, {Vand erPlas}, {Bradley},
  {P{\'e}rez-Su{\'a}rez}, {de Val-Borro}, {Aldcroft}, {Cruz}, {Robitaille},
  {Tollerud}, {Ardelean}, {Babej}, {Bach}, {Bachetti}, {Bakanov}, {Bamford},
  {Barentsen}, {Barmby}, {Baumbach}, {Berry}, {Biscani}, {Boquien}, {Bostroem},
  {Bouma}, {Brammer}, {Bray}, {Breytenbach}, {Buddelmeijer}, {Burke},
  {Calderone}, {Cano Rodr{\'\i}guez}, {Cara}, {Cardoso}, {Cheedella}, {Copin},
  {Corrales}, {Crichton}, {D'Avella}, {Deil}, {Depagne}, {Dietrich}, {Donath},
  {Droettboom}, {Earl}, {Erben}, {Fabbro}, {Ferreira}, {Finethy}, {Fox},
  {Garrison}, {Gibbons}, {Goldstein}, {Gommers}, {Greco}, {Greenfield},
  {Groener}, {Grollier}, {Hagen}, {Hirst}, {Homeier}, {Horton}, {Hosseinzadeh},
  {Hu}, {Hunkeler}, {Ivezi{\'c}}, {Jain}, {Jenness}, {Kanarek}, {Kendrew},
  {Kern}, {Kerzendorf}, {Khvalko}, {King}, {Kirkby}, {Kulkarni}, {Kumar},
  {Lee}, {Lenz}, {Littlefair}, {Ma}, {Macleod}, {Mastropietro}, {McCully},
  {Montagnac}, {Morris}, {Mueller}, {Mumford}, {Muna}, {Murphy}, {Nelson},
  {Nguyen}, {Ninan}, {N{\"o}the}, {Ogaz}, {Oh}, {Parejko}, {Parley}, {Pascual},
  {Patil}, {Patil}, {Plunkett}, {Prochaska}, {Rastogi}, {Reddy Janga},
  {Sabater}, {Sakurikar}, {Seifert}, {Sherbert}, {Sherwood-Taylor}, {Shih},
  {Sick}, {Silbiger}, {Singanamalla}, {Singer}, {Sladen}, {Sooley},
  {Sornarajah}, {Streicher}, {Teuben}, {Thomas}, {Tremblay}, {Turner},
  {Terr{\'o}n}, {van Kerkwijk}, {de la Vega}, {Watkins}, {Weaver}, {Whitmore},
  {Woillez}, {Zabalza}, \& {Astropy Contributors}}]{astropy}
{Astropy Collaboration}, {Price-Whelan}, A.~M., {Sip{\H{o}}cz}, B.~M., {et~al.}
  2018, \aj, 156, 123, \dodoi{10.3847/1538-3881/aabc4f}

\bibitem[{{Ballesteros-Paredes} {et~al.}(1999){Ballesteros-Paredes},
  {V{\'a}zquez-Semadeni}, \& {Scalo}}]{BallesterosParedes_1999}
{Ballesteros-Paredes}, J., {V{\'a}zquez-Semadeni}, E., \& {Scalo}, J. 1999,
  \apj, 515, 286, \dodoi{10.1086/307007}

\bibitem[{{Barranco} \& {Goodman}(1998)}]{Barranco_1998}
{Barranco}, J.~A., \& {Goodman}, A.~A. 1998, \apj, 504, 207,
  \dodoi{10.1086/306044}

\bibitem[{{Beaumont} {et~al.}(2015){Beaumont}, {Goodman}, \&
  {Greenfield}}]{glue}
{Beaumont}, C., {Goodman}, A., \& {Greenfield}, P. 2015, in Astronomical
  Society of the Pacific Conference Series, Vol. 495, Astronomical Data
  Analysis Software an Systems XXIV (ADASS XXIV), ed. A.~R. {Taylor} \&
  E.~{Rosolowsky}, 101

\bibitem[{{Beaumont} {et~al.}(2013){Beaumont}, {Offner}, {Shetty}, {Glover}, \&
  {Goodman}}]{Beaumont_2013}
{Beaumont}, C.~N., {Offner}, S.~S.~R., {Shetty}, R., {Glover}, S.~C.~O., \&
  {Goodman}, A.~A. 2013, \apj, 777, 173, \dodoi{10.1088/0004-637X/777/2/173}

\bibitem[{{Benson} {et~al.}(1998){Benson}, {Caselli}, \& {Myers}}]{Benson_1998}
{Benson}, P.~J., {Caselli}, P., \& {Myers}, P.~C. 1998, \apj, 506, 743,
  \dodoi{10.1086/306276}

\bibitem[{{Benson} \& {Myers}(1983)}]{Benson_1983}
{Benson}, P.~J., \& {Myers}, P.~C. 1983, \apj, 270, 589, \dodoi{10.1086/161151}

\bibitem[{{Benson} \& {Myers}(1989)}]{Benson_1989}
---. 1989, \apjs, 71, 89, \dodoi{10.1086/191365}

\bibitem[{{Bertoldi} \& {McKee}(1992)}]{Bertoldi_1992}
{Bertoldi}, F., \& {McKee}, C.~F. 1992, \apj, 395, 140, \dodoi{10.1086/171638}

\bibitem[{{Bonnor}(1956)}]{Bonnor_1956}
{Bonnor}, W.~B. 1956, \mnras, 116, 351, \dodoi{10.1093/mnras/116.3.351}

\bibitem[{{Bourke} {et~al.}(1995){Bourke}, {Hyland}, \&
  {Robinson}}]{Bourke_1995}
{Bourke}, T.~L., {Hyland}, A.~R., \& {Robinson}, G. 1995, \mnras, 276, 1052,
  \dodoi{10.1093/mnras/276.4.1052}

\bibitem[{{Caselli} {et~al.}(2002){Caselli}, {Benson}, {Myers}, \&
  {Tafalla}}]{Caselli_2002}
{Caselli}, P., {Benson}, P.~J., {Myers}, P.~C., \& {Tafalla}, M. 2002, \apj,
  572, 238, \dodoi{10.1086/340195}

\bibitem[{{Cox} \& {Pilachowski}(2000)}]{Cox_2000}
{Cox}, A.~N., \& {Pilachowski}, C.~A. 2000, Physics Today, 53, 77,
  \dodoi{10.1063/1.1325201}

\bibitem[{{Crapsi} {et~al.}(2007){Crapsi}, {Caselli}, {Walmsley}, \&
  {Tafalla}}]{Crapsi_2007}
{Crapsi}, A., {Caselli}, P., {Walmsley}, M.~C., \& {Tafalla}, M. 2007, \aap,
  470, 221, \dodoi{10.1051/0004-6361:20077613}

\bibitem[{{Dame} \& {Thaddeus}(1985)}]{Dame_1985}
{Dame}, T.~M., \& {Thaddeus}, P. 1985, \apj, 297, 751, \dodoi{10.1086/163573}

\bibitem[{{de Geus} {et~al.}(1989){de Geus}, {de Zeeuw}, \&
  {Lub}}]{deGeus_1989}
{de Geus}, E.~J., {de Zeeuw}, P.~T., \& {Lub}, J. 1989, \aap, 216, 44

\bibitem[{{Dullemond} {et~al.}(2012){Dullemond}, {Juhasz}, {Pohl}, {Sereshti},
  {Shetty}, {Peters}, {Commercon}, \& {Flock}}]{radmc3d}
{Dullemond}, C.~P., {Juhasz}, A., {Pohl}, A., {et~al.} 2012, {RADMC-3D: A
  multi-purpose radiative transfer tool}, Astrophysics Source Code Library.
\newblock \doeprint{1202.015}

\bibitem[{{Dunham} {et~al.}(2015){Dunham}, {Allen}, {Evans},
  {Broekhoven-Fiene}, {Cieza}, {Di Francesco}, {Gutermuth}, {Harvey},
  {Hatchell}, {Heiderman}, {Huard}, {Johnstone}, {Kirk}, {Matthews}, {Miller},
  {Peterson}, \& {Young}}]{Dunham_2015}
{Dunham}, M.~M., {Allen}, L.~E., {Evans}, II, N.~J., {et~al.} 2015, \apjs, 220,
  11, \dodoi{10.1088/0067-0049/220/1/11}

\bibitem[{{Ebert}(1955)}]{Ebert_1955}
{Ebert}, R. 1955, \zap, 37, 217

\bibitem[{{Elmegreen}(2000)}]{Elmegreen_2000}
{Elmegreen}, B.~G. 2000, \apj, 530, 277, \dodoi{10.1086/308361}

\bibitem[{{Enoch} {et~al.}(2008){Enoch}, {Evans}, {Sargent}, {Glenn},
  {Rosolowsky}, \& {Myers}}]{Enoch_2008}
{Enoch}, M.~L., {Evans}, II, N.~J., {Sargent}, A.~I., {et~al.} 2008, \apj, 684,
  1240, \dodoi{10.1086/589963}

\bibitem[{{Evans}(1999)}]{Evans_1999}
{Evans}, II, N.~J. 1999, \araa, 37, 311, \dodoi{10.1146/annurev.astro.37.1.311}

\bibitem[{{Federrath}(2013)}]{Federrath_2013}
{Federrath}, C. 2013, \mnras, 436, 1245, \dodoi{10.1093/mnras/stt1644}

\bibitem[{{Fischera} \& {Martin}(2012)}]{Fischera_2012}
{Fischera}, J., \& {Martin}, P.~G. 2012, \aap, 547, A86,
  \dodoi{10.1051/0004-6361/201219728}

\bibitem[{{Friesen} {et~al.}(2010){Friesen}, {Di Francesco}, {Shimajiri}, \&
  {Takakuwa}}]{Friesen_2010}
{Friesen}, R.~K., {Di Francesco}, J., {Shimajiri}, Y., \& {Takakuwa}, S. 2010,
  \apj, 708, 1002, \dodoi{10.1088/0004-637X/708/2/1002}

\bibitem[{{Friesen} {et~al.}(2009){Friesen}, {Di Francesco}, {Shirley}, \&
  {Myers}}]{Friesen_2009}
{Friesen}, R.~K., {Di Francesco}, J., {Shirley}, Y.~L., \& {Myers}, P.~C. 2009,
  \apj, 697, 1457, \dodoi{10.1088/0004-637X/697/2/1457}

\bibitem[{{Friesen} {et~al.}(2017){Friesen}, {Pineda}, {co-PIs}, {Rosolowsky},
  {Alves}, {Chac{\'o}n-Tanarro}, {How-Huan Chen}, {Chun-Yuan Chen}, {Di
  Francesco}, {Keown}, {Kirk}, {Punanova}, {Seo}, {Shirley}, {Ginsburg},
  {Hall}, {Offner}, {Singh}, {Arce}, {Caselli}, {Goodman}, {Martin}, {Matzner},
  {Myers}, {Redaelli}, \& {The GAS Collaboration}}]{GAS_DR1}
{Friesen}, R.~K., {Pineda}, J.~E., {co-PIs}, {et~al.} 2017, \apj, 843, 63,
  \dodoi{10.3847/1538-4357/aa6d58}

\bibitem[{{Fuller} \& {Myers}(1992)}]{Fuller_1992}
{Fuller}, G.~A., \& {Myers}, P.~C. 1992, \apj, 384, 523, \dodoi{10.1086/170894}

\bibitem[{{Galli} {et~al.}(2018){Galli}, {Loinard}, {Ortiz-L{\'e}on},
  {Kounkel}, {Dzib}, {Mioduszewski}, {Rodr{\'\i}guez}, {Hartmann}, {Teixeira},
  {Torres}, {Rivera}, {Boden}, {Evans}, {Brice{\~n}o}, {Tobin}, \&
  {Heyer}}]{Galli_2018}
{Galli}, P. A.~B., {Loinard}, L., {Ortiz-L{\'e}on}, G.~N., {et~al.} 2018, \apj,
  859, 33, \dodoi{10.3847/1538-4357/aabf91}

\bibitem[{{Garay} \& {Lizano}(1999)}]{Garay_1999}
{Garay}, G., \& {Lizano}, S. 1999, \pasp, 111, 1049, \dodoi{10.1086/316416}

\bibitem[{{Ginsburg} \& {Mirocha}(2011)}]{pyspeckit}
{Ginsburg}, A., \& {Mirocha}, J. 2011, {PySpecKit: Python Spectroscopic
  Toolkit}, Astrophysics Source Code Library.
\newblock \doeprint{1109.001}

\bibitem[{{Goodman} {et~al.}(1998){Goodman}, {Barranco}, {Wilner}, \&
  {Heyer}}]{Goodman_1998}
{Goodman}, A.~A., {Barranco}, J.~A., {Wilner}, D.~J., \& {Heyer}, M.~H. 1998,
  \apj, 504, 223, \dodoi{10.1086/306045}

\bibitem[{{Goodman} {et~al.}(1993){Goodman}, {Benson}, {Fuller}, \&
  {Myers}}]{Goodman_1993}
{Goodman}, A.~A., {Benson}, P.~J., {Fuller}, G.~A., \& {Myers}, P.~C. 1993,
  \apj, 406, 528, \dodoi{10.1086/172465}

\bibitem[{{Gurney} {et~al.}(2008){Gurney}, {Plume}, \&
  {Johnstone}}]{Gurney_2008}
{Gurney}, M., {Plume}, R., \& {Johnstone}, D. 2008, Publications of the
  Astronomical Society of the Pacific, 120, 1193, \dodoi{10.1086/593074}

\bibitem[{{Hacar} {et~al.}(2013){Hacar}, {Tafalla}, {Kauffmann}, \&
  {Kov{\'a}cs}}]{Hacar_2013}
{Hacar}, A., {Tafalla}, M., {Kauffmann}, J., \& {Kov{\'a}cs}, A. 2013, \aap,
  554, A55, \dodoi{10.1051/0004-6361/201220090}

\bibitem[{{Hildebrand}(1983)}]{Hildebrand_1983}
{Hildebrand}, R.~H. 1983, \qjras, 24, 267

\bibitem[{{Hilton} \& {Lahulla}(1995)}]{Hilton_1995}
{Hilton}, J., \& {Lahulla}, J.~F. 1995, \aaps, 113, 325

\bibitem[{{Jeans}(1902)}]{Jeans_1902}
{Jeans}, J.~H. 1902, Philosophical Transactions of the Royal Society of London
  Series A, 199, 1, \dodoi{10.1098/rsta.1902.0012}

\bibitem[{{Jijina} {et~al.}(1999){Jijina}, {Myers}, \& {Adams}}]{Jijina_1999}
{Jijina}, J., {Myers}, P.~C., \& {Adams}, F.~C. 1999, The Astrophysical Journal
  Supplement Series, 125, 161, \dodoi{10.1086/313268}

\bibitem[{{Johnstone} {et~al.}(2001){Johnstone}, {Fich}, {Mitchell}, \&
  {Moriarty-Schieven}}]{Johnstone_2001}
{Johnstone}, D., {Fich}, M., {Mitchell}, G.~F., \& {Moriarty-Schieven}, G.
  2001, \apj, 559, 307, \dodoi{10.1086/322323}

\bibitem[{{Johnstone} {et~al.}(2000){Johnstone}, {Wilson}, {Moriarty-Schieven},
  {Joncas}, {Smith}, {Gregersen}, \& {Fich}}]{Johnstone_2000}
{Johnstone}, D., {Wilson}, C.~D., {Moriarty-Schieven}, G., {et~al.} 2000, \apj,
  545, 327, \dodoi{10.1086/317790}

\bibitem[{{J{\o}rgensen} {et~al.}(2008){J{\o}rgensen}, {Johnstone}, {Kirk},
  {Myers}, {Allen}, \& {Shirley}}]{Jorgensen_2008}
{J{\o}rgensen}, J.~K., {Johnstone}, D., {Kirk}, H., {et~al.} 2008, \apj, 683,
  822, \dodoi{10.1086/589956}

\bibitem[{{Kauffmann} {et~al.}(2008){Kauffmann}, {Bertoldi}, {Bourke}, {Evans},
  \& {Lee}}]{Kauffmann_2008}
{Kauffmann}, J., {Bertoldi}, F., {Bourke}, T.~L., {Evans}, II, N.~J., \& {Lee},
  C.~W. 2008, \aap, 487, 993, \dodoi{10.1051/0004-6361:200809481}

\bibitem[{{Kauffmann} {et~al.}(2010{\natexlab{a}}){Kauffmann}, {Pillai},
  {Shetty}, {Myers}, \& {Goodman}}]{Kauffmann_2010a}
{Kauffmann}, J., {Pillai}, T., {Shetty}, R., {Myers}, P.~C., \& {Goodman},
  A.~A. 2010{\natexlab{a}}, \apj, 712, 1137,
  \dodoi{10.1088/0004-637X/712/2/1137}

\bibitem[{{Kauffmann} {et~al.}(2010{\natexlab{b}}){Kauffmann}, {Pillai},
  {Shetty}, {Myers}, \& {Goodman}}]{Kauffmann_2010b}
---. 2010{\natexlab{b}}, \apj, 716, 433, \dodoi{10.1088/0004-637X/716/1/433}

\bibitem[{{Kelly} {et~al.}(2012){Kelly}, {Shetty}, {Stutz}, {Kauffmann},
  {Goodman}, \& {Launhardt}}]{Kelly_2012}
{Kelly}, B.~C., {Shetty}, R., {Stutz}, A.~M., {et~al.} 2012, \apj, 752, 55,
  \dodoi{10.1088/0004-637X/752/1/55}

\bibitem[{{Keto} \& {Myers}(1986)}]{Keto_1986}
{Keto}, E.~R., \& {Myers}, P.~C. 1986, \apj, 304, 466, \dodoi{10.1086/164181}

\bibitem[{{Kirk} {et~al.}(2010){Kirk}, {Pineda}, {Johnstone}, \&
  {Goodman}}]{Kirk_2010}
{Kirk}, H., {Pineda}, J.~E., {Johnstone}, D., \& {Goodman}, A. 2010, \apj, 723,
  457, \dodoi{10.1088/0004-637X/723/1/457}

\bibitem[{{Kirk} {et~al.}(2017{\natexlab{a}}){Kirk}, {Friesen}, {Pineda},
  {Rosolowsky}, {Offner}, {Matzner}, {Myers}, {Di Francesco}, {Caselli},
  {Alves}, {Chac{\'o}n-Tanarro}, {Chen}, {Chun-Yuan Chen}, {Keown}, {Punanova},
  {Seo}, {Shirley}, {Ginsburg}, {Hall}, {Singh}, {Arce}, {Goodman}, {Martin},
  \& {Redaelli}}]{Kirk_2017b}
{Kirk}, H., {Friesen}, R.~K., {Pineda}, J.~E., {et~al.} 2017{\natexlab{a}},
  \apj, 846, 144, \dodoi{10.3847/1538-4357/aa8631}

\bibitem[{{Kirk} {et~al.}(2017{\natexlab{b}}){Kirk}, {Dunham}, {Di Francesco},
  {Johnstone}, {Offner}, {Sadavoy}, {Tobin}, {Arce}, {Bourke}, {Mairs},
  {Myers}, {Pineda}, {Schnee}, \& {Shirley}}]{Kirk_2017a}
{Kirk}, H., {Dunham}, M.~M., {Di Francesco}, J., {et~al.} 2017{\natexlab{b}},
  \apj, 838, 114, \dodoi{10.3847/1538-4357/aa63f8}

\bibitem[{{Kun}(1998)}]{Kun_1998}
{Kun}, M. 1998, \apjs, 115, 59, \dodoi{10.1086/313076}

\bibitem[{{Lada} {et~al.}(2008){Lada}, {Muench}, {Rathborne}, {Alves}, \&
  {Lombardi}}]{Lada_2008}
{Lada}, C.~J., {Muench}, A.~A., {Rathborne}, J., {Alves}, J.~F., \& {Lombardi},
  M. 2008, \apj, 672, 410, \dodoi{10.1086/523837}

\bibitem[{{Ladd} {et~al.}(1994){Ladd}, {Myers}, \& {Goodman}}]{Ladd_1994}
{Ladd}, E.~F., {Myers}, P.~C., \& {Goodman}, A.~A. 1994, \apj, 433, 117,
  \dodoi{10.1086/174629}

\bibitem[{{Larson}(1981)}]{Larson_1981}
{Larson}, R.~B. 1981, \mnras, 194, 809, \dodoi{10.1093/mnras/194.4.809}

\bibitem[{{Lee} {et~al.}(1999){Lee}, {Myers}, \& {Tafalla}}]{Lee_1999}
{Lee}, C.~W., {Myers}, P.~C., \& {Tafalla}, M. 1999, \apj, 526, 788,
  \dodoi{10.1086/308027}

\bibitem[{{Li} {et~al.}(2015){Li}, {Yuen}, {Otto}, {Leung}, {Sridharan},
  {Zhang}, {Liu}, {Tang}, \& {Qiu}}]{Li_2015}
{Li}, H.-B., {Yuen}, K.~H., {Otto}, F., {et~al.} 2015, \nat, 520, 518,
  \dodoi{10.1038/nature14291}

\bibitem[{{Li} {et~al.}(2012){Li}, {Martin}, {Klein}, \& {McKee}}]{Li_2012}
{Li}, P.~S., {Martin}, D.~F., {Klein}, R.~I., \& {McKee}, C.~F. 2012, \apj,
  745, 139, \dodoi{10.1088/0004-637X/745/2/139}

\bibitem[{{Loren}(1989)}]{Loren_1989}
{Loren}, R.~B. 1989, \apj, 338, 902, \dodoi{10.1086/167244}

\bibitem[{{Mangum} \& {Shirley}(2015)}]{Mangum_2015}
{Mangum}, J.~G., \& {Shirley}, Y.~L. 2015, \pasp, 127, 266,
  \dodoi{10.1086/680323}

\bibitem[{{Maruta} {et~al.}(2010){Maruta}, {Nakamura}, {Nishi}, {Ikeda}, \&
  {Kitamura}}]{Maruta_2010}
{Maruta}, H., {Nakamura}, F., {Nishi}, R., {Ikeda}, N., \& {Kitamura}, Y. 2010,
  \apj, 714, 680, \dodoi{10.1088/0004-637X/714/1/680}

\bibitem[{{McKee} \& {Ostriker}(2007)}]{McKee_2007}
{McKee}, C.~F., \& {Ostriker}, E.~C. 2007, \araa, 45, 565,
  \dodoi{10.1146/annurev.astro.45.051806.110602}

\bibitem[{{McLaughlin} \& {Pudritz}(1996)}]{McLaughlin_1996}
{McLaughlin}, D.~E., \& {Pudritz}, R.~E. 1996, \apj, 469, 194,
  \dodoi{10.1086/177771}

\bibitem[{{McLaughlin} \& {Pudritz}(1997)}]{McLaughlin_1997}
---. 1997, \apj, 476, 750, \dodoi{10.1086/303657}

\bibitem[{{Montillaud} {et~al.}(2015){Montillaud}, {Juvela}, {Rivera-Ingraham},
  {Malinen}, {Pelkonen}, {Ristorcelli}, {Montier}, {Marshall}, {Marton},
  {Pagani}, {Toth}, {Zahorecz}, {Ysard}, {McGehee}, {Paladini}, {Falgarone},
  {Bernard}, {Motte}, {Zavagno}, \& {Doi}}]{Montillaud_2015}
{Montillaud}, J., {Juvela}, M., {Rivera-Ingraham}, A., {et~al.} 2015, \aap,
  584, A92, \dodoi{10.1051/0004-6361/201424063}

\bibitem[{{Motte} {et~al.}(1998){Motte}, {Andre}, \& {Neri}}]{Motte_1998}
{Motte}, F., {Andre}, P., \& {Neri}, R. 1998, \aap, 336, 150

\bibitem[{{Myers}(1983)}]{Myers_1983c}
{Myers}, P.~C. 1983, \apj, 270, 105, \dodoi{10.1086/161101}

\bibitem[{{Myers}(1998)}]{Myers_1998}
---. 1998, \apj, 496, L109, \dodoi{10.1086/311256}

\bibitem[{{Myers} \& {Benson}(1983)}]{Myers_1983b}
{Myers}, P.~C., \& {Benson}, P.~J. 1983, \apj, 266, 309, \dodoi{10.1086/160780}

\bibitem[{{Myers} {et~al.}(1991){Myers}, {Fuller}, {Goodman}, \&
  {Benson}}]{Myers_1991}
{Myers}, P.~C., {Fuller}, G.~A., {Goodman}, A.~A., \& {Benson}, P.~J. 1991,
  \apj, 376, 561, \dodoi{10.1086/170305}

\bibitem[{{Myers} {et~al.}(1983){Myers}, {Linke}, \& {Benson}}]{Myers_1983a}
{Myers}, P.~C., {Linke}, R.~A., \& {Benson}, P.~J. 1983, \apj, 264, 517,
  \dodoi{10.1086/160619}

\bibitem[{{Offner} \& {Arce}(2015)}]{Offner_2015}
{Offner}, S.~S.~R., \& {Arce}, H.~G. 2015, \apj, 811, 146,
  \dodoi{10.1088/0004-637X/811/2/146}

\bibitem[{{Onishi} {et~al.}(2002){Onishi}, {Mizuno}, {Kawamura}, {Tachihara},
  \& {Fukui}}]{Onishi_2002}
{Onishi}, T., {Mizuno}, A., {Kawamura}, A., {Tachihara}, K., \& {Fukui}, Y.
  2002, \apj, 575, 950, \dodoi{10.1086/341347}

\bibitem[{{Ortiz-Le{\'o}n} {et~al.}(2017){Ortiz-Le{\'o}n}, {Loinard},
  {Kounkel}, {Dzib}, {Mioduszewski}, {Rodr{\'{\i}}guez}, {Torres},
  {Gonz{\'a}lez-L{\'o}pezlira}, {Pech}, {Rivera}, {Hartmann}, {Boden}, {Evans},
  {Brice{\~n}o}, {Tobin}, {Galli}, \& {Gudehus}}]{OrtizLeon_2017}
{Ortiz-Le{\'o}n}, G.~N., {Loinard}, L., {Kounkel}, M.~A., {et~al.} 2017, \apj,
  834, 141, \dodoi{10.3847/1538-4357/834/2/141}

\bibitem[{{Padoan} {et~al.}(2014){Padoan}, {Federrath}, {Chabrier}, {Evans},
  {Johnstone}, {J{\o}rgensen}, {McKee}, \& {Nordlund}}]{Padoan_2014}
{Padoan}, P., {Federrath}, C., {Chabrier}, G., {et~al.} 2014, Protostars and
  Planets VI, 77, \dodoi{10.2458/azu_uapress_9780816531240-ch004}

\bibitem[{{Pattle} {et~al.}(2015){Pattle}, {Ward-Thompson}, {Kirk}, {White},
  {Drabek-Maunder}, {Buckle}, {Beaulieu}, {Berry}, {Broekhoven-Fiene},
  {Currie}, {Fich}, {Hatchell}, {Kirk}, {Jenness}, {Johnstone}, {Mottram},
  {Nutter}, {Pineda}, {Quinn}, {Salji}, {Tisi}, {Walker-Smith}, {di Francesco},
  {Hogerheijde}, {Andr{\'e}}, {Bastien}, {Bresnahan}, {Butner}, {Chen},
  {Chrysostomou}, {Coude}, {Davis}, {Duarte-Cabral}, {Fiege}, {Friberg},
  {Friesen}, {Fuller}, {Graves}, {Greaves}, {Gregson}, {Griffin}, {Holland},
  {Joncas}, {Knee}, {K{\"o}nyves}, {Mairs}, {Marsh}, {Matthews},
  {Moriarty-Schieven}, {Rawlings}, {Richer}, {Robertson}, {Rosolowsky},
  {Rumble}, {Sadavoy}, {Spinoglio}, {Thomas}, {Tothill}, {Viti}, {Wouterloot},
  {Yates}, \& {Zhu}}]{Pattle_2015}
{Pattle}, K., {Ward-Thompson}, D., {Kirk}, J.~M., {et~al.} 2015, \mnras, 450,
  1094, \dodoi{10.1093/mnras/stv376}

\bibitem[{{Pineda} {et~al.}(2010){Pineda}, {Goodman}, {Arce}, {Caselli},
  {Foster}, {Myers}, \& {Rosolowsky}}]{Pineda_2010}
{Pineda}, J.~E., {Goodman}, A.~A., {Arce}, H.~G., {et~al.} 2010, \apjl, 712,
  L116, \dodoi{10.1088/2041-8205/712/1/L116}

\bibitem[{{Pineda} {et~al.}(2011){Pineda}, {Goodman}, {Arce}, {Caselli},
  {Longmore}, \& {Corder}}]{Pineda_2011}
---. 2011, \apjl, 739, L2, \dodoi{10.1088/2041-8205/739/1/L2}

\bibitem[{{Pineda} {et~al.}(2015){Pineda}, {Offner}, {Parker}, {Arce},
  {Goodman}, {Caselli}, {Fuller}, {Bourke}, \& {Corder}}]{Pineda_2015}
{Pineda}, J.~E., {Offner}, S.~S.~R., {Parker}, R.~J., {et~al.} 2015, \nat, 518,
  213, \dodoi{10.1038/nature14166}

\bibitem[{{Planck Collaboration} {et~al.}(2014){Planck Collaboration},
  {Abergel}, {Ade}, {Aghanim}, {Alves}, {Aniano}, {Armitage-Caplan}, {Arnaud},
  {Ashdown}, {Atrio-Barandela}, \& et~al.}]{PlanckXI}
{Planck Collaboration}, {Abergel}, A., {Ade}, P.~A.~R., {et~al.} 2014, \aap,
  571, A11, \dodoi{10.1051/0004-6361/201323195}

\bibitem[{{Rebull} {et~al.}(2010){Rebull}, {Padgett}, {McCabe}, {Hillenbrand},
  {Stapelfeldt}, {Noriega-Crespo}, {Carey}, {Brooke}, {Huard}, {Terebey},
  {Audard}, {Monin}, {Fukagawa}, {G{\"u}del}, {Knapp}, {Menard}, {Allen},
  {Angione}, {Baldovin-Saavedra}, {Bouvier}, {Briggs}, {Dougados}, {Evans},
  {Flagey}, {Guieu}, {Grosso}, {Glauser}, {Harvey}, {Hines}, {Latter},
  {Skinner}, {Strom}, {Tromp}, \& {Wolf}}]{Rebull_2010}
{Rebull}, L.~M., {Padgett}, D.~L., {McCabe}, C.-E., {et~al.} 2010, \apjs, 186,
  259, \dodoi{10.1088/0067-0049/186/2/259}

\bibitem[{{Ridge} {et~al.}(2006){Ridge}, {Di Francesco}, {Kirk}, {Li},
  {Goodman}, {Alves}, {Arce}, {Borkin}, {Caselli}, {Foster}, {Heyer},
  {Johnstone}, {Kosslyn}, {Lombardi}, {Pineda}, {Schnee}, \&
  {Tafalla}}]{Ridge_2006}
{Ridge}, N.~A., {Di Francesco}, J., {Kirk}, H., {et~al.} 2006, \aj, 131, 2921,
  \dodoi{10.1086/503704}

\bibitem[{Robitaille {et~al.}(2017)Robitaille, Beaumont, Qian, Borkin, \&
  Goodman}]{glue_2017}
Robitaille, T., Beaumont, C., Qian, P., Borkin, M., \& Goodman, A. 2017,
  {glueviz v0.13.1: multidimensional data exploration glueviz v0.13.1:
  multidimensional data exploration}, \dodoi{10.5281/zenodo.1237692}.
\newblock \url{https://doi.org/10.5281/zenodo.1237692}

\bibitem[{{Rosolowsky} \& {Leroy}(2006)}]{Rosolowsky_2006}
{Rosolowsky}, E., \& {Leroy}, A. 2006, \pasp, 118, 590, \dodoi{10.1086/502982}

\bibitem[{{Rosolowsky} {et~al.}(2008{\natexlab{a}}){Rosolowsky}, {Pineda},
  {Foster}, {Borkin}, {Kauffmann}, {Caselli}, {Myers}, \&
  {Goodman}}]{Rosolowsky_2008a}
{Rosolowsky}, E.~W., {Pineda}, J.~E., {Foster}, J.~B., {et~al.}
  2008{\natexlab{a}}, \apjs, 175, 509, \dodoi{10.1086/524299}

\bibitem[{{Rosolowsky} {et~al.}(2008{\natexlab{b}}){Rosolowsky}, {Pineda},
  {Kauffmann}, \& {Goodman}}]{Rosolowsky_2008b}
{Rosolowsky}, E.~W., {Pineda}, J.~E., {Kauffmann}, J., \& {Goodman}, A.~A.
  2008{\natexlab{b}}, \apj, 679, 1338, \dodoi{10.1086/587685}

\bibitem[{{Schlafly} {et~al.}(2014){Schlafly}, {Green}, {Finkbeiner}, {Rix},
  {Bell}, {Burgett}, {Chambers}, {Draper}, {Hodapp}, {Kaiser}, {Magnier},
  {Martin}, {Metcalfe}, {Price}, \& {Tonry}}]{Schlafly_2014}
{Schlafly}, E.~F., {Green}, G., {Finkbeiner}, D.~P., {et~al.} 2014, \apj, 786,
  29, \dodoi{10.1088/0004-637X/786/1/29}

\bibitem[{{Schmidt}(1975)}]{Schmidt_1975}
{Schmidt}, E.~G. 1975, \mnras, 172, 401, \dodoi{10.1093/mnras/172.2.401}

\bibitem[{{Schneider} \& {Elmegreen}(1979)}]{Schneider_1979}
{Schneider}, S., \& {Elmegreen}, B.~G. 1979, \apjs, 41, 87,
  \dodoi{10.1086/190609}

\bibitem[{{Sch{\"o}ier} {et~al.}(2005){Sch{\"o}ier}, {van der Tak}, {van
  Dishoeck}, \& {Black}}]{Schoeier_2005}
{Sch{\"o}ier}, F.~L., {van der Tak}, F.~F.~S., {van Dishoeck}, E.~F., \&
  {Black}, J.~H. 2005, \aap, 432, 369, \dodoi{10.1051/0004-6361:20041729}

\bibitem[{{Seo} {et~al.}(2015){Seo}, {Shirley}, {Goldsmith}, {Ward-Thompson},
  {Kirk}, {Schmalzl}, {Lee}, {Friesen}, {Langston}, {Masters}, \&
  {Garwood}}]{Seo_2015}
{Seo}, Y.~M., {Shirley}, Y.~L., {Goldsmith}, P., {et~al.} 2015, \apj, 805, 185,
  \dodoi{10.1088/0004-637X/805/2/185}

\bibitem[{{Shetty} {et~al.}(2011){Shetty}, {Glover}, {Dullemond}, \&
  {Klessen}}]{Shetty_2011}
{Shetty}, R., {Glover}, S.~C., {Dullemond}, C.~P., \& {Klessen}, R.~S. 2011,
  \mnras, 412, 1686, \dodoi{10.1111/j.1365-2966.2010.18005.x}

\bibitem[{{Shetty} {et~al.}(2009{\natexlab{a}}){Shetty}, {Kauffmann}, {Schnee},
  \& {Goodman}}]{Shetty_2009a}
{Shetty}, R., {Kauffmann}, J., {Schnee}, S., \& {Goodman}, A.~A.
  2009{\natexlab{a}}, \apj, 696, 676, \dodoi{10.1088/0004-637X/696/1/676}

\bibitem[{{Shetty} {et~al.}(2009{\natexlab{b}}){Shetty}, {Kauffmann}, {Schnee},
  {Goodman}, \& {Ercolano}}]{Shetty_2009b}
{Shetty}, R., {Kauffmann}, J., {Schnee}, S., {Goodman}, A.~A., \& {Ercolano},
  B. 2009{\natexlab{b}}, \apj, 696, 2234, \dodoi{10.1088/0004-637X/696/2/2234}

\bibitem[{{Shu}(1977)}]{Shu_1977}
{Shu}, F.~H. 1977, \apj, 214, 488, \dodoi{10.1086/155274}

\bibitem[{{Spitzer}(1968)}]{Spitzer_1968}
{Spitzer}, Jr., L. 1968, {Dynamics of Interstellar Matter and the Formation of
  Stars}, ed. B.~M. {Middlehurst} \& L.~H. {Aller} (the University of Chicago
  Press), 1

\bibitem[{{Stanke} {et~al.}(2006){Stanke}, {Smith}, {Gredel}, \&
  {Khanzadyan}}]{Stanke_2006}
{Stanke}, T., {Smith}, M.~D., {Gredel}, R., \& {Khanzadyan}, T. 2006, \aap,
  447, 609, \dodoi{10.1051/0004-6361:20041331}

\bibitem[{{Straizys} {et~al.}(1992){Straizys}, {Cernis}, {Kazlauskas}, \&
  {Meistas}}]{Straizys_1992}
{Straizys}, V., {Cernis}, K., {Kazlauskas}, A., \& {Meistas}, E. 1992, Baltic
  Astronomy, 1, 149, \dodoi{10.1515/astro-1992-0203}

\bibitem[{{Tafalla} \& {Hacar}(2015)}]{Tafalla_2015}
{Tafalla}, M., \& {Hacar}, A. 2015, \aap, 574, A104,
  \dodoi{10.1051/0004-6361/201424576}

\bibitem[{{Tafalla} {et~al.}(2004){Tafalla}, {Myers}, {Caselli}, \&
  {Walmsley}}]{Tafalla_2004}
{Tafalla}, M., {Myers}, P.~C., {Caselli}, P., \& {Walmsley}, C.~M. 2004, \aap,
  416, 191, \dodoi{10.1051/0004-6361:20031704}

\bibitem[{{Tan} {et~al.}(2006){Tan}, {Krumholz}, \& {McKee}}]{Tan_2006}
{Tan}, J.~C., {Krumholz}, M.~R., \& {McKee}, C.~F. 2006, \apjl, 641, L121,
  \dodoi{10.1086/504150}

\bibitem[{{Ward-Thompson} {et~al.}(2006){Ward-Thompson}, {Nutter}, {Bontemps},
  {Whitworth}, \& {Attwood}}]{WardThompson_2006}
{Ward-Thompson}, D., {Nutter}, D., {Bontemps}, S., {Whitworth}, A., \&
  {Attwood}, R. 2006, \mnras, 369, 1201,
  \dodoi{10.1111/j.1365-2966.2006.10356.x}

\bibitem[{{Ward-Thompson} {et~al.}(2007){Ward-Thompson}, {Di Francesco},
  {Hatchell}, {Hogerheijde}, {Nutter}, {Bastien}, {Basu}, {Bonnell}, {Bowey},
  {Brunt}, {Buckle}, {Butner}, {Cavanagh}, {Chrysostomou}, {Curtis}, {Davis},
  {Dent}, {van Dishoeck}, {Edmunds}, {Fich}, {Fiege}, {Fissel}, {Friberg},
  {Friesen}, {Frieswijk}, {Fuller}, {Gosling}, {Graves}, {Greaves}, {Helmich},
  {Hills}, {Holland}, {Houde}, {Jayawardhana}, {Johnstone}, {Joncas}, {Kirk},
  {Kirk}, {Knee}, {Matthews}, {Matthews}, {Matzner}, {Moriarty-Schieven},
  {Naylor}, {Padman}, {Plume}, {Rawlings}, {Redman}, {Reid}, {Richer},
  {Shipman}, {Simpson}, {Spaans}, {Stamatellos}, {Tsamis}, {Viti}, {Weferling},
  {White}, {Whitworth}, {Wouterloot}, {Yates}, \& {Zhu}}]{WardThompson_2007}
{Ward-Thompson}, D., {Di Francesco}, J., {Hatchell}, J., {et~al.} 2007, \pasp,
  119, 855, \dodoi{10.1086/521277}

\bibitem[{{Wood} {et~al.}(1994){Wood}, {Myers}, \& {Daugherty}}]{Wood_1994}
{Wood}, D. O.~S., {Myers}, P.~C., \& {Daugherty}, D.~A. 1994, The Astrophysical
  Journal Supplement Series, 95, 457, \dodoi{10.1086/192107}

\bibitem[{{Young} {et~al.}(2006){Young}, {Enoch}, {Evans}, {Glenn}, {Sargent},
  {Huard}, {Aguirre}, {Golwala}, {Haig}, {Harvey}, {Laurent}, {Mauskopf}, \&
  {Sayers}}]{Young_2006}
{Young}, K.~E., {Enoch}, M.~L., {Evans}, Neal~J., I., {et~al.} 2006, \apj, 644,
  326, \dodoi{10.1086/503327}

\bibitem[{{Zdanavi{\v c}ius} {et~al.}(2011){Zdanavi{\v c}ius}, {Maskoli{\=
  u}nas}, {Zdanavi{\v c}ius}, {Strai{\v z}ys}, \&
  {Kazlauskas}}]{Zdanavicius_2011}
{Zdanavi{\v c}ius}, K., {Maskoli{\= u}nas}, M., {Zdanavi{\v c}ius}, J.,
  {Strai{\v z}ys}, V., \& {Kazlauskas}, A. 2011, Baltic Astronomy, 20, 317

\end{thebibliography}

\end{document}